\documentclass[10pt,a4paper]{article}
\usepackage[text={145mm,223mm},centering]{geometry}
\usepackage{bbm}
\usepackage[final]{graphicx}
\usepackage{amsmath,amscd}
\usepackage{amsfonts,amsbsy}
\usepackage{amssymb}
\usepackage{xcolor}

\def\empile#1\over#2{\mathrel{\mathop{\kern 0pt#1}\limits_{#2}}}
\def\bs{\boldsymbol}

\def\ka{\kappa}
\def\oka{\overline{\kappa}}

\newcommand{\slv}{\raise.15ex\hbox{$/$}\kern-.53em\hbox{$v$}}
\newcommand{\slF}{\raise.15ex\hbox{$/$}\kern-.53em\hbox{$F$}}
\newcommand{\slL}{\raise.15ex\hbox{$/$}\kern-.53em\hbox{$L$}}
\newcommand{\slP}{\raise.15ex\hbox{$/$}\kern-.53em\hbox{$P$}}
\newcommand{\slp}{\raise.15ex\hbox{$/$}\kern-.53em\hbox{$p$}}
\newcommand{\slq}{\raise.15ex\hbox{$/$}\kern-.53em\hbox{$q$}}
\newcommand{\slR}{\raise.15ex\hbox{$/$}\kern-.53em\hbox{$R$}}
\newcommand{\slQ}{\raise.15ex\hbox{$/$}\kern-.53em\hbox{$Q$}}
\newcommand{\slK}{\raise.15ex\hbox{$/$}\kern-.53em\hbox{$K$}}
\newcommand{\slk}{\raise.15ex\hbox{$/$}\kern-.53em\hbox{$k$}}
\newcommand{\slD}{\raise.15ex\hbox{$/$}\kern-.73em\hbox{$D$}}
\newcommand{\slC}{\raise.15ex\hbox{$/$}\kern-.53em\hbox{$C$}}
\newcommand{\slA}{\raise.15ex\hbox{$/$}\kern-.53em\hbox{$A$}}
\newcommand{\slSigma}{\raise.15ex\hbox{$/$}\kern-.53em\hbox{$\Sigma$}}
\newcommand{\slpartial}{\raise.15ex\hbox{$/$}\kern-.53em\hbox{$\partial$}}
\newcommand{\slcalP}{\raise.15ex\hbox{$/$}\kern-.63em\hbox{$\cal P$}}

\def\a{{\boldsymbol a}}

\def\b{{\boldsymbol b}}
\def\p{{\boldsymbol p}}
\def\q{{\boldsymbol q}}

\def\k{{\boldsymbol k}}

\def\x{{\boldsymbol x}}

\def\r{{\boldsymbol r}}

\def\v{{\boldsymbol v}}

\def\b{{\boldsymbol b}}

\def\wh#1{{\widehat{#1}}}

\def\sbk#1{{\big[#1\big]}}
\def\abk#1{{\big<#1\big>}}

\catcode`\@=11

\newcount\@tempcntc
\def\@citex[#1]#2{\if@filesw\immediate\write\@auxout{\string\citation{#2}}\fi
  \@tempcnta\z@\@tempcntb\m@ne\def\@citea{}\@cite{%
        \@for\@citeb:=#2\do%
    {\@ifundefined{b@\@citeb}%
        {\@citeo\@tempcntb\m@ne\@citea%
                \def\@citea{,\penalty\@m\ }{\bf ?}\@warning%
                {Citation `\@citeb' on page \thepage \space undefined}}%
        {\setbox\z@\hbox{\global\@tempcntc0\csname b@\@citeb\endcsname\relax}
     \ifnum\@tempcntc=\z@ \@citeo\@tempcntb\m@ne%
       \@citea\def\@citea{,\penalty\@m}%
       \hbox{\csname b@\@citeb\endcsname}%
     \else%
      \advance\@tempcntb\@ne%
      \ifnum\@tempcntb=\@tempcntc%
      \else\advance\@tempcntb\m@ne\@citeo%
      \@tempcnta\@tempcntc\@tempcntb\@tempcntc\fi\fi}}\@citeo}{#1}}%

\def\@citeo{\ifnum\@tempcnta>\@tempcntb\else\@citea
  \def\@citea{,\penalty\@m}%
  \ifnum\@tempcnta=\@tempcntb\the\@tempcnta\else
   {\advance\@tempcnta\@ne\ifnum\@tempcnta=\@tempcntb \else
\def\@citea{--}\fi
    \advance\@tempcnta\m@ne\the\@tempcnta\@citea\the\@tempcntb}\fi\fi}

\catcode`\@=12

\begin{document}

\title{\bf Gluon production in the Color Glass Condensate from BCFW recursion} \author{\large{\textbf{\sc Fran\c cois Gelis}}\\{\ }\\
Institut de Physique Théorique\\
Université Paris-Saclay, CEA-CNRS\\
91191 Gif sur Yvette, France} \maketitle

\begin{abstract}
  In the Color Glass Condensate framework, the colliding projectiles
  are described as classical color currents. Gluon production at
  leading order in the coupling is obtained from the retarded solution
  of the classical Yang-Mills equations, with these currents acting as
  sources. However, while the final gluon spectrum is gauge invariant,
  the classical color field $A^\mu$ from which it is obtained is gauge dependent. This makes
  the intermediate steps of this approach unnecessarily complicated,
  as some effort is spent to also determine the gauge dependent part
  of $A^\mu$, which is then discarded when one calculates a gauge
  invariant observable.

  In this work, we use a generalization of BCFW recursion applicable to off-shell gluons in order to
  calculate directly the gauge independent gluon production amplitude. This approach allows all manipulations to be performed in terms of gauge invariant amplitudes instead of gauge dependent Feynman diagrams, and therefore provides a gauge agnostic way to obtain the result.
\end{abstract}

\newpage
\tableofcontents

\section{Introduction}
A crucial property of the strong nuclear force is asymptotic freedom, i.e., the fact that it becomes weak at short distances. The counterpart of that is that its strength increases at large distance. A consequence of this is confinement; the quarks and gluons cannot exist in isolation, but aggregate into color singlet bound states, for which the strong nuclear force is screened at large distance. This property complicates the use of quantum chromodynamics, since the elementary degrees of freedom  of QCD (quarks and gluons) differ from the asymptotic states that exist in nature (hadronic bound states). {\sl Factorization theorems} \cite{Collins:1989gx,Catani:1990eg,Collins:1985ue,Bauer:2001yt} show, under certain assumptions, that the information about the parton content of hadrons can be encapsulated into universal distributions that do not depend under the reaction process under consideration.

In their simplest form, these distributions provide the single parton density as a function of the fraction $x$ of the longitudinal momentum of the hadron, and of a transverse resolution scale $Q$. Because these functions are inclusive single parton distributions, they do not contain information about multi-parton correlations inside the hadron. This information plays no role as  long as the parton density in hadrons stays rather low, but becomes important when the density grows. In particular, it is known that the parton density grows with the energy of the hadron (equivalently, if the hadron  energy is held fixed, it increases as we lower the momentum fraction $x$), due to the enlarged phase-space for gluon radiation by splitting. Eventually, the density reaches a large enough value where the recombination of partons becomes equally likely as their splitting. In this regime, known as {\sl saturation} \cite{Gribov:1983ivg,Mueller:1985wy}, a sort of equilibrium between mergings and splittings sets in, and the growth of the parton density is tamed.

In the saturation regime, the ordinary factorization framework is no longer applicable, because the large parton density makes multi-parton processes important. In order to calculate reactions when this happens, we need a framework that provides more information about the distribution of multiple partons in the hadron. The Color Glass Condensate \cite{Iancu:2000hn,Ferreiro:2001qy,Iancu:2003xm,Weigert:2005us,Gelis:2010nm} was designed for this purpose. In the CGC framework, the hadron  content is described as a collection of color charges moving collectively in the direction of the hadron \cite{McLerran:1993ni,McLerran:1993ka,McLerran:1994vd}. These charges have a spatial density $\rho_a(x)$, and their motion creates a color current in the direction of motion $v^\mu$ of the hadron, $J^\mu_a\equiv \rho_a v^\mu$. The CGC does not give $\rho_a$ itself, but its probability distribution, i.e., a density functional $W[\rho]$. There is an analogue of the factorization theorems valid in the low density regime: the calculation of a loop correction in the CGC framework generically produces logarithms of the energy; these logarithms are the same for all inclusive quantities, and can be resummed into {\sl energy dependent} distributions $W[\rho]$ \cite{Gelis:2008rw,Gelis:2008ad,Gelis:2008sz}. This energy dependence is driven by the JIMWLK equation.

When we study the collision of two hadrons (or nuclei) in the CGC framework, there are two charge densities, $\rho_{1a}$ and $\rho_{2a}$, and two directions, $v_1^\mu$ and $v_2^\mu$. The color current is the sum $J^\mu_a= \rho_{1a}v_1^\mu + \rho_{2a}v_2^\mu$. At tree level, observables can be expressed in terms of the color field $A^\mu$ that solves the classical Yang-Mills equations \cite{Gelis:2006yv,Gelis:2006cr}, $[D_\mu,F^{\mu\nu}+J^\nu=0]$ (with the additional constraint that the current must be conserved, $[D_\mu,J^\mu]=0$). The Yang-Mills equations are non-linear in $A^\mu$, which makes them particularly difficult to solve. Two main approaches have been used in the literature. A first approach is to formulate the general non-linear problem on a lattice representation of space and then to solve for the time evolution numerically \cite{Krasnitz:1999wc,Krasnitz:1998ns,Krasnitz:2000gz,Krasnitz:2001qu,Krasnitz:2002mn,Lappi:2003bi}. Analytic solutions have not been found for this general case, but have been obtained when one of the two densities is weak (e.g., $\rho_1\ll\rho_2$). In this situation, one is looking for the solution at order one in $\rho_1$ and to all orders in $\rho_2$ \cite{Dumitru:2001ux,Blaizot:2004wu,Gelis:2005pt}. 

It is crucial to note that $A^\mu$ and $J^\mu$ are both gauge dependent. This gauge dependence enters in the definition of the initial condition\footnote{In particular, an empty initial state does not necessarily mean that $A^\mu=0$; the color field could be any pure gauge.} (i.e., the data provided at a time long before the collision happens) and in the subsequent evolution, because one also has a certain gauge freedom at every step of the time evolution. The Yang-Mills equations in the dilute-dense approximation have been solved in several gauges (Fock-Schwinger gauge, Landau Gauge, light-cone gauge), leading to solutions that are very different looking, and the complexity of the derivation also depends a lot on the gauge.

Nevertheless, when these seemingly different solutions are used to calculate an observable such as the gluon yield, the results are the same, as expected \cite{Blaizot:2004wu,Gelis:2005pt}. This means that, when solving the Yang-Mills equations, a considerable effort is put into obtaining gauge dependent terms that eventually do not contribute to physical quantities, and it could be very beneficial to have a way to get directly these gauge invariant physical quantities. A similar problem arises in the calculation of scattering amplitudes. In the traditional perturbation theory, scattering amplitudes are obtained by applying LSZ reduction formulas to sums of Feynman graphs. These graphs are usually gauge dependent until their external legs are put on-shell and contracted with physical polarization vectors. It was realized that the physical scattering amplitudes can be considerably simpler than the Feynman graphs they are obtained from \cite{Dixon:1996wi,Elvang:2013cua}. In particular, the Parke-Taylor formula \cite{Parke:1986gb} is a one-term expression for MHV amplitudes (amplitudes with two helicities of one kind, and all other helicities of the other kind) with an arbitrary number of external gluons. Initially a conjecture, the Parke-Taylor formula was first proven by a recursion on Feynman graphs (the Berends-Giele recursion \cite{Berends:1987me}), and eventually a proof was obtained by a recursion that involves only on-shell scattering amplitudes (the BCFW recursion \cite{Britto:2005fq}). This realization paved the way to {\sl on-shell methods}, that allow to calculate any tree level gauge theory amplitude without Feynman graphs, in an approach where amplitudes are obtained recursively from amplitudes with fewer external legs. The BCFW recursion starts with the $3$-gluon amplitudes. They can be obtained from the Feynman rule for the $3$-gluon vertex, but even that is not necessary: their momentum dependence is completely constrained by {\sl little group scaling} \cite{Elvang:2013cua} and the Yang-Mills Lagrangian is only needed to find the numerical prefactor in the amplitude. Once the $3$-point amplitudes are known, all other tree amplitudes are obtained by the BCFW recursion.

In the CGC at tree level, one also has tree graphs with only gluons. Unfortunately, the gluons attached to the densities $\rho_{1,2}$ are {\sl off-shell}, which seems to prevent the application of the BCFW method. The situation is further complicated by the fact that the densities themselves are dynamical: they evolve depending on the color field, in such a way that color conservation $[D_\mu,J^\mu]=0$ is satisfied.
Extensions to the usual BCFW method have been proposed to cope with these complications \cite{vanHameren:2014iua,vanHameren:2012if,Bury:2017jxo,Bury:2015dla,vanHameren:2015bba,Kutak:2016goj}. In this paper, our goal is to apply these extensions in order to calculate gluon production in the dilute-dense approximation of the CGC. The final results will obviously not be new, but the method we employ is new in this context: by using BCFW recursion, we provide a gauge-agnostic derivation of this quantity that completely bypasses the use of Feynman diagrams and manipulates only gauge invariant objects. Moreover, even though we will spend a lot of time explaining the method, its application to gluon production is quite simple and lean, and arguably competes favorably with the traditional way of solving the Yang-Mills equations, even in light-cone gauge where it is the simplest (with the added benefit of not having to fix the gauge, so that in a sense our calculation can be viewed as a proof of gauge invariance).

The paper is organized as follows.  Section \ref{sec:YMsol} discusses the classical solutions of Yang-Mills equations, with emphasis on their gauge dependence and on their diagrammatic representation. In Section \ref{sec:spectrum}, we express the gluon production amplitude in terms of solutions of the Yang-Mills equations. In particular, we discuss the appropriate way of formulating the LSZ reduction formula in order to obtain a gauge invariant result. Section \ref{sec:onshell} exposes the tools we use in the rest of the paper: the color decomposition of tree amplitudes, the spinor-helicity formalism to represent on-shell momenta and its extension to express the off-shell momenta attached to gluon densities, the BCFW recursion and the extensions needed to deal with the kind of amplitudes encountered in the CGC.  In Section \ref{sec:3point}, we review the $3$-point amplitudes that will be used as starting points of the recursion. This includes the well-known on-shell amplitudes given by the Parke-Taylor formula, but also amplitudes with one or two off-shell gluons. Before calculating gluon production, we study first a simpler problem to familiarize ourselves with the BCFW approach: we calculate in Section \ref{sec:scatt} the scattering amplitude of a gluon off the color field produced by sources $\rho_2$ (to all orders in the source). Gluon production by sources $\rho_1$ and $\rho_2$, in the dilute-dense approximation, is calculated in Section \ref{sec:prod}.
Some more technical material is relegated into several appendices.

\section{Solutions of classical Yang-Mills equations}
\label{sec:YMsol}
\subsection{Equations of motion}
The classical Yang-Mills equations read
\begin{align}
  \big(D_\mu\big)_{ab} F^{\mu\nu}_b +J^\nu_a =0,
  \label{eq:ym}
\end{align}
where $J^\nu_a$ is a color current (or a sum of two currents when
there are two projectiles) that acts as a source. This current
implicitly depends on the gauge potential through a conservation
equation:
\begin{align}
  \big(D_\mu\big)_{ab} J^\mu_b =0.
  \label{eq:cons}
\end{align}
Besides these equations, boundary conditions are necessary in order to
uniquely determine the gauge potential $A^\mu_a$ that solves the
problem.

\subsection{Initial condition}
When describing the collision of two projectiles, the initial state of
the system only contains the color charges of the colliding objects,
i.e., only the current $J^\nu_a$. Thus, the initial condition for the
color field is the null field $A^\mu=0$, or a pure gauge
$A^\mu=\tfrac{i}{g}\Omega^\dagger\partial^\mu\Omega$ if one allows for
an arbitrary gauge choice at the initial time.

The current conservation (\ref{eq:cons}) describes how the color
current evolves as the projectiles propagate in the presence of color
fields. One generally makes an ansatz according to
which the current $J^\mu$ stays proportional to a vector $v^\mu$ that
indicates the direction of propagation of the projectile\footnote{In a
collision of two projectiles, $J^\mu$ is a sum of two currents
$J^\mu_1$ and $J^\mu_2$, that stay proportional to $v_1^\mu$ and
$v_2^\mu$, respectively.},
\begin{align}
J^{\mu}_a(x) \equiv v^\mu \rho_a(x),
\end{align}
where $\rho_a$ is a density of color charges. The vector $v^\mu$ must be
time-like or light-like. The implicit assumption behind this ansatz is
that the projectiles do not recoil during the collision, i.e., they
have a very large energy compared to the typical momentum transferred
in the collision. With this ansatz,
current conservation reads
\begin{align}
  \big(v\cdot D\big)_{ab} \rho_b =0.
  \label{eq:cons-eik}
\end{align}

This equation must be endowed with some input that specifies the
charge density when the color fields are zero.  When $A^\mu=0$,
eq.~(\ref{eq:cons-eik}) simplifies into
\begin{align}
  (v\cdot\partial)\rho_a^{(0)} =0.
  \label{eq:cons-vac}
\end{align}
(The superscript $(0)$ indicates that $\rho_a^{(0)}$ is the charge
density at zeroth order in $A^\mu$.)  Thus, current conservation
imposes that $\rho_a^{(0)}(x)$ does not depend on the coordinate along
$v^\mu$ (the coordinate which is Fourier conjugate to the momentum
component $v\cdot p$), and leaves totally unconstrained its dependence
on the remaining three coordinates\footnote{Equivalently, the Fourier
transform of $\rho_a^{(0)}(x)$, $\rho_a^{(0)}(p)$,  is proportional to
$\delta(v\cdot p)$.}. More precisely, define $x_v$ the coordinate such
that\footnote{For a static source, we have $v^0=1, v^{1,2,3}=0$ and
$x_v = x^0$. For a source moving along the light-cone, we have
$v^+=1,v^{-,1,2}=0$ and $v\cdot\partial = \partial^- =
\tfrac{\partial}{\partial x^+}$; hence $x_v=x^+$.}
\begin{align}
  v\cdot\partial \equiv \frac{\partial\ }{\partial x_v},\qquad (v\cdot\partial) x_v =1.
\end{align}
The remaining three coordinates, that we denote $\vec{x}$, satisfy
$(v\cdot \partial)\vec{x}=0$ ($\vec{x}$ may be viewed as a set of
``spatial'' coordinates, while $x_v$ is a temporal coordinate). In
terms of these coordinates, current conservation simply implies the
independence of the source with respect to the coordinate $x_v$,
\begin{align}
  \rho_a^{(0)}(x_v,\vec{x}) = \rho_a^{(0)}(\vec{x}),
\end{align}
and we need an initial condition that specifies the function
$\rho_a^{(0)}(\vec{x})$. Note that if we set the initial conditions on
a surface $x_v=\mbox{const}$, we must provide the values of $A^\mu$
and $(v\cdot\partial)A^\mu$ (because the Yang-Mills equations contain
second derivatives).

\subsection{Gauge dependence}
The objects we have introduced are not gauge invariant. Under a gauge
transformation $\Omega(x)$, they transform as
\begin{align}
  \rho  &\quad\to\quad \Omega^\dagger \rho\, \Omega,\nonumber\\
  A^\mu&\quad\to\quad \Omega^\dagger A^\mu \Omega +\frac{i}{g}\Omega^\dagger\partial^\mu \Omega,\nonumber\\
  F^{\mu\nu} &\quad\to\quad \Omega^\dagger F^{\mu\nu} \Omega,
\end{align}
where we denote $A^\mu\equiv A^\mu_a t^a$,$F^{\mu\nu}\equiv
F^{\mu\nu}_a t^a, \rho\equiv \rho_a t^a$, with $t^a$ the generators of
the ${\mathfrak{su}}(N)$ algebra.

Since the problem we are considering is an initial value problem in
which we specify data on the surface $x_v=-\infty$, which then evolves
according to the equations of motion (\ref{eq:ym}) and (\ref{eq:cons})
(this is not entirely exact; as we shall see shortly, an additional input is necessary to define uniquely the solution), let us consider
first the effect of gauge transforming the initial data with an
$\Omega_{\rm ini}(\vec{x})$ that depends only on the ``spatial'' coordinates
(such a ``time independent'' gauge transformation obeys $(n\cdot
\partial)\Omega_{\rm ini}=0$.). The initial data is transformed according to
\begin{align}
  \rho  &\quad\to\quad \Omega_{\rm ini}^\dagger \rho\, \Omega_{\rm ini},\nonumber\\
  A^\mu&\quad\to\quad \Omega_{\rm ini}^\dagger A^\mu \Omega_{\rm ini} +\frac{i}{g}\Omega_{\rm ini}^\dagger\partial^\mu \Omega_{\rm ini},\nonumber\\
  (n\cdot\partial)A^\mu&\quad\to\quad \Omega_{\rm ini}^\dagger \big((n\cdot\partial)A^\mu\big) \Omega_{\rm ini}.
\end{align}
One can then verify that the equations of motion extend this time
independent gauge transformation to the sources and fields at all
subsequent values of $x_v$. In other words, a time independent gauge
transformation applied to the initial data extends to a global time
independent transformation applied in all space-time.

Let us now consider in more detail the evolution of $\rho$ and
$A^\mu$ during an infinitesimal timestep. In this discussion, we use
$x^0$ as the time variable, but the same discussion could be done with
any time-like coordinate, such as the $x_v$ introduced earlier. By
separating temporal and spatial components, the Yang-Mills equations
read
\begin{align}
  \big[D_i,F^{0i}\big]=J^0,\quad
  \partial_0 F^{0i}-ig\big[A_0,F^{0i}\big] +[D_j,F^{ji}\big]+J^i=0.
\end{align}
Note that $F^{0i}$ is the canonical momentum conjugate of $A^i$.  The
first of these equations is Gauss law, a constraint on $F^{0i}$ which
is automatically fulfilled at all times provided it is true in the
initial condition.  Using the second of these equations, the evolution
of $F^{0i}$ during a timestep $dx^0$ reads
\begin{align}
  F^{0i}\quad\to\quad
  F^{0i}-dx^0\big(\big[D_j,F^{ji}\big] +J^i\big)+\underline{ig dx^0\big[A_0,F^{0i}\big]}.
  \label{eq:evol-F}
\end{align}
From the definition of $F^{0i}$, we get the time derivative of $A^i$,
\begin{align}
  \partial_0 A^i
  =
  F^{0i} +\partial^i A^0 +ig\big[A^0,A^i\big],
\end{align}
which implies the following evolution for $A^i$,
\begin{align}
  A^i\quad\to\quad
  A^i+dx^0 F^{0i} +\underline{dx^0\big(\partial^i A^0 -ig\big[A^i,A^0\big]\big)}.
  \label{eq:evol-A}
\end{align}
Note that eqs.~(\ref{eq:evol-F}) and (\ref{eq:evol-A}) contain the
field $A^0$ (in the underlined terms), while there is no dynamical equation telling how $A^0$
evolves. This feature is due to the fact that we are trying to solve
the Yang-Mills equations without gauge fixing: the undetermined $A^0$
reflects the fact that, at each time step of the evolution, we are
free to choose the gauge in that particular time-slice. Indeed, one
may check that the $A^0$ dependence in eqs.~(\ref{eq:evol-F}) and
(\ref{eq:evol-A}) is equivalent to applying an infinitesimal gauge
transformation
\begin{align}
  \omega\equiv \exp(ig \theta),\quad\mbox{with\ }\theta=-dx^0 A^0
\end{align}
to $A^i$ and $F^{0i}$. The undetermined $A^0$ also affects the
evolution of the source $\rho$ by the same gauge transformation.  When
combining the effects of the $A^0$'s in all the successive
time-slices, we obtain a global time-dependent gauge transformation,
\begin{align}
  \omega^\dagger(x)\equiv {\rm T}\,\exp\Big(ig\int_{-\infty}^{x^0}dy^0\,A^0(y^0,\vec{x})\Big),
\end{align}
that should be combined with the gauge transformation
$\Omega_{\rm ini}(\vec{x})$ applied to the initial condition. The overall gauge
transformation allowed on the solutions of Yang-Mills equations
therefore reads
\begin{align}
  \Omega^\dagger(x^0,\vec{x})
  =
  \Bigg[{\rm T}\,\exp\Big(ig\int_{-\infty}^{x^0}dy^0\,A^0(y^0,\vec{x})\Big)\Bigg]\Omega_{\rm ini}^\dagger(\vec{x}),
  \label{eq:GT}
\end{align}
resulting from the freedom to choose the gauge in which the initial
condition is given and the $A^0$'s at each step of the time evolution.

\subsection{Diagrammatic expansion of the solution}
Until now, we have discussed the solution of Yang-Mills equations and
its use in constructing the gluon production amplitude without
specifying at all a gauge. We have also pointed out that the initial
gauge field should be a pure gauge, so that the gluon content of the
system is initially empty.

Among the set of possible initial gauge transformations, $\Omega_{\rm
  ini}(\vec{x})$, the gauge such that the initial gauge field is zero
stands out. Indeed, with this initial gauge, an initially null gauge
field is sourced by the color currents, and its Fourier modes at the
final time give the gluon production amplitude.

In this section, we discuss the diagrammatic expansion of the solution
of Yang-Mills equations in this particular gauge where the initial
condition of the field is zero. The solution in this gauge can be
represented as a sum of tree diagrams. The ``root'' of the tree is the
point $x$ at which the gauge potential is evaluated, while each
``leaf'' is attached to an instance of the color current $J^\nu_a$:
\setbox1\hbox to 4cm{\includegraphics[width=40mm]{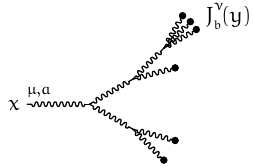}}
\begin{equation}
  A^\mu_a(x)
  =
  \sum_{{\mbox{\scriptsize trees}}}
  \raise -9mm\box1.
\end{equation}
In this diagrammatic representation, the wavy lines are gluon
propagators and the black dots represent the current $J^\nu_a$ (in
this representation, the current is ``dressed'' by the gauge
potential, i.e., it is the solution of the conservation equation
(\ref{eq:cons})).  The rules for the vertices and lines of these trees
are the usual Feynman rules of Yang-Mills theory, except that the
propagators should be retarded. Note that one must fix the gauge for
the gluon propagator to be uniquely defined, which is equivalent to
the necessity of choosing an $A^0$ at each time-step when solving
directly the Yang-Mills equations.

Now, write $J^\mu_a(x)=v^\mu\rho_a(x)$. The color density $\rho_a$,
that obeys $(v\cdot D)_{ab}\rho_b=0$, can be organized as a series in
powers of the gauge field,
\begin{align}
  \rho_a(x)\equiv
  \sum_{n=0}^\infty
  \rho_a^{(n)}(x),\quad\mbox{with\ }
  \rho_a^{(n)}\sim {\cal O}(A^n).
\end{align}
When solving the conservation equation, the zeroth order density
$\rho_a^{(0)}$ is used as input from which all higher orders are
derived, according to the following recursion
\begin{align}
  (v\cdot\partial) \rho_a^{(n)}
  =
  ig (v\cdot A)_{ab}\, \rho_b^{(n-1)}.
  \label{eq:j-recur}
\end{align}
The only constraint on the zeroth order reads
\begin{equation}
  (v\cdot\partial) \rho_a^{(0)} =0,
\end{equation}
meaning that $\rho_a^{(0)}$ does not depend on the coordinate along the
direction set by $v^\mu$ (the physical interpretation of this
condition is that, in the absence of external fields, the projectile
is invariant along its direction of motion). One may solve
eq.~(\ref{eq:j-recur}) in momentum space, where we have
\begin{align}
  \rho_a^{(n)}(p)
  =
  -g f^{abc}
  \int \frac{d^4q}{(2\pi)^4}\, \frac{v\cdot A_c(q)}{v\cdot p}\,\rho_b^{(n-1)}(p-q),
\end{align}
which can easily be expressed in terms of $\rho_a^{(0)}$ by induction.

Diagrammatically, one may represent this solution as follows,
\setbox1\hbox to 6cm{\includegraphics[width=60mm]{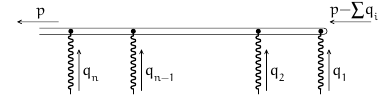}}
\begin{equation}
  \rho_a(p)
  =
\sum_{n=0}^\infty
  \raise -9.5mm\box1.
\end{equation}
(Momentum conservation applies at the vertices in this
representation.)  The various building blocks that appear in this
diagrammatic representation correspond to the following extra Feynman rules:
\setbox1\hbox to  6mm{\includegraphics[width= 6mm]{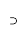}}
\setbox2\hbox to 15mm{\includegraphics[width=15mm]{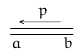}}
\setbox3\hbox to 15mm{\includegraphics[width=15mm]{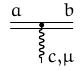}}
\begin{equation}
  \raise -4.1mm\box1 = \rho^{(0)},\quad
  \raise -3.2mm\box2 = \frac{\delta_{ab}}{v\cdot p},\quad
  \raise -7.2mm\box3 = -g f^{abc} v^\mu.
\end{equation}
The double lines represent gluon propagators in the eikonal
approximation, and the vertex connecting a gluon to a double line is
the corresponding eikonal vertex (in particular, its Lorentz structure
depends only on the direction of motion $v^\mu$ of  the fast
moving charge).

In the following, we may refer to $\rho^{(0)}$ as a ``source'', since
it is this object that seeds the full series of terms in the density
$\rho$ ($\rho^{(0)}$ may be viewed as the color charge density of a
projectile long before the collision, before it enters in regions of
spacetime where the color field is non-zero). Note that in momentum
space, the source of a projectile moving in the direction $v^\mu$ is
proportional to
\begin{equation}
  \rho_a^{(0)}(p)\propto \delta(v\cdot p).
\end{equation}
(This property is in general not true for the $\rho_a^{(n)}(p)$ with $n>0$.)
By combining all these rules, we can construct the solution of
Yang-Mills equations in terms of the sources $\rho^{(0)}$ (note that
there is a different $\rho^{(0)}$ for each projectile in a collision --
we will distinguish them diagrammatically by using different shades of
gray and different orientations) as a sum of all the tree graphs such
as \setbox1\hbox to 70mm{\includegraphics[width=70mm]{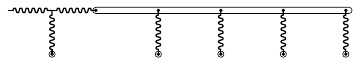}}
\begin{equation}
  A^\mu_a
  =
  \sum_{{\mbox{\scriptsize trees}}}
  \raise -9mm\box1.
  \label{eq:trees}
\end{equation}

When applying the LSZ formula to this gauge potential, we obtain the
gluon production amplitude ${\cal M}_{\p\lambda a}$, which we may write
in terms of the sources as
\begin{align}
  {\cal M}_{\p\lambda a}\equiv
  \sum_{n=2}^\infty
  \int\Bigg[\prod_{j=1}^n \frac{d^4p_j}{(2\pi)^4}\rho_{_{b_j}}(p_j)\Bigg]\,
  (2\pi)^4\delta(p+\sum_{j=1}^n p_j)\,
  {\cal M}_{1+n}^{a b_1\cdots b_n}(0^\lambda 1^* 2^*\cdots n^*).
\end{align}
In this representation, ${\cal M}_{1+n}^{a b_1\cdots b_n}$ is an
$n+1$-gluon tree amplitude. The produced gluon (labeled $0$) is
on-shell, with momentum $\p$, polarization (or helicity, if we use the
helicity basis) $\lambda$ and color $a$. The $n$ other gluons are
off-shell (which is indicated by a star next to their label) and
attached to sources, with colors $b_1,\cdots,b_n$.

\section{Gluon spectrum from classical Yang-Mills}
\label{sec:spectrum}
\subsection{Naive LSZ formula}
Let us now discuss how a gauge invariant gluon spectrum may be defined
from the gauge dependent solution of the classical Yang-Mills
equations. A naive approach is to apply the usual LSZ reduction
formula to the classical field, in order to define the amplitude to
produce a gluon with momentum $\k$, polarization $\lambda$ and color
$a$,
\begin{align}
  {\cal M}_{\p\lambda a}\equiv i\int d^3\x\,e^{ip\cdot x} \stackrel{\longleftrightarrow}{\partial_0} A^\mu_a(x) \epsilon_{\lambda,\mu}^*(\p).
\label{eq:LSZ-naive}
\end{align}
Let us assume now that the classical gauge field admits the following
Fourier decomposition,
\begin{align}
  A^\mu_a(x)
  =
  \sum_{\lambda}\int\frac{d^3\k}{(2\pi)^32k}\,
  \Big(
  \alpha_{\k\lambda a}^* \epsilon^{\mu *}_\lambda(\k) e^{ik\cdot x}
  +
  \alpha_{\k\lambda a} \epsilon^{\mu}_\lambda(\k) e^{-ik\cdot x}
  \Big).
  \label{eq:A-modes}
\end{align}
This is the usual mode decomposition of the free field operator, where
we replace the creation and annihilation operators
$a^\dagger_{\k\lambda a}$ and $a_{\k\lambda a}$ by the complex  numbers
$\alpha_{\k\lambda a}^*$ and $\alpha_{\k\lambda a}$.
If we apply (\ref{eq:LSZ-naive}) to (\ref{eq:A-modes}), we obtain
\begin{align}
  {\cal M}_{\p\lambda a}=\alpha_{\p\lambda a},
\end{align}
which indeed appears to extract the desired Fourier component of the
field.

\subsection{Caveats}
There are several caveats with the naive LSZ formula (\ref{eq:LSZ-naive}):

\paragraph{1.} The mode decomposition (\ref{eq:A-modes}), as a sum of plane waves,
implicitly assumes that the non-linear terms in the Yang-Mills
equations are negligible (so that plane waves are indeed solutions),
at least in the region of asymptotically large times $x^0\to
+\infty$. This is not true in all gauges. Indeed, a gauge
transformation $\Omega$ produces a non-homogeneous term
$\tfrac{i}{g}\Omega^\dagger \partial^\mu\Omega$ in $A^\mu$. If $\Omega(x)$ has
rapid variations with $x^\mu$, this term is not small and the
hypothesis of having small non-linear terms is not fulfilled.

However, the extra term $\tfrac{i}{g}\Omega^\dagger\partial^\mu \Omega$ has no
effect inside the LSZ formula (\ref{eq:LSZ-naive}). Indeed, this
formula is designed to extract the residue of the pole at $p^2=0$ in
the Fourier transform of $A^\mu(x)$, but a term in
$\tfrac{i}{g}\Omega^\dagger\partial^\mu \Omega$ generically does not have a pole
at $p^2=0$. Therefore, the LSZ formula we used above does not see the
inhomogeneous term produced in a gauge transformation.

\paragraph{2.} Gauge transformations also affect the linear part (\ref{eq:A-modes})
of the gauge field. If we apply the LSZ formula (\ref{eq:LSZ-naive})
without change, then the extracted coefficients are gauge
dependent. In order to circumvent this problem, note that the formula
(\ref{eq:LSZ-naive}) consists in decomposing the field on the plane
wave basis, i.e., on a basis made of solutions of the linearized
Yang-Mills equations,
\begin{align}
        \big(\square
        g_{\mu\nu}-\partial_\mu\partial_\nu\big)a^\nu_a(x)=0.
\end{align}
This linearization assumes that the gauge field is small. However, if
we apply a gauge transformation $\Omega$ to the system, a null
background $A^\mu=0$ is transformed into the (possibly large) pure
gauge $\tfrac{i}{g}\Omega^\dagger\partial^\mu\Omega$. The 
wave equation for the propagation over this pure gauge background reads
\begin{align}
        \big({\cal D}^2 g_{\mu\nu}-{\cal D}_\mu{\cal D}_\nu\big)_{ab}a^\nu_b(x)=0,
\end{align}
where ${\cal D}^\mu\equiv\partial^\mu+\Omega^\dagger
\partial^\mu\Omega$ is the covariant derivative in the presence of the
background field produced by the gauge transformation. The solutions
of this equation are simply the gauge transformed plane waves. If we
define the LSZ reduction formula in terms of these gauge transformed
plane waves, the resulting amplitude is gauge invariant.

This modification of the LSZ formula may be made explicit by
introducing the following inner product between gauge fields,
\begin{align}
        \big(\alpha\big|\beta\big)\equiv \int\limits_{x^0={\rm const}}d^3\x\;
        \alpha_{\mu a}(x) \stackrel{\longleftrightarrow}{{\cal D}_{0,ab}}\beta^\mu_b(x).
        \label{eq:inner-prod}
\end{align}
Note that here we use the covariant derivative ${\cal D}_0$, as
opposed to the ordinary derivative $\partial_0$ in the naive LSZ
formula. The reason of this modification is to make the inner product
gauge invariant also when we apply a time dependent gauge
transformation.

Consider a gauge transformation
\begin{align}
        \alpha_{\mu a}&\quad\to\quad \Omega^\dagger_{ac} \alpha_{\mu c}=  \alpha_{\mu c}\Omega_{ca},\nonumber\\
        \beta^\mu_b&\quad\to\quad \Omega^\dagger_{bd} \beta^\mu_d.
\end{align}
(In the first line, we have used the fact that $\Omega_{ab}$ in the
adjoint representation is a real valued matrix.) Note that, if
$\alpha$ or $\beta$ is the full gauge field, its gauge transformation
should also contain the non-homogeneous term
$\tfrac{i}{g}\Omega^\dagger \partial^\mu\Omega$. We have not written
this term, because it does not have a pole at $p^2=0$ and therefore
does not contribute to the LSZ formula.

Under this transformation, the inner product (\ref{eq:inner-prod})
transforms as
\begin{align}
        \big(\alpha\big|\beta\big)
        \quad\to\quad
        \int\limits_{x^0={\rm const}}d^3\x\;
        \big(\alpha_{\mu c}(x)\Omega_{ca}(x)\big) \stackrel{\longleftrightarrow}{{\cal D}_{0,ab}}\big(\Omega_{bd}^\dagger(x)\beta^\mu_d(x)\big)
\end{align}
Let us now restrict ourselves to gauge transformations of the form
(\ref{eq:GT}), whose time dependence is precisely generated by the
choice of $A^0$ in each timestep. For such a gauge transformation, we
have
\begin{align}
        \stackrel{\longrightarrow}{\cal D}_{0,ab}\Omega^\dagger_{bd}(x) =0,\quad \Omega_{ca}(x) \stackrel{\longleftarrow}{\cal D}_{0,ab}=0,
\end{align}
      which implies
\begin{align}
        \big(\alpha_{\mu c}(x)\Omega_{ca}(x)\big)&
        \stackrel{\longleftrightarrow}{{\cal D}_{0,ab}}
        \big(\Omega_{bd}^\dagger(x)\beta^\mu_d(x)\big)\nonumber\\
        =&
        \big(\alpha_{\mu c}(x)\Omega_{ca}(x)\big)
        \big(\underbrace{\stackrel{\longrightarrow}{\cal D}_{0,ab}\Omega_{bd}^\dagger(x)}_{=0}\big)\beta^\mu_d(x)\nonumber\\
        +&
        \alpha_{\mu c}(x)\underbrace{\Omega_{ca}(x)\delta_{ab}\Omega_{bd}^\dagger(x)}_{=\delta_{cd}}(\partial_0\beta^\mu_d(x))\nonumber\\
        -&
        \alpha_{\mu c}(x)\big(\underbrace{\Omega_{ca}(x)\stackrel{\longleftarrow}{{\cal D}_{0,ab}}}_{=0}\big)
        \big(\Omega_{bd}^\dagger(x)\beta^\mu_d(x)\big)\nonumber\\
        -&
        \big(\partial_0\alpha_{\mu c}(x))\underbrace{\Omega_{ca}(x)\delta_{ab}
          \Omega_{bd}^\dagger(x)}_{=\delta_{cd}}\beta^\mu_d(x)
        =
        \alpha_{\mu c}(x)
        \stackrel{\longleftrightarrow}{{\cal \partial}_{0}}
        \beta^\mu_c(x).
\end{align}
Therefore, the inner product defined in eq.~(\ref{eq:inner-prod}) is
invariant under all the gauge transformations (\ref{eq:GT}) that may
affect solutions of the Yang-Mills equations.

\subsection{Improved LSZ formula}
The previous discussion suggests that we should generalize the
definition of the gluon production amplitude to
\begin{align}
  {\cal M}_{\p\lambda a}\equiv i \big(a_{\p\lambda a}\big|A\big),
  \label{eq:LSZ-GI}
\end{align}
where $A^\mu$ is the classical solution of the Yang-Mills equations,
and $a_{\p\lambda a}^\mu$ is a plane wave of momentum $\p$,
polarization $\lambda$ and color $a$. In a trivial background, this
plane wave simply reads
\begin{align}
  a_{\p\lambda a}^\mu(x)=t^b \underbrace{e^{ip\cdot x} \epsilon^\mu_\lambda(\p) \delta_{ab}}_{\equiv a_{\p\lambda a}^{\mu b}(x)}. 
\end{align}
In the notation $a_{\p\lambda a}^{\mu b}(x)$, $a$ is the color of the
wave at $x^0=-\infty$, and $b$ is its color at the current position
$x$. The factor $\delta_{ab}$ means that its color does not change
when it propagators over a null background. After a gauge
transformation $\Omega$, this plane wave becomes
\begin{align}
  a_{\p\lambda a}^\mu(x)=t^b \Omega^\dagger_{bb'}(x) e^{ip\cdot x} \epsilon^\mu_\lambda(\p) \delta_{ab'}, 
\end{align}
and now the initial color $a$ and the current color $b$ at the point
$x$ are no longer the same, but are related by the gauge
transformation $\Omega^\dagger_{ba}(x)$.

\section{Tools for on-shell methods}
\label{sec:onshell}
On-shell methods are a set of recently developed techniques for calculating amplitudes with physical gluons without using the traditional perturbative expansion. These methods emerged from the realization that amplitudes with on-shell gluons endowed with physical polarizations are considerably simpler than the expressions obtained from Feynman diagrams (which in principle work equally for off-shell gluons, or for unphysical polarizations), and from the desire to obtain more directly these simpler expressions. This approach starts with the separation of the color degrees of freedom from the kinematical factors, described in Section \ref{sec:color-decomp}. Since our goal is to use an extension of these methods to deal with the off-shell gluons produced by external color sources, we discuss in Section \ref{sec:off-shell} a convenient parameterization for off-shell momenta. Finally, we present in Section \ref{sec:BCFW} the Britto-Cachazo-Feng-Witten (BCFW) recursion that allows to calculate amplitudes by a recursion on the number of external lines, where we focus mostly on its extension to calculate amplitudes with some off-shell gluons.

\subsection{Color decomposition of tree amplitudes}
\label{sec:color-decomp}
The color structure of the tree graphs that appear in the sum of 
eq.~(\ref{eq:trees}) is identical to that of purely gluonic tree 
amplitudes in QCD, since the trivalent eikonal vertex brings the same 
factor $-gf^{abc}$ as a normal $3$-gluon vertex. From this observation, it 
follows that the color dependence of these more general tree graphs 
can be decomposed as a sum of traces of generators in the fundamental 
representation multiplied by a color-ordered amplitude that does not 
carry any color:
\begin{align}
  {\cal M}^{ab_1 b_2 \cdots b_n}_{1+n}(0^\lambda 1^{*} 2^{*}\cdots n^{*})
  =
  2
  \smash{\sum_{\sigma\in{\mathfrak S}_n}}
      {\rm tr}\,(t^a t^{b_{\sigma_1}}\cdots t^{b_{\sigma_n}})\,
      {\cal A}_{1+n}(0^h \sigma_1^{*} \sigma_2^{*}\cdots \sigma_n^{*}).
\end{align}
The color-ordered amplitude ${\cal A}_{1+n}$ can be obtained by 
the following color-ordered Feynman rules:
\setbox5\hbox to 11mm{\resizebox*{11mm}{!}{\includegraphics{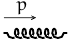}}}%
\setbox1\hbox to 25mm{\resizebox*{25mm}{!}{\includegraphics{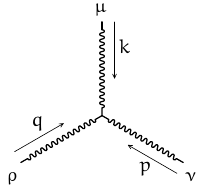}}}%
\setbox2\hbox to 23mm{\resizebox*{23mm}{!}{\includegraphics{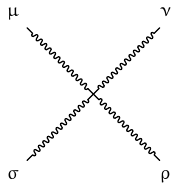}}}%
\setbox4\hbox to 15mm{\includegraphics[width=15mm]{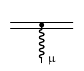}}%
\begin{eqnarray*}
  \raise -11mm\box1&=&
  \begin{aligned}
    g\,&\big\{g^{\mu\nu}\,(k-p)^\rho\\
    &\!\!\!\!+g^{\nu\rho}\,(p-q)^\mu+g^{\rho\mu}\,(q-k)^\nu\big\}
  \end{aligned}
  \\
  &&\vphantom{\big\{}\nonumber\\
\raise -11.5mm\box2&=&
                       \begin{aligned} 
                         &-i\,g^2\,(2\,g^{\mu\rho}g^{\nu\sigma}\\
                         &\quad\qquad -g^{\mu\sigma}g^{\nu\rho}-g^{\mu\nu}g^{\rho\sigma})\\
                       \end{aligned}
  \\
  &&\vphantom{\big\{}\nonumber\\
  \raise -7.2mm\box4 &=& g v^\mu
\end{eqnarray*}
(We have written only the vertices; the propagators are just the 
full-color ones with the color factor $\delta_{ab}$ removed.) Although we list 
these Feynman rules for completeness here, this is not the approach we 
have in mind for calculating the amplitudes. Instead, our goal is to 
use recursion in order to bootstrap them from the simplest ones (for 
which we may perhaps need the Feynman rules).

\subsection{Parameterization of off-shell momenta}
\label{sec:off-shell}
\subsubsection{Spinor representation of on-shell momenta}
Let us first start by a brief reminder of the spinor representation for on-shell momenta. Consider a $4$-momentum $p^\mu$. There is a one-to-one correspondence between momenta and complex $2\times 2$ matrices:
\begin{align}
p_ \mu\quad\to\quad P\equiv p_\mu \sigma^\mu,\quad\mbox{with\ \ $\sigma^\mu\equiv (1,\sigma^i)$},
\end{align}
where the $\sigma^i$ are the three Pauli matrices. An alternate representation is obtained by defining $\overline{P}\equiv -p_\mu\overline{\sigma}^\mu$ with $\overline{\sigma}^\mu\equiv(1,-\sigma^i)$. This mapping can be done for any momentum, even off-shell. 
However, note that 
\begin{align}
p_\mu p^\mu= {\rm det}\,(P).
\end{align}
When $p^2=0$, its matrix representation has a null determinant. Since the matrix $P$ is $2\times 2$, its rank is equal to $1$ and it can be factorized as a direct product,
\begin{align}
p^2=0\quad\Longleftrightarrow
P=\big|\p\big]\big<\p\big|,\quad
\overline{P}=\big|\p\big>\big[\p\big|,
\end{align}
where $\big|\p\big],\big|\p\big>,\big<\p\big|,\big[\p\big|$ are 2-component objects\footnote{In most applications, we do not need an explicit representation of $\big|\p\big],\big|\p\big>,\big<\p\big|,\big[\p\big|$, since all the common manipulations rely on generic properties of these spinors. Note also that the representation of an on-shell momentum is not unique, since the momentum is unchanged under the rescaling $\big|\p\big>\to \xi \big|\p\big>, \big[\p\big]\to \xi^{-1}\big[\p\big|$. This invariance, known as {\sl little group scaling}, completely constrains the momentum dependence of any 3-point amplitude with massless particles of any helicity, leaving only a constant prefactor to be determined from the Lagrangian.} (they are in fact Weyl spinors, but this property will not be needed). For two on-shell momenta $p_\mu$ and $q_\mu$, we have
\begin{align}
(p+q)^2=2p\cdot q=
\abk{\p\q}\sbk{\p\q}.
\end{align}
The brackets  $\abk{\p\q}$ and $\sbk{\p\q}$ are antisymmetric. Moreover, when the momenta have real valued components, we have $\abk{\p\q}=\sbk{\p\q}^*$.

Since the spinors introduced above have two components, they live in a 2-dimensional vector space. This implies that any collection of three such spinors is linearly dependent. This fact leads to the Schouten identities,
\begin{align}
\big|\p\big>\abk{\q\r}
+
\big|\q\big>\abk{\r\p}
+
\big|\r\big>\abk{\p\q}
=0,\quad
\big|\p\big]\sbk{\q\r}
+
\big|\q\big]\sbk{\r\p}
+
\big|\r\big]\sbk{\p\q}.
\label{eq:schouten}
\end{align}

For an on-shell gluon of momentum $\p$, its polarization vectors for helicities $\pm$ in the spinor representation can be written as
\begin{align}
  \epsilon_-^\mu(\p;\q)
  \equiv
  \frac{\big<\p\big|\overline{\sigma}^\mu\big|\q\big]}{\sqrt{2}\sbk{\p\q}},
    \quad
    \epsilon_+^\mu(\p;\q)
    \equiv
    \frac{\big<\q\big|\overline{\sigma}^\mu\big|\p\big]}{\sqrt{2}\abk{\q\p}},
      \label{eq:pols}
\end{align}
where $\q$ is an ''auxiliary vector'' (also on-shell). Varying the auxiliary vector $\q$ only
    produces terms proportional to $p^\mu$ (i.e., changes of $\q$ are
    like gauge transformations for the polarization vectors). In calculations of gauge invariant amplitudes, leaving $\q$ unspecified and checking that it drops out of the final result can provide a way of spotting possible mistakes. Note that in amplitudes with several external on-shell gluons, the auxiliary vectors can be chosen independently for each gluon.

\subsubsection{Momenta produced by a source moving along a light-like direction}
\label{sec:offshell}
Let us now present a convenient parameterization of the momentum
$p^\mu$ carried by a source $\rho^{(0)}$, describing a projectile that
moves along the direction $v^\mu$ (this is also the momentum of an  off-shell gluon directly emitted by the source). We shall restrict to cases where
the projectiles move at the speed of light, so that the vector $v^\mu$
is a light-like vector, $v_\mu v^\mu=0$.

Recall also that we have $v\cdot p=0$, because
$(v\cdot\partial)\rho^{(0)}(x)=0$. Because of this restriction\footnote{In Appendix \ref{sec:generic}, we extend this representation to momenta having two non-zero longitudinal components.}, $p^\mu$
can be described in terms of three non-zero coordinates.  We seek a
decomposition of $p^\mu$ into a component proportional to $v^\mu$
(this brings one coordinate) and a ``transverse'' component
$p_\perp^\mu$ described with two additional coordinates:
\begin{align}
  p^\mu\equiv x v^\mu + p_\perp^\mu.
\end{align}
By contracting this decomposition with $v_\mu$, we obtain $v\cdot
p_\perp=0$. However, this orthogonality property is not sufficient to
determine uniquely the $2$-dimensional transverse momentum $p_\perp^\mu$.

In order to make the decomposition more explicit, let us introduce an
auxiliary light-like vector $q^\mu$, and choose the longitudinal
coordinate $x$ to be
\begin{align}
  x=x_q\equiv \frac{q\cdot p}{q\cdot v}\quad\Longrightarrow\quad
  p^\mu =
  \frac{q\cdot p}{q\cdot v} v^\mu + p_\perp^\mu.
\end{align}
Now, in addition to $v\cdot p_\perp=0$, we also have $q\cdot p_\perp
=0$. In other words, introducing the auxiliary vector $q^\mu$ makes $p_\perp^\mu$ belong to the 2-dimensional subspace orthogonal to both $v^\mu$ and $q^\mu$, which allows to define it uniquely. (In the generic case where $v^\mu$ and $q^\mu$ are
not collinear, these two orthogonality conditions indeed define a
unique two-dimensional subspace.) Moreover, a basis of this subspace is
provided by the polarization vectors $\epsilon^\mu_\pm(\v;\q)$. In these polarization vectors,
$\v$ plays the role of the ``gluon momentum''\footnote{Note that, since the vector
$v^\mu$ is a dimensionless vector, the spinors $\big|\v\big>$ and
$\big|\v\big]$ are themselves dimensionless. This should be contrasted
with the spinors associated with a genuine momentum, such as
$\big|\q\big>$ and $\big|\q\big]$, whose dimension is
  $(\mbox{mass})^{1/2}$.}.

In terms of this basis, it is
convenient to parameterize $p_\perp^\mu$ as
\begin{align}
  p_\perp^\mu \equiv
  -\frac{\kappa}{\sqrt{2}}\,\epsilon_-^\mu(\v;\q)
  +\frac{\bar\kappa}{\sqrt{2}}\,\epsilon_+^\mu(\v;\q).
\end{align}
The coordinates $\kappa$ and $\bar\kappa$ are in general independent,
except when $p_\perp^\mu$ is real, in which case we have
$\bar\kappa=\kappa^*$. This is seen by writing
\begin{align}
  p_\perp^{\mu\,*}
  &=
  -\frac{\kappa^*}{\sqrt{2}}\,
  \underbrace{\big(\epsilon_-^\mu(\v;\q)\big)^*}_{=\,-\epsilon_+^\mu(\v;\q)}
  +\frac{{\bar\kappa}^*}{\sqrt{2}}\,
  \underbrace{\big(\epsilon_+^\mu(\v;\q)\big)^*}_{=\,-\epsilon_-^\mu(\v;\q)}\nonumber\\
  &=
  \frac{\kappa^*}{\sqrt{2}}\,\epsilon_+^\mu(\v;\q)
  -\frac{{\bar\kappa}^*}{\sqrt{2}}\,\epsilon_-^\mu(\v;\q),
\end{align}
and by equating this to $p_\perp^\mu$ (assuming it is a real vector).

The explicit determination of the coordinates $\kappa,\bar\kappa$ can be achieved by using
the following properties of the polarization vectors,
\begin{align}
  \epsilon_+(\v;\q)\cdot\epsilon_+(\v;\q)=\epsilon_-(\v;\q)\cdot\epsilon_-(\v;\q)=0,\quad
  \epsilon_+(\v;\q)\cdot\epsilon_-(\v;\q)=1,
\end{align}
that imply
\begin{align}
  \kappa &= -\sqrt{2}\; p\cdot\epsilon_+(\v;\q)
  = \frac{\big<\q\big|\overline{P}\big|\v\big]}{\abk{\q\v}},\nonumber\\
    \bar\kappa &= \sqrt{2}\; p\cdot\epsilon_-(\v;\q)
    =-\frac{\big<\v\big|\overline{P}\big|\q\big]}{\sbk{\v\q}},
      \label{eq:kappa}
\end{align}
where we denote $\overline{P}\equiv -p_\mu\overline{\sigma}^\mu$ (when
$p^\mu$ is light-like, this factorizes as
$\overline{P}=\big|\p\big>\big[\p\big|$).
  Note also that
  \begin{align}
    \kappa \bar\kappa
    =
    \frac{\big<\q\big|\overline{P}\big|\v\big]\big<\v\big|\overline{P}\big|\q\big]}{\abk{\q\v}\sbk{\q\v}}
  =
  \frac{2\,{\rm tr}\big(Q\overline{P} V \overline{P}\big)}{2(q\cdot v)},
  \end{align}
  where we have introduced $Q\equiv q_\mu\sigma^\mu$ and $V\equiv
  v_\mu\sigma^\mu$, with $\sigma^\mu\equiv(1,{\bs\sigma}^i)$. The
  evaluation of the trace is standard, and we have\footnote{For two
  off-shell momenta $p_a^\mu$ and $p_b^\mu$ produced by sources sharing a common direction $v^\mu$, this identity generalizes into $\kappa_a\overline{\kappa}_b+\overline{\kappa}_a\kappa_b=-2\,p_a\cdot p_b$.}
  \begin{align}
    \kappa\bar{\kappa}
    =
    \frac{2\big(2(q\cdot p)\overbrace{(v\cdot p)}^{=\,0}-(q\cdot v) p^2\big)}{2(q\cdot v)}
    =-p^2.
  \end{align}
Note that when the momentum $p^\mu$ is real, we have
$p^2=-\kappa\kappa^* = -|\kappa|^2 \le 0$, in agreement with the fact that
a real momentum $p^\mu$ that satisfies $v\cdot p=0$ must be space-like (or light-like) when $v^\mu$ is light-like.

Remarkably, while both $p_\perp^\mu$ and the polarization vectors
depend on the choice of auxiliary vector $q^\mu$, they do so in such a way that
the coordinates $\kappa,\bar\kappa$ are independent of the auxiliary
vector.  In order to see this, we can use the Schouten identities,
\begin{align}
  \frac{\big|\v\big]\big[\r\big|}{\sbk{\v\r}}
  +
  \frac{\big|\r\big]\big[\v\big|}{\sbk{\r\v}}=-1_+,
  \quad
  \frac{\big|\v\big>\big<\r\big|}{\abk{\v\r}}
  +
  \frac{\big|\r\big>\big<\v\big|}{\abk{\r\v}}=-1_-,
\end{align}
where $r^\mu$ is another arbitrary light-like vector not collinear
with $v^\mu$ (the subscripts $\pm$ in $1_\pm$ are just a reminder that these two
equations give the identity in two different spaces). By inserting
$1_-$ into $\kappa$, we obtain
\begin{align}
  \kappa
  =
  \frac{\big<\q\big|1_-\overline{P}\big|\v\big]}{\abk{\q\v}}
    =
    -\Bigg(
    \frac{\abk{\q\v}\big<\r\big|\overline{P}\big|\v\big]}{\abk{\q\v}\abk{\v\r}}
      +
      \frac{\abk{\q\r}\!\!\!\!\!\!\overbrace{\big<\v\big|\overline{P}\big|\v\big]}^{{\rm tr}(V\overline{P})=2 v\cdot p=0}\!\!\!\!\!\!}{\abk{\q\v}\abk{\r\v}}
        \Bigg)
        =\frac{\big<\r\big|\overline{P}\big|\v\big]}{\abk{\r\v}},
\end{align}
which is the same as the original expression of $\kappa$, but with the
auxiliary vector $r^\mu$ instead of $q^\mu$. The verification is same
for $\bar\kappa$, by inserting $1_+$.

\subsection{BCFW recursion}
\label{sec:BCFW}
Our goal is to use the Britto-Cachazo-Feng-Witten (BCFW) \cite{Britto:2005fq} recursion in
order to evaluate the off-shell amplitudes that appear in the
expansion of the gluon production amplitude. The method starts by
applying a shift to two external momenta $p_i^\mu\to
\wh{p}_i^\mu\equiv p_i^\mu +z e^\mu, p_j^\mu\to\wh{p}_j^\mu\equiv
p_j^\mu-z e^\mu$, where $z$ is a complex variable. In the standard
case of on-shell amplitudes, the direction $e^\mu$ of the shift must
be chosen so that the shifted external momenta remain on-shell at all
$z$. Then one uses elementary complex analysis to write the original
amplitude as a sum over poles\footnote{What makes this approach possible is the fact that any tree amplitude is a rational function of the external momenta, that has only simple poles (for non-exceptional configurations of the momenta). Therefore, the knowledge of the residues of these poles is sufficient to reconstruct the full amplitude.} in $z$, each residue taking the form of
a product of two on-shell amplitudes with fewer external legs.

\subsubsection{Shifts of two on-shell momenta}
Let us first recall how the recursion works when the external lines to
which the shift is applied are both on-shell.  Firstly, note that
overall momentum conservation in the amplitude is not upset, since
$\wh{p}_i+\wh{p}_j=p_i+p_j$ at any $z$.  The vector $e^\mu$ is
constrained by requiring that the shifted momenta $\wh{p}_{i,j}$
remain light-like at all $z$. This requires
\begin{align}
  p_i\cdot e = p_j\cdot e = e\cdot e =0.
\end{align}
(Observe that these three conditions can be simultaneously satisfied
only if $e^\mu$ is a complex momentum.) Since $e^2=0$, we know that
$E\equiv e_\mu\sigma^\mu$ factorizes as
$E=\big|e\big]\big<e\big|$. The remaining two conditions can be solved
  by choosing
  \begin{align}
    \big|e\big> = \big|i\big>,\quad
        \big|e\big]=\big|j\big]\quad
        \Longrightarrow\quad
        E=\big|j\big]\big<i\big|,\;\overline{E}=\big|i\big>\big[j\big|.
  \end{align}
  (There is a second solution obtained by exchanging the roles of $i$
  and $j$.) 
 Here, we have adopted a compact notation for the spinors associated to on-shell external momenta:
 \begin{align}
 \big|i\big>\equiv \big|\p_i\big>, \quad \big|i\big]\equiv\big|\p_i\big].
 \end{align}
 Since the shifted momenta $\wh{p}_{i,j}$ are still
  light-like, we can define the corresponding spinors,
  \begin{align}
    &\wh{p}_{i\mu}\sigma^\mu\equiv \big|\wh{i}\big]\big<\wh{i}\big|\quad
      \mbox{with\ }
      \big|\wh{i}\big>=\big|i\big>,\quad
      \big|\wh{i}\big]=\big|i\big]+z\big|j\big],\nonumber\\
    &\wh{p}_{j\mu}\sigma^\mu\equiv \big|\wh{j}\big]\big<\wh{j}\big|\quad
      \mbox{with\ }
      \big|\wh{j}\big>=\big|j\big>-z\big|i\big>,\quad
      \big|\wh{j}\big]=\big|j\big].
  \end{align}
The polarization vectors of the lines $i$ and $j$ become
$z$-dependent, according to eqs.~(\ref{eq:pols}). Before generalizing
this to off-shell momenta, note also that for any pair of spinors
$\big<a\big|,\big|b\big]$, we have
  \begin{align}
    \big<a\big|\overline{E}\big|b\big]
      =
\abk{ai}\sbk{jb}
      =
    -\tfrac{1}{2}
      \big<i\big|\overline{\sigma}^\mu\big|j\big]\big<a\big|\overline{\sigma}_\mu\big|b\big]
        ,
  \end{align}
  which implies that the shift vector $e^\mu$ defined above can also be written as
  \begin{align}
    e^\mu
    =
    \tfrac{1}{2}\big<i\big|\overline{\sigma}^\mu\big|j\big].
  \end{align}

\subsubsection{Shifts involving off-shell momenta}
  \paragraph{Shift of an on-shell and an off-shell momenta}
Consider now an external momentum $p_i$ flowing from a source
$\rho^{(0)}_i$ corresponding to a projectile moving in the direction
$v_i^\mu$. In this discussion, we assume that the other shifted line,
$j$, is an on-shell external gluon. We have $p_i^2\not=0$, while
$v_i^2=p_j^2=0$. We still shift the momenta $p_{i,j}$ according to
\begin{align}
  p_i^\mu \to \wh{p}_i^\mu\equiv p_i^\mu+z\,e^\mu,\quad
  p_j^\mu\to \wh{p}_j^\mu\equiv p_j-z\,e^\mu,\quad
   e^\mu
    =
    \tfrac{1}{2}\big<\v_i\big|\overline{\sigma}^\mu\big|j\big],
      \label{eq:shift1}
\end{align}
where the spinor $\big<\v_i\big|$ is defined from the light-like
direction $v_i^\mu$ of the source that produces the off-shell gluon of
momentum $p_i$ ($v_{i\mu}\sigma^\mu\equiv \big|\v_i\big]\big<\v_i\big|$)
  rather than from $p_i$ itself (since $p_i^2\not=0$,
  $p_{i\mu}\sigma^\mu$ cannot be factorized as a direct product of
  spinors). With this shift vector, we still have $v_i\cdot
  \wh{p}_i=v_i\cdot p_i=0$, and therefore the shifted off-shell momentum
  $\wh{p}_i^\mu$ is a valid argument for the source
  $\rho_i^{(0)}$.

  Recall that the off-shell momentum $p_i^\mu$ can be decomposed as
  \begin{align}
    p_i^\mu
    =
    \frac{q\cdot p_i}{q\cdot v_i}\,v_i^\mu
    -
    {\kappa_i}
    \frac{\big<\v_i\big|\overline{\sigma}^\mu\big|\q\big]}{{2}\sbk{\v_i\q}}
      +
      {\overline{\kappa}_i}
      \frac{\big<\q\big|\overline{\sigma}^\mu\big|\v_i\big]}{{2}\abk{\q \v_i}}
        .
        \label{eq:off-shell-k}
  \end{align}
  Given the shift vector $e^\mu$ in eq.~(\ref{eq:shift1}), a
  significant simplification occurs if we choose the auxiliary vector
  $\q \equiv \p_j$ (recall that the transverse coordinates
  $\kappa_i,\overline{\kappa}_i$ do not depend on this choice), since
  with this choice the shift vector becomes proportional to the
  polarization vector that multiplies $\kappa_i$ so that the shifted
  momentum $\wh{p}_i^\mu$ reads
  \begin{align}
    \wh{p}_i^\mu
    =
    p_i^\mu
    +z e^\mu
    =
    \frac{p_j\cdot p_i}{p_j\cdot v_i}\,v_i^\mu
    -
    \big(\underbrace{{\kappa_i}-z\sbk{\v_ij}}_{\equiv\,\wh{\kappa}_i}\big)
    \frac{\big<\v_i\big|\overline{\sigma}^\mu\big|j\big]}{{2}\sbk{\v_ij}}
      +
      {\underbrace{\overline{\kappa}_i}_{\equiv\,\wh{\overline{\kappa}}_i}}
      \frac{\big<j\big|\overline{\sigma}^\mu\big|\v_i\big]}{{2}\abk{j\v_i}}
        .
  \end{align}
Thus, this shift combined with the above choice of auxiliary vector is
equivalent to shifting the coordinate $\kappa_i$ and the spinor
$\big|{j}\big>$, while $\overline{\kappa}_i$ and $\big|j\big]$ remain unchanged:
  \begin{align}
    &
    \wh{\kappa}_i = {\kappa_i}-z\sbk{\v_ij},\quad \wh{\overline{\kappa}}_i=\overline{\kappa}_i,\nonumber\\
    &
    \big|\wh{j}\big> = \big|j\big> - z\big|\v_i\big>,\quad
    \big|\wh{j}\big] = \big|j\big].
  \end{align}
Note also that the shift does not alter the direction $v_i^\mu$ of the
source that produces the off-shell gluon; it only modifies one of the
transverse components of its momentum.

To close this subsection, let us quote without proof the changes to the external lines if we have $p_i^2=0$ and $p_j^2\not=0$ instead, with a  shift defined as $e^\mu=\tfrac{1}{2}\big<i\big|\overline{\sigma}^\mu\big|\v_j\big]$. In this case the spinors and transverse components that described the external lines change as follows,
\begin{align}
&\big|\wh{i}\big>=\big|i\big>,\quad
      \big|\wh{i}\big]=\big|i\big]+z\big|\v_j\big],\nonumber\\
      &\wh{\kappa}_j=\kappa_j,\quad \wh{\overline{\kappa}}_j=\overline{\kappa}_j-z\abk{i\v_j}.
\end{align}

\paragraph{Shift of two off-shell momenta}
Consider finally the case where both lines $i$ and $j$ are attached to
two sources $\rho^{(0)}$ (the two sources may not belong to the same
projectile, and therefore they can have distinct light-like directions
$v_i^\mu$ and $v_j^\mu$).  We shift the momenta $p_{i,j}^\mu$
according to
\begin{align}
  p_i^\mu \to \wh{p}_i^\mu\equiv p_i^\mu+z\,e^\mu,\quad
  p_j^\mu\to \wh{p}_j^\mu\equiv p_j-z\,e^\mu,\quad
   e^\mu
    =
    \tfrac{1}{2}\big<\v_i\big|\overline{\sigma}^\mu\big|\v_j\big].
      \label{eq:shift2}
\end{align}
Thus, we have $v_i\cdot e =
    v_j\cdot e=0$ and the shifted momenta $\wh{p}_i^\mu$ and
    $\wh{p}_j^\mu$ are still allowed arguments for their respective
    sources. Note that we are allowed to use different auxiliary
    vectors $q_i^\mu$ and $q_j^\mu$ when parameterizing $p_i^\mu$ and
    $p_j^\mu$, respectively. The most convenient choice is $\q_i = 
    \v_j$, $\q_j =  \v_i$, so that the shifted momenta can be
    written as
    \begin{align}
      \wh{p}_i^\mu
    &=
    p_i^\mu+z e^\mu
    =
    \frac{v_j\cdot p_i}{v_j\cdot v_i}\,v_i^\mu
    -
    \big(\underbrace{{\kappa_i}-z\sbk{\v_i\v_j}}_{\equiv\,\wh{\kappa}_i}\big)
    \frac{\big<\v_i\big|\overline{\sigma}^\mu\big|\v_j\big]}{{2}\sbk{\v_i\v_j}}
      +
      {\underbrace{\overline{\kappa}_i}_{\equiv\,\wh{\overline{\kappa}}_i}}
      \frac{\big<\v_j\big|\overline{\sigma}^\mu\big|\v_i\big]}{{2}\abk{\v_j\v_i}}
        ,\nonumber\\
      \wh{p}_j^\mu
      &=
      p_j^\mu-z e^\mu
      =
      \frac{v_i\cdot p_j}{v_i\cdot v_j}\,v_j^\mu
      -
      \underbrace{{\kappa_j}}_{\equiv\,\wh{\kappa}_j}
      \frac{\big<\v_j\big|\overline{\sigma}^\mu\big|\v_i\big]}{{2}\sbk{\v_j\v_i}}
        +
        \big(\underbrace{\overline{\kappa}_j-z\abk{\v_i\v_j}}_{\equiv\,\wh{\overline{\kappa}}_j}\big)
        \frac{\big<\v_i\big|\overline{\sigma}^\mu\big|\v_j\big]}{{2}\abk{\v_i\v_j}}
          .
    \end{align}
    From these equations, we read directly how the shift affects the transverse components of $p_i^\mu$ and $p_j^\mu$:
    \begin{align}
      &\wh{\kappa}_i = {\kappa_i}-z\sbk{\v_i\v_j},\quad \wh{\overline{\kappa}}_i = {\overline{\kappa}}_i,\nonumber\\
      &\wh{\kappa}_j = {\kappa_j},\quad \wh{\overline{\kappa}}_j = {\overline{\kappa}}_j-z\abk{\v_i\v_j}.
    \end{align}
    
\subsubsection{Summary} For later reference, we summarize in the following table the changes to the external momenta for a shift $e^\mu=\tfrac{1}{2}\big<i,\v_i\big|\overline{\sigma}^\mu\big|j,\v_j\big]$, depending on whether the lines $i,j$ are on-shell or not.
\begin{align*}
\begin{tabular}{|c|c||c|c|c|c|}
\hline
$\vphantom{\Big]}k_i^2$ &  $k_j^2$ &&&&
\\
\hline
$\vphantom{\Big]}0$
&
$0$
& 
$\big|\wh{i}\big>=\big|i\big>$
&
$\big|\wh{i}\big]=\big|i\big]+z\big|j\big]$
&
$\big|\wh{j}\big>=\big|j\big>-z\big|i\big>$
&
$\big|\wh{j}\big]=\big|j\big]$
\\
\hline
$\vphantom{\Big]}\not=0$
&
$0$
& 
$\wh{\kappa}_i=\kappa_i-z\sbk{\v_ij}$
&
$\wh{\overline{\kappa}}_i=\overline{\kappa}_i$
&
$\big|\wh{j}\big>=\big|j\big>-z\big|\v_i\big>$
&
$\big|\wh{j}\big]=\big|j\big]$
\\
\hline
$\vphantom{\Big]}0$
&
$\not=0$
& 
$\big|\wh{i}\big>=\big|i\big>$
&
$\big|\wh{i}\big]=\big|i\big]+z\big|\v_j\big]$
&
$\wh{\kappa}_j=\kappa_j$
&
$\wh{\overline{\kappa}}_j=\overline{\kappa}_j-z\abk{i\v_j}$
\\
\hline
$\vphantom{\Big]}\not=0$
&
$\not=0$
& 
$\wh{\kappa}_i=\kappa_i-z\sbk{\v_i\v_j}$
&
$\wh{\overline{\kappa}}_i=\overline{\kappa}_i$
&
$\wh{\kappa}_j=\kappa_j$
&
$\wh{\overline{\kappa}}_j=\overline{\kappa}_j-z\abk{\v_i\v_j}$
\\
\hline
\end{tabular}
\end{align*}
Note that, in the special case where a shift is applied to two off-shell gluons produced by sources that have the same direction $v_i^\mu$, then $\abk{\v_i\v_j}=\sbk{\v_i\v_j}=0$ in the above equations and the transverse components $\kappa_i, \overline{\kappa}_i,\kappa_j,\overline{\kappa}_j$ all remain unchanged. Therefore, the denominators of the corresponding propagators do not change (since $p_i^2=-\kappa_i\overline{\kappa}_i$ and $p_j^2=-\kappa_j\overline{\kappa}_j$) and remain independent of $z$. We show in appendix \ref{sec:high-z} that such a shift leads to a shifted amplitude that decreases as $z^{-1}$ and is therefore a valid shift for the application of BCFW.

Let us also establish the behavior at large $z$ of the polarization vectors for on-shell external lines shifted by $e^\mu=\tfrac{1}{2}\big<i\big|\overline{\sigma}^\mu\big|j\big]$,
\begin{align}
&
e^\mu_+(\wh{i},\q)=\frac{\big<\q\big|\overline{\sigma}^\mu\big|\wh{i}\big]}{\sqrt{2}\,\abk{\q\wh{i}}}\sim z, 
&
e^\mu_-(\wh{i},\q)=\frac{\big<\wh{i}\big|\overline{\sigma}^\mu\big|\q\big]}{\sqrt{2}\,\sbk{\wh{i}\q}}\sim \frac{1}{z},
\nonumber\\ 
&
e^\mu_+(\wh{j},\q)=\frac{\big<\q\big|\overline{\sigma}^\mu\big|\wh{j}\big]}{\sqrt{2}\,\abk{\q\wh{j}}}\sim \frac{1}{z}, 
&
e^\mu_-(\wh{j},\q)=\frac{\big<\wh{j}\big|\overline{\sigma}^\mu\big|\q\big]}{\sqrt{2}\,\sbk{\wh{j}\q}}\sim z.
\end{align}

\subsubsection{Recursion}
\paragraph{General strategy}
By shifting the momenta of a pair of external lines (and of all
the internal lines that connect them inside the tree amplitude) as
described in the preceding section, an amplitude ${\cal
  A}(\cdots)$ is turned into a $z$-dependent deformation:
\begin{align}
  {\cal A}(\cdots)\quad\empile{\longrightarrow}\over{{\mbox{\scriptsize shift}}}\quad
  {\cal A}(\cdots;z).
\end{align}
Note that, since the $z$ dependence of this object comes from the
momenta carried by the propagators, the vertices and the polarization
vectors of a tree amplitude, this dependence is rational. Its only 
singularities are therefore poles at discrete values of $z$.

Then, one considers the following integral,
\begin{align}
  \oint_{{\cal C}_R}\frac{dz}{2i\pi}\,\frac{{\cal A}(\cdots;z)}{z},
\end{align}
where ${\cal C}_R$ is a circle of radius $R$ (we
will eventually take the limit $R\to \infty$) in the complex plane.  We
shall assume that the shifted amplitude goes to zero\footnote{Note
that this property is not always true, and must be checked prior
to using this approach.} when $|z|\to \infty$.  When this is the
case, the above integral is zero when the radius $R$ goes
to infinity because the integrand decreases
faster than the circumference of the circle:
\begin{align}
  \oint_{{\cal C}_\infty}\frac{dz}{2i\pi}\,\frac{{\cal A}(\cdots;z)}{z}=0.
\end{align}
On the other hand, we may also evaluate this integral with the
theorem of residues, by summing over all the poles of the
integrand. The integrand has an obvious pole at $z=0$, whose
residue is the unshifted amplitude we wish to
calculate. Therefore, we may write:
\begin{align}
  {\cal A}(\cdots)
  =
  -\sum_{\mbox{\scriptsize other poles $z_*$}}{\rm Residue}\,\left[\frac{{\cal A}(\cdots;z)}{z}\right]_{z=z_*},
  \label{eq:bcfw0}
\end{align}
where the sum in the right hand side runs over all the other poles
of the integrand. These poles, that come from the vanishing of
denominators inside the amplitude as $z$ is varied, can be
classified in several categories, as discussed below.

\paragraph{Poles from internal gluon lines}
Let us first consider a generic internal gluon traversed by the
shift, with a momentum $K_{_I}^\mu$ before the shift is
applied. By ``generic'', we mean a gluon not carrying the whole
momentum of a source (which requires a separate
discussion -- see later). This case is the standard use of BCFW
shifts for calculating on-shell amplitudes. Firstly, let us write
the amplitude ${\cal A}(\cdots;z)$ in a way that exhibits the
gluon propagator that will produce the pole:
\setbox1\hbox to 28mm{\includegraphics[width=28mm]{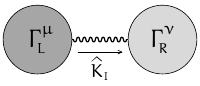}}
\begin{equation}
  {\cal A}(\cdots;z)
  =\raise -5.8mm\box1=
  \Gamma_{_L}^\mu(\cdots;-\wh{K}_{_I})\frac{-i g_{\mu\nu}}{\wh{K}_{_I}^2}\Gamma_{_R}^\nu(+\wh{K}_{_I};\cdots).
\end{equation}
In this notation, the ellipses denote external on-shell lines
contracted with a physical polarization vector, or off-shell propagators contracted with the direction $v^\mu$ of the corresponding source. The objects $\Gamma_{_L}^\mu$ and
$\Gamma_{_R}^\nu$ are the subgraphs on the left and on the right
of the exhibited gluon, respectively. In addition to the external
lines represented by the ellipses, they also depend on a dangling Lorentz
index and on the shifted internal momentum $\wh{K}_{_I}$ (we use the
convention of all-incoming momenta, hence $+\wh{K}_{_I}$ on the
right and $-\wh{K}_{_I}$ on the left). Finally, the factor in the
middle is the propagator of the gluon under consideration. Its
numerator is written in Feynman gauge, but the result
does not depend on this thanks to the non-Abelian Ward identities
satisfied by the factors on the left and on the right.

Since the shift vector satisfies $e\cdot e=0$, the denominator,
\begin{align}
  \wh{K}_{_I}^2 \equiv (k_{_I}+z e)^2 = K_{_I}^2 + 2z e\cdot K_{_I},
\end{align}
is in fact linear in $z$ and therefore has a unique zero
located at
\begin{align}
  z_* = -\frac{K_{_I}^2}{2e\cdot K_{_I}}.
\end{align}
It is straightforward to calculate its residue. Using the
decomposition of $-g_{\mu\nu}$ on the basis of the polarization
vectors $\epsilon^\mu_{\pm}(\wh{K}_{_I};\q)$ (up to terms
proportional to $\wh{K}_{_I}^\mu$ or $\wh{K}_{_I}^\nu$, that give
zero thanks to the Ward identities), we find that this type of pole
gives the following contribution in eq.~(\ref{eq:bcfw0}):
\begin{align}
  &-{\rm Residue}\,\left[\frac{{\cal A}(\cdots;z)}{z}\right]_{z=z_*}\nonumber\\
  &=
  \Gamma_{_L}^\mu(\cdots;-\wh{K}_{_I})\frac{i}{{K}_{_I}^2}
  \Big(\underbrace{\sum_{h=\pm}
    \epsilon_{-h,\mu}(-\wh{K}_{_I};\q)
    \epsilon_{h,\nu}(\wh{K}_{_I};\q)}_{-g_{\mu\nu}\oplus \wh{K}_{_I,\mu}\oplus\wh{K}_{_I,\nu}}
  \Big)
  \Gamma_{_R}^\nu(\wh{K}_{_I};\cdots)\nonumber\\
  &=
  \sum_{h=\pm} {\cal A}_{_L}(\cdots;-\wh{K}_{_I}^{-h})\frac{i}{{K}_{_I}^2}
      {\cal A}_{_R}(\wh{K}_{_I}^{h};\cdots),
\end{align}
where in the final expression the amplitudes ${\cal A}_{_{L,R}}$ on the
left and on the right are amplitudes with an extra on-shell gluon
of momentum $\pm\wh{K}_{_I}$ and helicity $\pm h$. These
amplitudes have fewer external legs than the original amplitude,
and this approach leads to a recursion on the number of external
legs. The crucial point is that this is a recursion for amplitudes (i.e., gauge invariant objects) and not for Feynman graphs (that are in general gauge dependent).

\paragraph{Gluons carrying the whole momentum of a source}
We need to treat differently the situation where a shifted gluon
carries the whole momentum of a source to which it is
attached. This occurs when the shift affects an off-shell external
line, for the gluon directly attached to this source.

\paragraph{Zero in $\wh{p}_i^2$:}\label{sec:C} Let us start with the case where this gluon is attached
to the line $i$ and therefore carries the momentum $\wh{k}_i$
after the shift. Such a contribution to the shifted amplitude
${\cal A}(\cdots;z)$ can be written as
\setbox1\hbox to 25mm{\includegraphics[width=25mm]{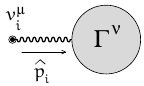}}
\begin{equation}
  \raise -5.8mm\box1=     v_i^\mu \frac{-ig_{\mu\nu}}{\wh{p}_i^2}\Gamma^\nu(\wh{p}_i;\cdots),
\end{equation}
where $\Gamma^\nu$ is the rest of the amplitude after we have
removed the propagator of the off-shell gluon under consideration (here
also, we have written the numerator of the gluon propagator in
Feynman gauge). The denominator has a simple expression in terms
of the transverse coordinates,
\begin{align}
  \wh{p}_i^2 = -\wh{\kappa}_i \wh{\overline{\kappa}}_i
  =
  -(\kappa_i-z\sbk{\v_ij})\overline{\kappa}_i.
\end{align}
(Here, we assume that the line $j$ is an on-shell gluon, otherwise we would have $\sbk{\v_i\v_j}$ instead of $\sbk{\v_i j}$.)
This contribution has a simple pole at $z_*=\kappa_i/\sbk{\v_ij}$, at
which the shifted gluon momentum reads
\begin{align}
  \wh{p}_i^\mu\empile{=}\over{z=z_*}
  \frac{q_i\cdot p_i}{q_i\cdot v_i}\,v_i^\mu
  +
  \overline{\kappa}_i
  \frac{\big<j\big|\overline{\sigma}^\mu\big|\v_i\big]}{{2}\abk{j\v_i}}.
\end{align}
Using the Fierz identity
$\big<j\big|\overline{\sigma}^\mu\big|\v_i\big]\overline{\sigma}_\mu
  =-2\big|j\big>\big[\v_i\big|$ and $-v_{i}^\mu\overline{\sigma}_\mu=\big|\v_i\big>\big[\v_i\big|$, we also have the following identities at $z=z_*$,
\begin{align}
  &-\wh{p}_{i\mu}\overline{\sigma}^\mu
  =
  \Big(\underbrace{\frac{q_i\cdot p_i}{q_i\cdot v_i}\,\big|\v_i\big>-\frac{\overline{\kappa}_i}{\abk{\v_ij}}\,\big|j\big>}_{\big|\wh{i}\big>}\Big)
  \underbrace{\big[\v_i\big|\vphantom{\frac{\overline{\kappa}_i}{\abk{\v_ij}}}}_{\big[\wh{i}\big|},\nonumber\\
          &v_i\cdot \epsilon_-(-\wh{p}_i;q_i)=\frac{\overline\kappa_i}{\sqrt{2}},\quad
          v_i\cdot \epsilon_+(-\wh{p}_i;q_i) = 0.
\end{align}
(The fact that $-\wh{p}_{i\mu}\overline{\sigma}^\mu$ factorizes as a direct product of two
spinors is consistent with the fact that $\wh{p}_i^\mu$ is
on-shell at the pole.) The contractions of $v_i^\mu$ with the polarization vectors do not depend on the auxiliary vector $q_i$ thanks to the fact that $v_i\cdot \wh{p}_i=0$ (indeed, varying $q_i$ in $\epsilon_\pm(\wh{p}_i,q_i)$ modifies the polarization vector by a term proportional to $\wh{p}_i^\mu$). Note that the spinors $\big|\v_i\big>, \big|\v_i\big]$ parameterize the direction $v_i^\mu$ of the source attached to the off-shell gluon of momentum $p_i^\mu$ ($-v_i^\mu\overline{\sigma}_\mu=\big|\v_i\big>\big[\v_i\big|$), while $|\wh{i}\big>,\big|\wh{i}]$ parameterize the shifted gluon momentum, via $-\wh{p}_i^\mu\overline{\sigma}_\mu=\big|\wh{i}\big>\big[\wh{i}\big|$. The contribution of this pole to
eq.~(\ref{eq:bcfw0}) reads
\begin{align}
  -{\rm Residue}\,\left[\frac{{\cal A}(\cdots;z)}{z}\right]_{z=z_*}
  &\!\!\!\!\!=
  -\frac{i}{\kappa_i\overline{\kappa}_i}\sum_{h=\pm} \big(v_i\cdot\epsilon_{-h}(-\wh{p}_i;q_i)\big) \epsilon_{+h,\nu}(\wh{p}_i;q_i)\Gamma^\nu(\wh{p}_i;\cdots)\nonumber\\
  &=
  -\frac{i}{\sqrt{2}\,\kappa_i}
  \underbrace{\epsilon_{+,\nu}(\wh{p}_i;q_i)\Gamma^\nu(\wh{p}_i;\cdots)}_{{\cal A}(\wh{p}_i^+;\cdots)}.
  \label{eq:C}
\end{align}
We see that the contribution of this term is, up to a simple
prefactor, an amplitude where the direction $v_i^\mu$ of the source
and the gluon propagator directly attached to it have been replaced by
the polarization vector of an on-shell gluon of positive helicity,
thereby producing an amplitude where this external line is now
on-shell.

\paragraph{Zero in $\wh{p}_j^2$:}\label{sec:D} This case is very similar to the previous one, except that the pole comes from a gluon carrying the whole momentum of the off-shell line $j$. The shifted amplitude reads
\setbox1\hbox to 25mm{\includegraphics[width=25mm]{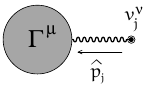}}
\begin{equation}
  \raise -5.8mm\box1=\Gamma^\mu(\cdots;\wh{p}_j) \frac{-ig_{\mu\nu}}{\wh{p}_j^2} v_j^\nu. 
\end{equation}
Now, the denominator reads
\begin{align}
  \wh{p}_j^2 = -\wh{\kappa}_j\wh{\overline{\kappa}}_j
  =
  -\kappa_j (\overline{\kappa}_j-z\abk{i\v_j}),
\end{align}
and the pole is at $z_*=\overline{\kappa}_j/\abk{i\v_j}$. At this value of $z$, we have
\begin{align}
  &\wh{p}_j^\mu \empile{=}\over{z=z_*}  \frac{q_j\cdot p_j}{q_j\cdot v_j}\,v_j^\mu
  -
  {\kappa_j}
  \frac{\big<\v_j\big|\overline{\sigma}^\mu\big|i\big]}{{2}\sbk{\v_ji}},\nonumber\\
    &-\wh{p}_{j\mu}\overline{\sigma}^\mu=
    \underbrace{\big|\v_j\big>\vphantom{\frac{\kappa_j}{\sbk{\v_ji}}}}_{\big|\wh{j}\big>}
    \Big(
    \underbrace{\frac{q_j\cdot p_j}{q_j\cdot v_j}\,\big[\v_j\big|-\frac{\kappa_j}{\sbk{\v_ji}}\,\big[i\big|
        }_{\big[\wh{j}\big|}
          \Big),\nonumber\\
          &v_j\cdot \epsilon_-(-\wh{p}_j;q_j)=0,\quad
          v_j\cdot \epsilon_+(-\wh{p}_j;q_j) = -\frac{\kappa_j}{\sqrt{2}},
          \label{eq:D1}
\end{align}
which leads to the following contribution in eq.~(\ref{eq:bcfw0}): \begin{align}
  -{\rm Residue}\,\left[\frac{{\cal A}(\cdots;z)}{z}\right]_{z=z_*}
  =
  \frac{i}{\sqrt{2}\,\overline{\kappa}_j}
  \underbrace{\epsilon_{-,\mu}(\wh{p}_j;q_j)\Gamma^\mu(\cdots;\wh{p}_j)}_{{\cal A}(\cdots;\wh{p}_j^-)}.
  \label{eq:D}
\end{align}

\paragraph{Shift to two sources moving in the same direction}
\label{sec:vv}
In view of subsequent applications, let us discuss here the special
case where a shift is applied to two off-shell gluons $i$ and $j$
produced by sources sharing a common light-like direction $v_i^\mu$. The shift
vector is
$e^\mu=\tfrac{1}{2}\big<\v_i\big|\overline{\sigma}^\mu\big|\v_j\big]$, with
        spinors $\big<\v_i\big|$ and $\big|\v_j\big]$ that refer to the same light-cone direction $v_i^\mu$. We thus have
      \begin{align}
        -e^\mu\overline{\sigma}_\mu
        =
        \big|\v_i\big>\big[\v_i\big|=-v_i^\mu\overline{\sigma}_\mu,
      \end{align}
      and in fact $e^\mu=v_i^\mu$.

Let us compute the squares of the shifted off-shell momenta $\wh{p}_i$
and $\wh{p}_j$,
\begin{align}
 &\wh{p}_i^2=(p_i+z v_i)^2=p_i^2+2zp_i\cdot v_i + z^2 v_i^2 = p_i^2,\nonumber\\ 
 &\wh{p}_j^2=(p_j-z v_i)^2=p_j^2-2zp_j\cdot v_i + z^2 v_i^2 = p_j^2,
\end{align}
since $p_i\cdot v_i = p_j\cdot v_i=0$ and $v_i^2=0$. Therefore, with such a
shift, the gluon propagators directly attached to the sources are
independent of $z$ and do not have poles; the poles produced by
this shift can only occur to internal gluon propagators (not attached
to sources).

\paragraph{Poles in eikonal propagators}
The eikonal propagators may potentially have poles when the momentum
they carry is affected by the shift. In the case of the collision of
two projectiles, there are two color currents $J_1^\mu$ and $J_2^\mu$,
of directions $v_1^\mu$ and $v_2^\mu$, respectively, and the eikonal propagators have one of the following two forms,
\begin{align}
\frac{1}{p\cdot v_1}\,\quad\mbox{or\ \ }\frac{1}{p\cdot v_2}.
\end{align}
If an eikonal propagator $1/(p\cdot v)$ is traversed by a momentum shift $ze^\mu$, its denominator becomes
\begin{align}
\wh{p}\cdot v=p\cdot v-z e\cdot v.
\end{align}
When $e\cdot v\not =0$, the shifted denominator vanishes at
\begin{align}
z_*=\frac{p\cdot v}{e\cdot v}.
\end{align} 
The contribution of this pole to the recursion formula is
\begin{align}
-\frac{1}{z_*}\,{\rm Residue}\,\Big[\frac{1}{p\cdot v-z\,e\cdot v}\Big]_{z=z_*}
=
\frac{1}{p\cdot v},
\end{align}
i.e., the unshifted eikonal propagator.  Before the shift, the momentum $p^\mu$ was generically off-shell with a non-zero $p\cdot v$. At the pole, the shifted momentum is still off-shell but has now a vanishing longitudinal component $\wh{p}\cdot v$.

A frequently encountered special case is when the shift direction $e^\mu$ is one of source directions $v_{1,2}^\mu$, for instance $e^\mu=v_2^\mu$. In this case, the shifted eikonal denominators are particularly simple
\begin{align}
\wh{p}\cdot v_1 = p\cdot v_1 -zv_1\cdot v_2 = p\cdot v_1-z,
\quad
\wh{p}\cdot v_2= p\cdot v_2-z v_2\cdot v_2 = p\cdot v_2.
\end{align}
The first type of eikonal propagator has a pole at $z_*=p\cdot v_1$, while the second type is $z$-independent and does not have any pole. At the pole, $\wh{p}\cdot v_1 =0$ and $\wh{p}\cdot v_2=p\cdot v_2$, and the virtuality of the shifted momentum is
\begin{align}
\wh{p}\,^2=2(\wh{p}\cdot v_1)(\wh{p}\cdot v_2)-\p_\perp^2=-\p_\perp^2.
\end{align}
In other words, with this shift, only eikonal propagators coming from a source $\rho_1^{(0)}$ can have a pole. Moreover, at the pole, the residue is the unshifted eikonal propagator and the momentum $\wh{p}^\mu$ is the momentum of a gluon directly produced by a source $\rho_1^{(0)}$.

Note also that this kind of pole never appears if we choose a shift $e^\mu$ which is orthogonal to all the directions $v^\mu$ that can enter in the amplitude under consideration. For the study of scattering off an external field produced by sources of direction $v_2^\mu$  (studied in section \ref{sec:scatt}), the following $e^\mu$'s,
\begin{align}
e^\mu=\tfrac{1}{2}\big<\v_2\big|\overline{\sigma}^\mu\big|i\big]\quad
\mbox{or\ \ }
e^\mu=\tfrac{1}{2}\big<i\big|\overline{\sigma}^\mu\big|\v_2\big],
\end{align}
both fulfill this property for any other external line  $i$. In the case of gluon production from off-shell gluons that come from sources in two directions $v_1^\mu$ and $v_2^\mu$, the shift should be picked among the following two choices,
\begin{align}
e^\mu=\tfrac{1}{2}\big<\v_2\big|\overline{\sigma}^\mu\big|\v_1\big]\quad
\mbox{or\ \ }
e^\mu=\tfrac{1}{2}\big<\v_1\big|\overline{\sigma}^\mu\big|\v_2\big],
\end{align}
if one wishes to avoid poles in eikonal propagators.

\section{Three-point tree amplitudes}
\label{sec:3point}
When applying the BCFW recursion to calculate amplitudes, the first
step is usually to find the 3-point amplitudes, from which amplitudes
with more external legs can then be obtained recursively. In this
section, we derive the 3-gluon amplitudes that will be used later.

\subsection{Three on-shell gluons}
Let us first recall the Parke-Taylor formulas for
amplitudes with three on-shell gluons:
\begin{align}
  {\cal A}_3(0^+1^+2^-)=-g\sqrt{2}\frac{\sbk{01}^3}{\sbk{12}\sbk{20}},\quad
  {\cal A}_3(0^-1^-2^+)=g\sqrt{2}\frac{\abk{01}^3}{\abk{12}\abk{20}}.
\end{align}
The amplitudes with all-equal helicities are zero, and the other
non-zero amplitudes are obtained by relabeling the external lines
in the above formulas. A noteworthy aspect of these 3-point amplitudes is that all their kinematical dependence can be obtained from the fact that massless on-shell 3-point amplitudes cannot mix angle and square brackets, and from little group scaling. The only use of the Yang-Mills Lagrangian in determining these amplitudes is to find the correct prefactor, $\pm g\sqrt{2}$. As we shall see, all the other amplitudes we discuss in the rest of this paper will be obtained with the BCFW recursion, and the Lagrangian will not be used again.

\subsection{Two on-shell and one off-shell gluons}
\label{sec:2on-1off}
Consider now amplitudes with two on-shell and one off-shell
external lines, ${\cal A}_3(0^{h_0}1^{h_1}2^*)$, where the momentum $p_2^\mu$ comes from a source moving in the direction $v_2^\mu$, and therefore obeys $p_2\cdot v_2=0$. In this configuration,
it is convenient to apply a shift on the external lines $1$ and $2$,
for which the unique contribution is of the type given in
eqs.~(\ref{eq:C}) or (\ref{eq:D}) (depending on whether the off-shell
external line $2$ is the one we denoted $i$ or $j$).

\paragraph{\underline{${\cal A}_3(0^+1^+2^*)$}:}
To evaluate this amplitude, the easiest\footnote{Since this shift vector contains the spinor $\big<\v_2\big|$, it is orthogonal to $v_2^\mu$ and therefore cannot produce poles in eikonal propagators.} is to use the shift
$e^\mu = \tfrac{1}{2}\big<\v_2\big|\overline{\sigma}^\mu\big|1\big]$
(i.e., with the notations of the previous sections, the
shifted line $i$ is $2$, and the shifted line $j$ is $1$). The
only contribution is a pole of type (\ref{eq:C}), which gives
\begin{align}
  {\cal A}_3(0^+1^+2^*) = -\frac{i}{\sqrt{2}\,\kappa_2}\,\underbrace{{\cal A}_3(0^+\wh{1}^+\wh{2}^+)}_{=\,0}=0.
\end{align}
The amplitude ${\cal A}_3(0^+\wh{1}^+\wh{2}^+)$ in the right hand side
is a standard on-shell amplitude, which is zero because all its
helicities are positive. Likewise, we have also ${\cal
  A}(0^-1^-2^*)=0$.

\paragraph{\underline{${\cal A}_3(0^+1^-2^*)$}:} 
A proper behavior when $|z|\to\infty$ is guaranteed by the shift
$e^\mu= \tfrac{1}{2}\big<1\big|\overline{\sigma}^\mu\big|\v_2\big]$, that
gives a contribution of type (\ref{eq:D}):
\begin{align}
  {\cal A}_3(0^+1^-2^*)
  =
  \frac{i}{\sqrt{2}\,\overline{\kappa}_2}\,{\cal A}_3(0^+\wh{1}^-\wh{2}^-)
  =
  \frac{ig}{\overline{\kappa}_2}\frac{\abk{\wh{1}\wh{2}}^3}{\abk{\wh{2}0}\abk{0\wh{1}}}
  =
  \frac{ig}{\overline{\kappa}_2}\frac{\abk{1\v_2}^3}{\abk{\v_20}\abk{01}}.
  \label{eq:3p-a}
\end{align}
It turns out that this amplitude admits an equivalent expression in terms of square brackets:
\begin{align}
  {\cal A}_3(0^+1^-2^*)=
  \frac{ig}{\kappa_2}\frac{\sbk{\v_20}^3}{\sbk{01}\sbk{1\v_2}}.
  \label{eq:3p-b}
\end{align}
The equivalence of these two expressions is shown in appendix \ref{sec:app-3p-ab}.

\paragraph{\underline{${\cal A}_3(0^-1^+2^*)$}:} Using the shift $e^\mu=\tfrac{1}{2}\big<\v_2\big|\overline{\sigma}^\mu\big|1\big]$ in order to have a proper behavior at $|z|\to\infty$, we obtain a contribution of type (\ref{eq:C}),
\begin{align}
  {\cal A}_3(0^-1^+2^*) =
  \frac{ig}{\kappa_2}\frac{\sbk{1\v_2}^3}{\sbk{\v_20}\sbk{01}},
\end{align}
which is also equal to
\begin{align}
  {\cal A}_3(0^-1^+2^*)
  =
  \frac{ig}{\overline{\kappa}_2}\frac{\abk{\v_20}^3}{\abk{01}\abk{1\v_2}}.
\end{align}
Note that we could also have obtained these expressions from ${\cal
  A}_3(0^- 1^+ 2^*)=-{\cal A}_3(1^+ 0^- 2^*)$ (this follows from the
fact that $3$-gluon amplitudes are fully symmetric; but since the color
factor $f^{abc}$ is antisymmetric, their kinematical part must also be
antisymmetric).

\subsection{One on-shell and two off-shell gluons}
\label{sec:1on-2off}
\paragraph{\underline{${\cal A}_3(0^+1^*2^*)$}:} Using a shift\footnote{Here, we have two source directions $v_1^\mu$ and $v_2^\mu$. The chosen shift direction is orthogonal to both. Alternatively, we could have chosen $e^\mu=\tfrac{1}{2}\big<\v_1\big|\overline{\sigma}^\mu\big|\v_2\big]$ to reach the same result.} $e^\mu = \tfrac{1}{2}\big<\v_2\big|\overline{\sigma}^\mu\big|\v_1\big]$, we get two contributions of type (\ref{eq:C}) and (\ref{eq:D}), respectively:
\begin{align}
  {\cal A}_3(0^+1^*2^*)
  &=
  -\frac{i}{\sqrt{2}\,\kappa_2}\underbrace{{\cal A}_3(0^+\wh{1}^*\wh{2}^+)}_{=\,0}
  +\frac{i}{\sqrt{2}\,\overline{\kappa}_1}{\cal A}_3(0^+\wh{1}^-\wh{2}^*)
  \nonumber\\
  &=
  \frac{i}{\sqrt{2}\,\overline{\kappa}_1}
  \frac{ig}{\wh{\overline{\kappa}}_2}
  \frac{\abk{\wh{1}\wh{\v}_2}^3}{\abk{\wh{\v}_20}\abk{0\wh{1}}}
  .
\end{align}
In this expression, the spinors $\big|\wh{\v}_2\big>,\big|\wh{\v}_2\big]$
    parameterize the direction of the source $2$ after the
    shift. However, the direction of motion of the sources is not
    affected by the shift, and we therefore have
    $\big|\wh{\v}_2\big>=\big|\v_2\big>,\big|\wh{\v}_2\big]=\big|\v_2\big]$. Moreover,
    for this shift, we have
    $\wh{\overline{\kappa}}_2=\overline{\kappa}_2$.  And from the
    penultimate equation in eqs.~(\ref{eq:D1}), we also get
    $\big|\wh{1}\big>=\big|\v_1\big>$. Therefore, this amplitude reads\footnote{The Yang-Mills Feynman rules
produce 3 distinct graphs for this amplitude (this number is reduced in some gauges), while the BCFW recursion produces
a single term, which is directly gauge invariant. The derivation of this amplitude in perturbation theory is sketched in Appendix \ref{sec:A3-eik-der}.}
\begin{align}
  {\cal A}_3(0^+1^*2^*)
  =-\frac{g}{\sqrt{2}\,\overline{\kappa}_1{\overline{\kappa}}_2}
  \frac{\abk{{\v_1}{\v_2}}^3}{\abk{{\v_2}0}\abk{0{\v_1}}}
  .     
\end{align}
  
\paragraph{\underline{${\cal A}_3(0^-1^*2^*)$}:} It is convenient to use a shift $e^\mu=\tfrac{1}{2}\big<\v_1\big|\overline{\sigma}^\mu\big|\v_2\big]$, that gives
\begin{align}
  {\cal A}_3(0^-1^*2^*)
  &=
  -\frac{i}{\sqrt{2}\,\kappa_1}\,{\cal A}_3(0^-\wh{1}^+\wh{2}^*)
  +\frac{i}{\sqrt{2}\,\overline{\kappa}_2}\,\underbrace{{\cal A}_3(0^-\wh{1}^* \wh{2}^-)}_{=\,0}\nonumber\\
  &=\frac{g}{\sqrt{2}\,\kappa_1\wh{\kappa}_2}\,\frac{\sbk{\wh{1}\wh{\v}_2}^3}{\sbk{\wh{\v}_20}\sbk{0\wh{1}}}=\frac{g}{\sqrt{2}\,\kappa_1\kappa_2}\,\frac{\sbk{{\v_1}{\v_2}}^3}{\sbk{{\v_2}0}\sbk{0{\v_1}}}.
\end{align}
In the last step, we have used the fact that
$\big|\wh{\v}_2\big]=\big|\v_2\big]$ because the direction of
motion of sources is not affected by the shift, $\wh{\kappa}_2=\kappa_2$ for the shift used here,
and $\big|\wh{1}\big]=\big|\v_1\big]$ for contributions of type (\ref{eq:C}).

Note that all the above formulae are valid only for light-like
directions $v_1^\mu$ and $v_2^\mu$ of the sources, since it was
assumed that $v_{1\mu}\sigma^\mu=\big|\v_1\big]\big<\v_1\big|$ and
  $v_{2\mu}\sigma^\mu=\big|\v_2\big]\big<\v_2\big|$.

\section{Scattering of a gluon off an external field}
\label{sec:scatt}
In this section, we study the scattering of a gluon off an external field produced by sources of direction $v_2^\mu$. The incoming and outgoing on-shell gluons are labeled $0$ and $1$. The gluons labeled $2,3,\cdots$ are all off-shell (with $p_2\cdot v_2=p_3\cdot v_2=\cdots=0$) and the spinors $\big|\v_2\big>,\big|\v_3\big>,\cdots$ are all equal and encode the common direction $v_2^\mu$,
\begin{align}
-\v_{2\mu} \overline{\sigma}^\mu=
\big|\v_2\big>\big[\v_2\big|
=
\big|\v_3\big>\big[\v_3\big|
=\cdots
\end{align}

\subsection{Order one in the external field}
The amplitudes ${\cal A}_3(0^{h_0}1^{h_1}2^*)$ with two on-shell
gluons and one off-shell gluon contribute to the scattering of a gluon
off an external field, at lowest order in this external field.  These
amplitudes are zero if $h_0=h_1$ because eikonal scattering cannot
flip the helicity of the incoming gluon (recall that the two
helicities are defined with respect to incoming momenta).

Before proceeding to amplitudes with more off-shell gluons, let us
note that ${\cal A}_3(0^+1^- 2^*)$ can also be written as
\begin{align}
  {\cal A}_3(0^+1^- 2^*)
  =-\frac{ig}{\big|\kappa_2\big|^2} \frac{\abk{1\v_2}^2\sbk{0\v_2}^2}{\abk{0\v_2}\sbk{0\v_2}}
  =\frac{ig}{\big|\kappa_2\big|^2} \frac{\abk{1\v_2}^2\sbk{0\v_2}^2}{\abk{1\v_2}\sbk{1\v_2}}.
\end{align}
The first equality is obtained by using
$\kappa_2=\abk{01}\sbk{0\v_2}/\abk{1\v_2}$ (see appendix
\ref{sec:app-3p-ab}), and the second equality uses
$\abk{1\v_2}\sbk{1\v_2}+\abk{0\v_2}\sbk{0\v_2}=0$ (which follows from $v_2\cdot
(p_0+p_1)=0$).

Partial amplitudes with different color ordering can be obtained
simply by using the symmetries of 3-point amplitudes:
\begin{align}
  {\cal A}_3(0^+ 2^* 1^-) = {\cal A}_3(1^- 0^+ 2^*) = -{\cal A}_3(0^+ 1^- 2^*).
\end{align}
(The first equality uses the cyclic invariance of color ordered
amplitudes, and the second one uses the fact that 3-point partial
amplitudes are antisymmetric.)  We have now all the ingredients to
reconstruct the full-color amplitude at lowest order in the external
field,
\begin{align}
  {\cal M}_3(0_{a_0}^+ 1^-_{a_1} 2^*_{a_2})
  &=
  2 \Big\{
  {\rm tr}\,(t^{a_0}t^{a_1}t^{a_2})\,{\cal A}_3(0^+ 1^- 2^*)
  +
  {\rm tr}\,(t^{a_0}t^{a_2}t^{a_1})\,{\cal A}_3(0^+ 2^* 1^-)
  \Big\}\nonumber\\
  &=
  2\,
  {\rm tr}\,(t^{a_0}[t^{a_1},t^{a_2}])\,
  {\cal A}_3(0^+ 1^- 2^*)
  =
  \frac{gf^{a_0 a_1 a_2}}{\big|\kappa_2\big|^2} \frac{\abk{1\v_2}^2\sbk{0\v_2}^2}{\abk{0\v_2}\sbk{0\v_2}}.
\end{align}

\subsection{Order two in the external field}
The full-color scattering amplitude of a gluon off an external field
at second order in the external field reads
\begin{align}
  {\cal M}_4(0^{h_0}_{a_0}1^{h_1}_{a_1}2^*_{a_2} 3^*_{a_3})
  = &2\Big\{
  {\rm tr}\,(t^{a_0}t^{a_1}t^{a_2}t^{a_3})\,{\cal A}_4(0^{h_0}1^{h_1}2^*3^*)
  \nonumber\\&
  +
  {\rm tr}\,(t^{a_0}t^{a_1}t^{a_3}t^{a_2})\,{\cal A}_4(0^{h_0}1^{h_1}3^*2^*)
  \nonumber\\&
  +
  {\rm tr}\,(t^{a_0}t^{a_2}t^{a_1}t^{a_3})\,{\cal A}_4(0^{h_0}2^*1^{h_1}3^*)
  \nonumber\\&
  +
  {\rm tr}\,(t^{a_0}t^{a_2}t^{a_3}t^{a_1})\,{\cal A}_4(0^{h_0}2^*3^*1^{h_1})
  \nonumber\\&
  +
  {\rm tr}\,(t^{a_0}t^{a_3}t^{a_1}t^{a_2})\,{\cal A}_4(0^{h_0}3^*1^{h_1}2^*)
  \nonumber\\&
  +
  {\rm tr}\,(t^{a_0}t^{a_3}t^{a_2}t^{a_1})\,{\cal A}_4(0^{h_0}3^*2^*1^{h_1})
   \Big\}.
\end{align}

\paragraph{${\bs{ {\cal A}_4(0^+1^+ 2^* 3^*)}}$}
Consider first the case where the two on-shell gluons have the same
helicity (defined with respect to incoming momenta, which means that
the helicity of the gluon actually flips during the scattering). We
start by a shift on the lines ($2,3$), of the type discussed in
section \ref{sec:vv}. With this shift, the external off-shell propagators and the internal eikonal propagators do not have poles. Since this shift can only produce poles to
internal gluon propagators, we obtain
  \begin{align}
    {\cal A}_4(0^+1^+ 2^* 3^*)
    =\sum_{h=\pm}
    {\cal A}_3(0^+\wh{I}^{+h}\wh{3}^*)\frac{i}{K_{_I}^2}
    {\cal A}_3(-\wh{I}^{-h}\,1^+\wh{2}^*).
  \end{align}
Note now that ${\cal A}_3(0^+\wh{I}^{+h}\wh{3}^*)=0$ unless $h=-$,
while ${\cal A}_3(-\wh{I}^{-h}\,1^+\wh{2}^*)=0$ unless $h=+$.  We thus
conclude that the scattering amplitude with a helicity flip vanishes
at order 2 in the external field, ${\cal A}_4(0^+1^+ 2^* 3^*)=0$. We
also have ${\cal A}_4(0^-1^- 2^* 3^*)=0$. Moreover, all other
amplitudes with $0^+ 1^+$ or $0^-1^-$ are zero, regardless of their
relative ordering with the two off-shell gluons.

\paragraph{${\bs{ {\cal A}_4(0^+1^- 2^* 3^*)}}$}
Let us consider now the amplitude ${\cal A}_4(0^+1^- 2^* 3^*)$, where
the helicity of the on-shell gluon does not flip during the
scattering. We again use the shift
$e^\mu=\tfrac{1}{2}\big<\v_2\big|\overline{\sigma}^\mu\big|\v_3\big]=v^\mu$
  applied to the lines ($2,3$). This gives
\begin{align}
  {\cal A}_4(0^+1^- 2^* 3^*)
  =
  \sum_{h=\pm}
    {\cal A}_3(0^+\wh{I}^{+h}\wh{3}^*)\frac{i}{K_{_I}^2}
    {\cal A}_3(-\wh{I}^{-h}\,1^-\wh{2}^*).
\end{align}
(Our definition of the intermediate momentum $K_{_I}$ is such that $p_0+p_3+K_{_I}=0$.)
A non-zero contribution is obtained with $h=-$,
\begin{align}
  {\cal A}_4(0^+1^- 2^* 3^*)
  =
    {\cal A}_3(0^+\wh{I}^{-}\wh{3}^*)\frac{i}{K_{_I}^2}
    {\cal A}_3(-\wh{I}^{+}\,1^-\wh{2}^*).
\end{align}
The two amplitudes that appear in the non-zero term read\footnote{To flip the sign of a momentum in an amplitude, we use $\big|-p\big>=i\big|p\big>$, $\big|-p\big]=i\big|p\big]$.}
\begin{align}
  {\cal A}_3(0^+\wh{I}^-\wh{3}^*)
  &=
  \frac{ig}{\overline{\kappa}_3}
  \frac{{\big<{\wh{I}\,\wh{\v}_3}\big>}^3}{\big<\wh{\v}_30\big>\big<0\wh{I}\big>}
  =
   \frac{ig}{\overline{\kappa}_3}
   \frac{{\big<{\wh{I}\,{\v_3}}\big>}^3}{\big<{\v_3}0\big>\big<0\wh{I}\big>}
   ,\nonumber\\
  {\cal A}_3(-\wh{I}^+{1}^-{\wh{2}}^*)
  &=
  -\frac{ig}{\overline{\kappa}_2}
  \frac{\big<{1}{\wh{\v}_2}\big>^3}
       {\big<{\wh{\v}_2}\wh{I}\big>\big<\wh{I}{1}\big>}
  =
  -\frac{ig}{\overline{\kappa}_2}
  \frac{\big<{1}{{\v_2}}\big>^3}
       {\big<{{\v_2}}\wh{I}\big>\big<\wh{I}{1}\big>}
  .
\end{align}
The denominator of the intermediate propagator reads
\begin{align}
  K_{_I}^2=(p_0+p_3)^2=(p_1+p_2)^2.
\end{align}
Recall also that $\big|\v_2\big>=\big|\v_3\big>$ and
$\big|\v_2\big]=\big|\v_3\big]$ because the off-shell gluons $2$ and $3$
    are produced by sources that travel in the same light-cone
    direction $v_2^\mu$.  At this stage, the amplitude ${\cal A}_4(0^+1^- 2^*
    3^*)$ is given by
    \begin{align}
      {\cal A}_4(0^+1^- 2^* 3^*)
      =
      -i \frac{g^2}{\overline{\kappa}_2\overline{\kappa}_3 (p_0+p_3)^2}
      \frac{\abk{1\v_2}^3\abk{\wh{I}\v_2}^2}{\abk{\v_20}\abk{0\wh{I}}\abk{\wh{I}1}}.
      \label{eq:A4+-**}
    \end{align}
    The pole that gives this contribution occurs when
    \begin{align}
      0
      = (p_0+p_3)^2-2ze\cdot(p_0+p_3)
      =
      (p_0+p_3)^2-2ze\cdot p_0,
    \end{align}
    i.e., at $z_*=(p_0+p_3)^2/(2e\cdot p_0)$. At this $z_*$,
    $\wh{K}_{_I}^\mu$ is on-shell and we have
    \begin{align}
      \big|\wh{I}\big>\big[\wh{I}\big|=-\wh{K}_{_I}^\mu\overline{\sigma}_\mu
      =
      -\big|0\big>\big[0\big|
        +
      \big(z_*-\frac{q\cdot p_3}{q\cdot v_2}\big) \big|\v_2\big>\big[\v_2\big|
        +
        \frac{\kappa_3}{\sbk{\v_2\q}}\big|\v_2\big>\big[\q\big|
          +
          \frac{\overline{\kappa}_3}{\abk{\v_2\q}}\big|\q\big>\big[\v_2\big|.
    \end{align}
    A convenient choice for the auxiliary vector is $q=p_0$. This
    leads to
    \begin{align}
      \big|\wh{I}\big>\big[\wh{I}\big|
      &=-\big|0\big>\big[0\big|+\frac{p_3^2}{2p_0\cdot v_2} \big|\v_2\big>\big[\v_2\big|
      +
      \frac{\kappa_3}{\sbk{\v_20}}\big|\v_2\big>\big[0\big|
        +
        \frac{\overline{\kappa}_3}{\abk{\v_20}}\big|0\big>\big[\v_2\big|
          \nonumber\\
          &=
          -\Big(\big|0\big>+\frac{\kappa_3}{\sbk{0\v_2}}\big|\v_2\big>\Big)
          \Big(\big[0\big|+\frac{\overline{\kappa}_3}{\abk{0\v_2}}\big[\v_2\big|\Big).
    \end{align}
(We used $2p_0\cdot v_2 = \abk{0\v_3}\sbk{0\v_3}$ and
    $p_3^2=-\kappa_3\overline{\kappa}_3$).  We see that at $z=z_*$,
    $-\wh{K}_{_I}^\mu\overline{\sigma}_\mu$ factorizes as a direct
    product of spinors, as expected since this is the value of $z$ at
    which $\wh{K}_{_I}^\mu$ is on-shell. From this result, we read the
    value\footnote{There is some arbitrariness in how we split the
    prefactor $-1$ among $\big|\wh{I}\big>$ and
    $\big[\wh{I}\big|$. However, this choice does not matter since
      ${\cal A}_4$ has equal numbers of spinors $\big|\wh{I}\big>$ in
      the numerator and denominator. } of $\big|\wh{I}\big>$ at the
    pole:
    \begin{align}
      \big|\wh{I}\big>=\big|0\big>+\frac{\kappa_3}{\sbk{0\v_2}}\big|\v_2\big>,
    \end{align}
    which gives
    \begin{align}
      \abk{0\wh{I}}=\frac{\kappa_3}{\sbk{0\v_2}}\abk{0\v_2},
      \quad
      \abk{1\wh{I}}=\abk{10}+\frac{\kappa_3}{\sbk{0\v_2}}\abk{1\v_2},
      \quad
      \abk{\v_2\wh{I}}=\abk{\v_20}.
    \end{align}

    By substitution in eq.~(\ref{eq:A4+-**}), we obtain
    \begin{align}
      {\cal A}_4(0^+1^- 2^* 3^*)
      =
      i \frac{g^2}{\overline{\kappa}_2\big|\kappa_3\big|^2 (p_0+p_3)^2}
      \frac{\abk{1\v_2}^3\sbk{\v_20}^2}{\sbk{0\v_2}\abk{01}+\kappa_3\abk{\v_21}}.
    \end{align}
    Starting from momentum conservation $p_0+p_1+p_2+p_3=0$ expressed as
    \begin{align}
      \big|0\big>\big[0\big|
        +
      \big|1\big>\big[1\big|
        -
        \kappa_2\frac{\big|\v_2\big>\big[\q_2\big|}{\sbk{\v_2\q_2}}
          -
          \overline{\kappa}_2\frac{\big|\q_2\big>\big[\v_2\big|}{\abk{\v_2\q_2}}
            -
            \kappa_3\frac{\big|\v_2\big>\big[\q_3\big|}{\sbk{\v_2\q_3}}
          -
          \overline{\kappa}_3\frac{\big|\q_3\big>\big[\v_2\big|}{\abk{\v_2\q_3}}
            = \# \, \big|\v_2\big>\big[\v_2\big|,
            \label{eq:mom-cons}
    \end{align}
    where $\#$ denotes the longitudinal component of $p_2+p_3$ (we do
    not need its value), and inserting this identity inside
    $\big<1\big|\cdots\big|\v_2\big]$, we obtain
            \begin{align}
              \sbk{0\v_2}\abk{01}+\kappa_3\abk{\v_21}=\kappa_2 \abk{1\v_2}.
            \end{align}
            (One does not need to assign specific values to the
            auxiliary vectors $q_2$ and $q_3$ to obtain this, and we
            observe that they drop out from the result.)  Thus, the
            amplitude reads
            \begin{align}
              {\cal A}_4(0^+1^- 2^* 3^*)
      =
      i \frac{g^2\sbk{0\v_2}^2\abk{1\v_2}^2}{\big|\kappa_2\big|^2\big|\kappa_3\big|^2 (p_0+p_3)^2}.
            \end{align}
In this expression, we recognize $\big|\kappa_2\big|^2=p_{2\perp}^2$
and $\big|\kappa_3\big|^2=p_{3\perp}^2$, the virtualities of the two
off-shell gluons, $(p_0+p_3)^2$ the denominator of the propagator of
the gluon after the first scattering, and the numerator
$\abk{1\v_2}^2\sbk{\v_20}^2$ comes from the momentum dependence of the gluon
vertices.

\paragraph{${\bs{ {\cal A}_4(0^+3^* 2^*1^-)}}$}
Partial amplitudes with different orderings are calculated by the same
me\-thod, so we will skip most of the intermediate steps. We have
\begin{align}
  { {\cal A}_4(0^+3^* 2^*1^-)}
  &=
  {\cal A}_3(\wh{2}^*1^-\wh{I}^+)\frac{i}{K_{_I}^2}{\cal A}_3(-\wh{I}^-0^+\wh{3}^*)
  \nonumber\\
  &=
  i\frac{g^2\sbk{0\v_2}^2\abk{1\v_2}^2}{\big|\kappa_2\big|^2\big|\kappa_3\big|^2 (p_0+p_3)^2}.
\end{align}

\paragraph{${\bs{ {\cal A}_4(0^+3^* 1^- 2^*)}}$}
For this ordering, a BCFW shift on the lines $2^*$ and $3^*$ produces
two terms,
\begin{align}
  {\cal A}_4(0^+2^* 1^- 3^*)
  &=
  {\cal A}_3(0^+\wh{2}^*\wh{I}^-)\frac{i}{K_{_I}^2}{\cal A}_3(-\wh{I}^+ 1^- \wh{3}^*)
\nonumber\\
  &+
  {\cal A}_3(\wh{3}^*0^+\wh{J}^-)\frac{i}{K_{_J}^2}{\cal A}_3(-\wh{J}^+ \wh{2}^* 1^-).
\end{align}
The rest of the steps is identical, and we obtain

\begin{align}
  {\cal A}_4(0^+2^* 1^- 3^*)
  =
  -i\frac{g^2\sbk{0\v_2}^2\abk{1\v_2}^2}{\big|\kappa_2\big|^2\big|\kappa_3\big|^2}
  \Big[\frac{1}{(p_0+p_2)^2}+\frac{1}{(p_0+p_3)^2}\Big].
\end{align}
Note that this amplitude is symmetric under the exchange of the two off-shell lines, $2^*$ and $3^*$. Therefore, we also have
\begin{align}
  {\cal A}_4(0^+3^* 1^- 2^*)
  =
  {\cal A}_4(0^+2^* 1^- 3^*).
\end{align}

\paragraph{Full-color amplitude}
We can combine all the 4-point partial amplitudes obtained so far in
order to reconstruct the corresponding full-color amplitude. It can be
written as
\begin{align}
  {\cal M}_4(0^+_{a_0}1^-_{a_1}2^*_{a_2} 3^*_{a_3})
  &=
  2i\frac{g^2\sbk{0\v_2}^2\abk{1\v_2}^2}{\big|\kappa_2\big|^2\big|\kappa_3\big|^2}
  \Big[
    \frac{{\rm tr}\,([t^{a_0},t^{a_2}][t^{a_3},t^{a_1}])}{(p_0+p_2)^2}
    \nonumber\\
    &\qquad\qquad\qquad\qquad
    +
    \frac{{\rm tr}\,([t^{a_0},t^{a_3}][t^{a_2},t^{a_1}])}{(p_0+p_3)^2}
    \Big]
  \nonumber\\
  &= -\frac{i}{2}\frac{g^2\sbk{0\v_2}^2\abk{1\v_2}^2}{\big|\kappa_2\big|^2\big|\kappa_3\big|^2}
  \Big[
    \frac{f^{a_0a_2e}f^{ea_3a_1}}{(p_0+p_2)^2}
    +
    \frac{f^{a_0a_3e}f^{ea_2a_1}}{(p_0+p_3)^2}
    \Big]
  .
\end{align}
The two terms in this formula differ from one another simply by the exchange of the two off-shell gluons.
            
\subsection{Order $n$ in the external field}
\paragraph{Conjectured expression} The previous examples of partial amplitudes for the scattering of a gluon
off an external field suggest that they all have the following generic
form:
\begin{align}
  {\cal A}(0^+\underbrace{\alpha_1^*...\alpha_p^*}_{\mbox{\scriptsize$p$ gluons}} 1^-\underbrace{\beta_1^*... \beta_q^*}_{\mbox{\scriptsize$q$ gluons}} )
  =
  i \frac{\big(g\sbk{1\v_2}\abk{1\v_2}\big)^{p+q}}{\prod_a \big|\kappa_a\big|^2}\frac{\sbk{0\v_2}^2}{\sbk{1\v_2}^2}\,{\cal K}(p_0,\{\alpha\},\{\beta\}),
  \label{eq:ansatz-scatt}
\end{align}
where $n$ is the number of off-shell lines.  The product of the
$\big|\kappa_a\big|^2$ in the denominator extends to all the off-shell
lines. The spinors $\big|\v_2\big>$ and $\big|\v_2\big]$ encode
          the common light-like direction of all the sources, via
          $-v_{2\mu}\overline{\sigma}^\mu=\big|\v_2\big>\big[\v_2\big|$. The factor ${\cal K}(p_0,\{\alpha\},\{\beta\})$ contains the momentum dependence not
            explicitly written in the form of brackets and
            $\big|\kappa_a\big|^2$. This quantity is a sum of products
            of denominators, and it depends on the momentum $p_0$ of the on-shell gluon entering on the left and on the momenta of the off-shell ones (separated in two sets, depending on their position with respect to the on-shell gluons in the cyclic ordering). Note also that in eq.~(\ref{eq:ansatz-scatt}), the explicitly written prefactor does not depend on the position of the off-shell lines in the cyclic ordering, while the kinematical factor ${\cal K}(p_0,\{\alpha\},\{\beta\})$ does.

\paragraph{BCFW recursion}
            Let us first use BCFW recursion in order to prove the validity of
            this ansatz.  This will also provide a recursive
            way to calculate the kinematical factor ${\cal K}$. To define the shift, we choose two adjacent
            off-shell lines. Schematically, the BCFW recursion
            expresses an amplitude in terms of smaller amplitudes as follows,
            \setbox1\hbox to
            22.5mm{\includegraphics[width=22.5mm]{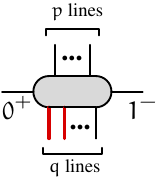}} \setbox2\hbox
            to 43mm{\includegraphics[width=43mm]{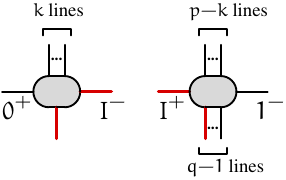}}
            \begin{equation}
              {\cal A}^{(p,q)}\equiv\raise -12.5mm\box1 = \sum_{k=0}^p\;\;\raise -13mm\box2
            \end{equation}
            In this representation, we
            have indicated in red the lines that carry a shifted
            momentum. The two factors in the right hand side are
            amplitudes with two on-shell gluons of helicities $+$ and
            $-$, and a number of off-shell gluons. Moreover, the
            numbers of off-shell lines in these factors, $k+1$ and
            $p+q-k-1$, respectively, are both strictly smaller
            than the $p+q$ off-shell lines in the original
            amplitude. This is therefore a recursion on the number of
            off-shell lines present in the amplitude.

            According to the above ansatz, the first amplitude in the right hand side
            is of the form
            \begin{align}
              {\cal A}(0^+\alpha_1^*... \alpha_k^*\,-\!\wh{I}^-\,\wh{\beta}_q^*)=i \frac{\big(g\sbk{\!-\wh{I}\v_2}\abk{\!-\wh{I}\v_2}\big)^{k+1}}{\prod_{a\in L} \big|\kappa_a\big|^2}\frac{\sbk{0\v_2}^2}{\sbk{\!-\wh{I}\v_2}^2}\,{\cal K}(p_0,\{\alpha_1... \alpha_k\},\{\wh{\beta}_q\}),
            \end{align}
            while the second amplitude reads
            \begin{align}
              {\cal A}(\wh{I}^+\alpha_{k+1}^*... \alpha_p^*1^- \beta_1^*... \wh{\beta}_{q-1}^*)
              &=
              i \frac{\big(g\sbk{1\v_2}\abk{1\v_2}\big)^{p+q-k-1}}{\prod_{a\in R} \big|\kappa_a\big|^2}
              \frac{\sbk{\wh{I}\v_2}^2}{\sbk{1\v_2}^2}\nonumber\\
              &\qquad\times{\cal K}(\wh{K}_{_I},\{\alpha_{k+1}... \alpha_p\},\{\beta_1...\wh{\beta}_{q-1}\}),
            \end{align}
            where the notation $a\in L$ (resp. $a\in R$) denote the
            set of all off-shell gluon indices in the left (resp. right)
            factors. The momentum of the intermediate gluon is
            \begin{align}
              K_{_I}=p_0+\sum_{a\in L}p_a,
            \end{align}
            and the corresponding shifted momentum reads
            \begin{align}
              \wh{K}_{_I}=p_0+\sum_{a\in L}p_a-z v_2.
            \end{align}
            The value of $z$ at which $\wh{K}_{_I}^2=0$ is given by
            \begin{align}
              z_*
              &=\frac{(p_0+\sum_{a\in L}p_a)^2}{2p_0\cdot v_2}
              =
              \frac{2p_0\cdot\sum_{a\in L}p_a+\sum_{a,b\in L}p_a\cdot p_b}{2p_0\cdot v_2}
              \nonumber\\
              &=
              \frac{2p_0\cdot\sum_{a\in L}p_a}{2p_0\cdot v_2}
              -\frac{\sum_{a,b\in L}\kappa_a\overline{\kappa}_b}{\sbk{0\v_2}\abk{0\v_2}}.
            \end{align}
            Using $q=p_0$ as auxiliary vector, we obtain
            \begin{align}
              \big|\wh{I}\big>\big[\wh{I}\big|
                =
              \big|0\big>\big[0\big|
                +\mbox{terms in $\big|0\big>\big[\v_2\big|$, $\big|\v_2\big>\big[0\big|$, or $\big|\v_2\big>\big[\v_2\big|$},
            \end{align}
            and at $z_*$ we have $ \big<\wh{I}\big|=\big<0\big|+\cdots
            \big<\v_2\big|$,$ \big[\wh{I}\big|=\big[0\big|+\cdots
                \big[\v_2\big|$, $ \big<\wh{I}\v_2\big>=\big<0\v_2\big>$, and $ \big[\wh{I}\v_2\big]=\big[0\v_2\big]$.
            Therefore, the amplitude we started from reads
            \begin{align}
              {\cal A}(0^+\{\alpha\}1^-\{\beta\})&=i
              \frac
                  {\big(g\sbk{1\v_2}\abk{1\v_2}\big)^{p+q}}
                  {\prod_{a} \big|\kappa_a\big|^2}
                  \frac{\sbk{0\v_2}^2}{\sbk{1\v_2}^2}\nonumber\\
                  &\times
                  \sum_{k=0}^p
                  \frac
                      {{\cal K}(p_0,\{\alpha_1...\alpha_k\},\{\wh{\beta}_q\})\;{\cal K}(\wh{K}_{_I},\{\alpha_{k+1}... \alpha_p\},\{\beta_1...\wh{\beta}_{q-1}\})}
                      {(p_0+\sum_{a\in L}p_a)^2}.
            \end{align}
            (We have used $\sbk{0\v_2}\abk{0\v_2}=-\sbk{1\v_2}\abk{1\v_2}$.) From this formula, we see that
            the amplitude ${\cal A}^{(p,q)}$ has the conjectured form,
            and we obtain a recursive formula for the kinematical
            factors,
            \begin{align}
              &{\cal K}(p_0,\{\alpha_1... \alpha_p\},\{\beta_1... \beta_q\})\nonumber\\
              &\quad=
              \sum_{k=0}^p
              \frac{
              {\cal K}(p_0,\{\alpha_1... \alpha_k\},\{\wh{\beta}_q\})\;
              {\cal K}(\wh{K}_{_I},\{\alpha_{k+1}... \alpha_p\},\{\beta_1...\wh{\beta}_{q-1}\})}
              {(p_0+\sum_{a\in L}p_a)^2}
              .
              \label{eq:K1}
            \end{align}
            Note that if $q<2$, the BCFW shift
            should be applied to the pair $(\alpha_1,\alpha_2)$ instead. In this case, the same method leads to
            \begin{align}
              &{\cal K}(p_0,\{\alpha_1... \alpha_p\},\{\beta_1... \beta_q\})\nonumber\\
              &\quad=
              \sum_{k=0}^q
              \frac{
              {\cal K}(p_0,\{\wh{\alpha}_1\},\{\beta_{k+1}... \beta_q\})\;
              {\cal K}(\wh{K}_{_I},\{\wh{\alpha}_2... \alpha_p\},\{\beta_1... \beta_k\})}
              {(p_0+\sum_{a\in L}p_a)^2}
              .
              \label{eq:K2}
            \end{align}

\paragraph{Initialization of the recursion}
            From the 3-point and 4-point functions already calculated, we know the following kinematical factors,
            \begin{align}
              &
              {\cal K}(p_0,\{\},\{\beta_1\})=1,\quad  {\cal K}(p_0,\{\alpha_1\},\{\})=-1,
              \nonumber\\
              &
                  {\cal K}(p_0,\{\},\{\beta_1\beta_2\})=\frac{1}{(p_0+p_{\beta_2})^2},\quad
                  {\cal K}(p_0,\{\alpha_1\alpha_2\},\{\})=\frac{1}{(p_0+p_{\alpha_1})^2},\nonumber\\
                  &
                   {\cal K}(p_0,\{\alpha_1\},\{\beta_1\})=-\frac{1}{(p_0+p_{\alpha_1})^2}-\frac{1}{(p_0+p_{\beta_1})^2}.
            \end{align}
            (The notation $\{\}$ denotes an empty list.)
            These results are sufficient, with the help of
            eqs.~(\ref{eq:K1}) and (\ref{eq:K2}), to obtain all other
            partial amplitudes. Note that, with this method, we obtain
            the partial amplitudes without having to calculate the
            numerous Feynman diagrams that would be necessary in
            conventional perturbation theory.

            \paragraph{Amplitudes ${\cal A}(0^+1^-\{\beta^*\})$} The partial amplitudes where the on-shell gluons $0$ and $1$ are adjacent in the cyclic ordering are the simplest. In this case, the recursion produces a single term,
            \begin{align}
            {\cal K}(p_0,\{\},\{\beta_1...\beta_{q+1}\})
            =
              \frac{
              {\cal K}(p_0,\{\},\{\wh{\beta}_{q+1}\})
              {\cal K}(p_0+\wh{p}_{\beta_{q+1}},\{\},\{\beta_1...\wh{\beta}_{q}\})}
              {(p_0+p_{\beta_{q+1}})^2}
              .
            \end{align}
            In the right hand side, the factor ${\cal K}(p_0,\{\},\{\wh{\beta}_{q+1}\})$ on the left is equal to one. The second ${\cal K}$ is made of denominators that are all of the form 
            \begin{align}
            (p_0+\wh{p}_{\beta_{q+1}}+\wh{p}_{\beta_q}+\cdots)^2
            =
            (p_0+{p}_{\beta_{q+1}}+{p}_{\beta_q}+\cdots)^2.
            \end{align}
            Therefore, we have
            \begin{align}
            {\cal K}(p_0+\wh{p}_{\beta_{q+1}},\{\},\{\beta_1...\wh{\beta}_{q}\})
            =
            {\cal K}(p_0+{p}_{\beta_{q+1}},\{\},\{\beta_1...{\beta}_{q}\}),
            \end{align}
            and the recursion simplifies into
            \begin{align}
            {\cal K}(p_0,\{\},\{\beta_1...\beta_{q+1}\})
            =
              \frac{
              {\cal K}(p_0+{p}_{\beta_{q+1}},\{\},\{\beta_1...{\beta}_{q}\})}
              {(p_0+p_{\beta_{q+1}})^2}
              .
            \end{align}
            The solution is trivial:
            \begin{align}
            {\cal K}(p_0,\{\},\{\beta_1...\beta_{q}\})
            =
            \frac{1}{(p_0+p_{\beta_q})^2(p_0+p_{\beta_q}+p_{\beta_{q-1}})^2...(p_0+p_{\beta_q}+..+p_{\beta_2})^2}
            \end{align}

\paragraph{Full-color amplitude}	    In order to reconstruct the full-color amplitude, one should in principle calculate the partial amplitudes with arbitrary orderings, multiply each of them by the trace of generators with the matching cyclic ordering, and sum over all possible orderings. This can be avoided by using an alternate color basis for decomposing the full amplitude \cite{DelDuca:1999rs}:
\begin{align}
&{\cal M}_{n+2}(0^+_{a_0}1^-_{a_1}2^*_{a_2}...(n+1)^*_{a_{n+1}})
\nonumber\\
&\qquad
=
-(-i)^{n} \!\!\!\!\!\!\!
\sum_{\sigma\in{\mathfrak S}(2...n+1)}\!\!\!\!\!
f^{a_0 a_{\sigma_2}d_1}
f^{d_1 a_{\sigma_3}d_2}
\cdots
f^{d_{n-1} a_{\sigma_{n+1}}a_1}
{\cal A}_{n+2}(0^+\sigma_2^*...\sigma_{n+1}^*1^-),
\label{eq:FC-scatt}
\end{align}
where ${\mathfrak S}(2...n+1)$ denotes the set of all permutations of $[2,3,...,n+1]$. The equivalence of the color decomposition in terms of traces and the one in terms of products of structure constants is a consequence of an identity known as the Kleiss-Kuijf relation\footnote{This identity can be established by working out the relationship between the color trace basis and the structure constants basis. Alternatively, one may calculate partial amplitudes with arbitrary orderings in order to check it explicitly. See Appendix \ref{sec:KK} for a discussion along this line.},
\begin{align}
{\cal K}(p_0,\{\alpha_1...\alpha_p\},\{\beta_1...\beta_q\})
=
(-1)^p\sum_{\{\sigma\}\in {\rm OP}(\{\alpha\}^t,\{\beta\})}
{\cal K}(p_0,\{\},\{\sigma\}),
\label{eq:kin}
\end{align}
where $\{\alpha\}^t$ is the list obtained by reversing the order of the elements of the list $\{\alpha\}$, and where
the notation ${\rm OP}(\{\alpha\}^t,\{\beta\})$ denotes the set of ``ordered permutations'' (also called ``mergings'') of the elements of the lists $\{\alpha\}^t$  and $\{\beta\}$, i.e., permutations where we shuffle the elements of $\{\alpha\}^t$ and $\{\beta\}$ without changing the relative orderings of the $\alpha$'s or of the $\beta$'s.

Note that the partial amplitude in the right hand side of eq.~(\ref{eq:FC-scatt}) is also
\begin{align}
{\cal A}_{n+2}(0^+\{\sigma^*\}1^-)
=
(-1)^n {\cal A}_{n+2}(0^+1^-\{\sigma^*\}^t).
\end{align}
By collecting the results obtained so far, one gets the following expression for the scattering amplitude with $n$ off-shell gluons,
\begin{align}
&{\cal M}_{n+2}(0^+_{a_0}1^-_{a_1}2^*_{a_2}...{n+1}^*_{a_{n+1}})
=
-i
\frac{\sbk{0\v_2}^2}{\sbk{1\v_2}^2}
\;
\frac{(ig\sbk{1\v_2}\abk{1\v_2})^n}{\big|\kappa_2\big|^2...\big|\kappa_{n+1}\big|^2}
\nonumber\\
&\quad\times
\sum_{\sigma\in{\mathfrak S}(2...n+1)}
\frac{f^{a_0 a_{\sigma_2}d_1}
f^{d_1 a_{\sigma_3}d_2}
\cdots
f^{d_{n-1} a_{\sigma_{n+1}}a_1}}{(p_0+p_{\sigma_2})^2(p_0+p_{\sigma_2}+p_{\sigma_3})^2\cdots (p_0+p_{\sigma_2}+...+p_{\sigma_n})^2}.
\label{eq:scatt-final}
\end{align}
Note that, thanks to the sum over the permutations $\sigma$, this expression is fully symmetric under the exchanges of the off-shell gluons. This was expected since they play the same role in the full amplitude. 

The various factors in this expression have a transparent interpretation.
The denominator $\big|\kappa_a\big|^2$ is the squared transverse momenta of an off-shell external gluon. The factor $\sbk{1\v_2}\abk{1\v_2}=2p_1\cdot v_2$ comes from  the momentum dependence of the 3-gluon vertex. And the $n-1$ denominators $(p_0+p_{\sigma_2}+\cdots)^2$ come from internal gluon propagators, between each scattering off the external field. Moreover, the prefactor $\sbk{0\v_2}^2/\sbk{1\v_2}^2$ is a phase common to all amplitudes, and it can thus be ignored. Indeed, we have
\begin{align}
\left|\frac{\sbk{0\v_2}^2}{\sbk{1\v_2}^2}\right|^2
=
\frac{\big(\sbk{0\v_2}\abk{0\v_2}\big)^2}
{\big(\sbk{1\v_2}\abk{1\v_2}\big)^2}
=
\left[\frac{2p_0\cdot v_2}{2p_1\cdot v_2}\right]^2
=(-1)^2=1.
\end{align}
One may see readily that these ingredients are the building blocks for the perturbative expansion of a Wilson line, which is indeed the correct answer for this scattering amplitude if not expanded in powers of the external field.
            
Let us recap here the only assumption that was used to derive these amplitudes: we assumed that the gauge in the initial time-slice was such that the initial value of the color field is zero (as opposed to a non-zero pure gauge). This constraint on the initial gauge also selects a specific initial value $\rho_2^{(0)}$ for the source $\rho_2$, out of infinitely many possibilities that differ by gauge transformations. No assumption was needed about the choice of gauge in the bulk, because all the work was performed at the level of gauge invariant amplitudes.
\begin{figure}[htbp]
\begin{center}
\includegraphics[width=\textwidth]{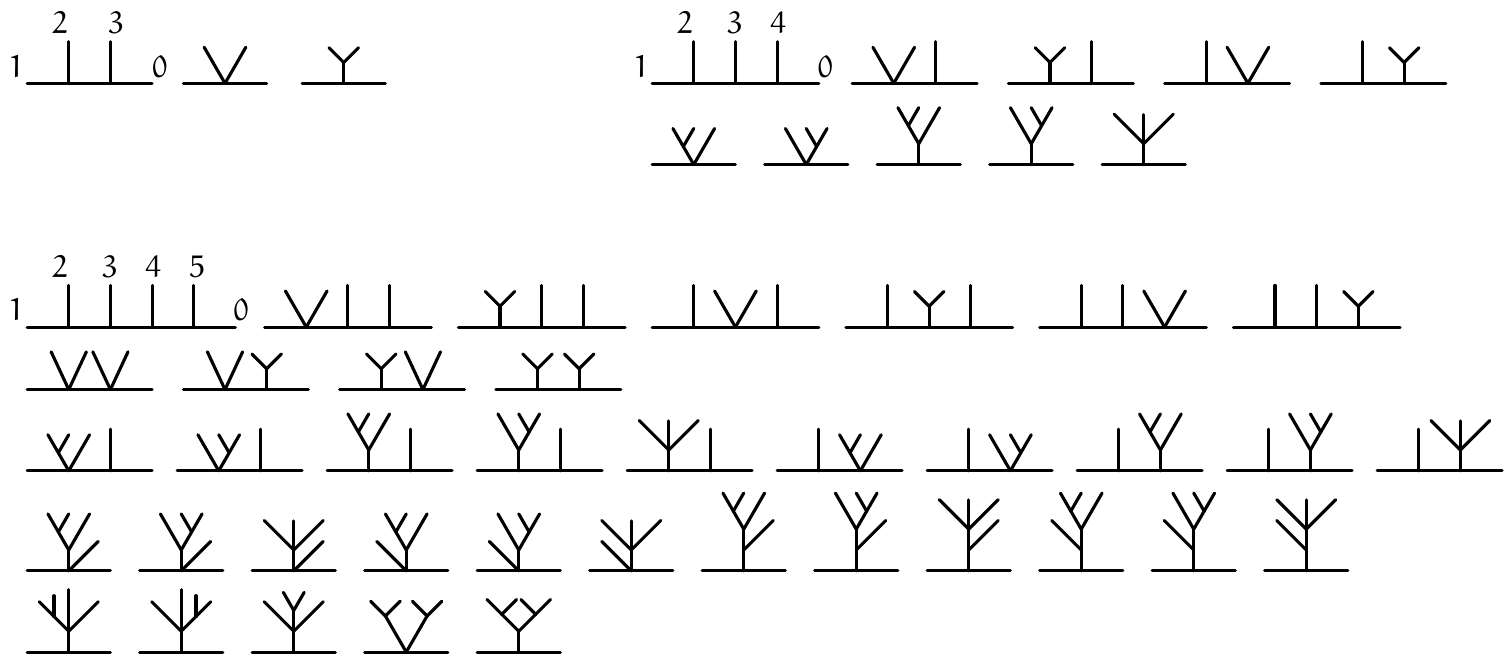}
\end{center}
\caption{\label{fig:scatt-ordered}Feynman graphs for the color ordered amplitudes corresponding to double, triple and quadruple scatterings off an external field. The graphs where a color source evolves due to current conservation are not shown here. The external lines are ordered clockwise: $0, 1$ (on-shell), $2,3,\cdots$ (off-shell, attached to the color sources).}
\end{figure}
In contrast, if derived in terms of Feynman graphs in a generic gauge, these amplitudes would have required the summation of a large number of color ordered graphs (3 for $n=2$, 10 for $n=4$, 38 for $n=5$, ...), as shown in Figure \ref{fig:scatt-ordered}. Moreover, the perturbative approach in terms of Feynman graphs requires that the gauge be fixed so that the gluon propagator is well defined, and is tractable only in gauges that significantly reduce the set of graphs that need to be calculated.

\section{Gluon production}
\label{sec:prod}
We consider now the production of a single on-shell gluon, labeled $0$.
An on-shell gluon cannot be produced from sources all sharing the same direction $v^\mu$. In order to produce a gluon, a ``collision'' that involves two projectiles with different directions $v_1^\mu$ and $v_2^\mu$ is necessary. In this section, we consider two projectiles whose directions are on the light-cone, $v_1^\mu=\delta^{\mu+}$ and $v_2^\mu=\delta^{\mu-}$, respectively. 
Although the solution of this problem to all orders in both sources has only been obtained numerically, the classical Yang-Mills equations at lowest order in the first source and all orders in the second source have been solved analytically in various gauges\footnote{The complexity of the solutions (and of their derivation) varies greatly with the choice of gauge fixing. But it has been checked  that these seemingly very different solutions (in Landau gauge, light-cone gauge and Fock-Schwinger gauge) lead to the same result when they are used to calculate gluon production.} \cite{Dumitru:2001ux,Blaizot:2004wu,Gelis:2005pt}. Our goal in this section is to recover gluon production in this approximation in a gauge agnostic way, by manipulations that only involve gauge independent amplitudes rather than Feynman graphs. We consider amplitudes with a single off-shell gluon (labeled $1$) produced by a source of direction $v_1^\mu$, and one or more off-shell gluons (labeled $2,3,\cdots$) produced by sources of direction $v_2^\mu$. We therefore have
\begin{align}
-v_{1\mu}\overline{\sigma}^\mu 
=
\big|\v_1\big>\big[\v_1\big|,\quad
-v_{2\mu}\overline{\sigma}^\mu 
=
\big|\v_2\big>\big[\v_2\big|
=
\big|\v_3\big>\big[\v_3\big|
=
\cdots
\end{align}

\subsection{Order $\rho_1\rho_2$}
To get started, let us calculate gluon production at order one in $\rho_1$ and order one in $\rho_2$. 
The relevant amplitudes have been obtained in Section \ref{sec:1on-2off},
\begin{align}
{\cal A}_3(0^+1^*2^*)
  =-\frac{g}{\sqrt{2}\,\overline{\kappa}_1{\overline{\kappa}}_2}
  \frac{\abk{{\v_1}{\v_2}}^3}{\abk{{\v_2}0}\abk{0{\v_1}}},\quad
{\cal A}_3(0^-1^*2^*)
  =\frac{g}{\sqrt{2}\,\kappa_1\kappa_2}\,\frac{\sbk{{\v_1}{\v_2}}^3}{\sbk{{\v_2}0}\sbk{0{\v_1}}}.
  \label{eq:A3-1}
\end{align}
Using eqs.~(\ref{eq:lipatov1}) and (\ref{eq:lipatov2}), we can rewrite them uniformly as
\begin{align}
{\cal A}_3(0^\pm 1^*2^*) = \frac{g}{\big|\kappa_1\big|^2\big|\kappa_2\big|^2}\;L\cdot\epsilon_\pm(\p_0,\q),
\label{eq:A3-2}
\end{align}
    where $L^\mu$ is the Lipatov vertex that appears explicitly in the solution of Yang-Mills equations (see appendices \ref{sec:YM} and \ref{sec:lipatov}).
    When written in this way, the equivalence between the results obtained from Yang-Mills equations and from BCFW becomes manifest. Moreover, this form of the amplitudes highlights the structure of the production process: the Lipatov effective vertex takes into account the direct production as well as effects of color source precession due to covariant current conservation, and the denominators $\big|\kappa_1\big|^2\big|\kappa_2\big|^2$ are those of the propagators for the two off-shell gluons in the initial state.  
    
More importantly, the derivation of these results using BCFW does not require that one fixes the gauge in the bulk (only the gauge in the initial time slice must be set so that the color field is zero in the far past). This approach gives in a straightforward manner the gauge invariant gluon production amplitude. In contrast, when solving the Yang-Mills equations, one must first choose a gauge, and a considerable effort is diverted into obtaining gauge-dependent parts of the solution that eventually drop out at the final stage when the solution is used in the LSZ formula.

\subsection{Order $\rho_1 \rho_2^2$}
Next, we consider the production of a gluon of positive helicity, $0^+$, from an off-shell gluon $1$ coming from a source $\rho_1$, and two off-shell gluons $2,3$ coming from sources $\rho_2$. The amplitude we wish to calculate is a 4-point amplitude ${\cal A}_4(0^+1^*2^*3^*)$. In order to perform this calculation, we apply a shift $e^\mu=\tfrac{1}{2}\big<\v_2\big|\overline{\sigma}^\mu\big|\v_2\big]$ to the lines 2 and 3. Note that $e^\mu=v_2^\mu$.
This implies that
\begin{align}
\wh{p}_2\cdot v_2=p_2\cdot v_2=0,\quad
\wh{p}_3\cdot v_2=p_3\cdot v_2=0,
\end{align}
and the shifted momenta $\wh{p}_{2,3}$ are still valid arguments for the source $\rho_2^{(0)}$. Moreover, the momenta of the gluons directly attached to sources 2 and 3 cannot have a pole, since
\begin{align}
\wh{p}_2^2=(p_2+zv_2)^2=p_2^2+2z\underbrace{v_2\cdot p_2}_{=\;0}=p_2^2,\quad
\wh{p}_3^2=(p_3-zv_2)^2=p_3^2-2z\underbrace{v_2\cdot p_3}_{=\;0}=p_2^2.
\end{align}
The only possible poles are in internal gluon propagators, and also in eikonal propagators of the form
\begin{align}
\frac{1}{\wh{p}\cdot v_1}
=
\frac{1}{p\cdot v_1-zv_2\cdot v_1}
=
\frac{1}{p\cdot v_1-z}.
\end{align}
Note that eikonal propagators $(\wh{p}\cdot v_2)^{-1}$ are in fact independent of $z$ and thus cannot have a pole. Therefore, BCFW recursion applied to the amplitude ${\cal A}_4(0^+1^*2^*3^*)$
gives two terms,
\begin{align}
{\cal A}_4(0^+1^*2^*3^*)
=
{\cal A}_3(0^+-\wh{I}^-\wh{3}^*)
\frac{i}{K_{_I}^2}
{\cal A}_3(\wh{I}^+1^*\wh{2}^*)
+
{\cal A}_ 3(0^+-\wh{I}^*\wh{3}^*)
\frac{1}{K_{_I}\cdot v_1}
{\cal A}_3(\wh{I}^*1^*\wh{2}^*).
\end{align}
In the two terms, $K_{_I}=p_0+p_3$ and  $\wh{K}_{_I}=p_0+p_3-zv_2$. However, the values of $z_*$ at the pole (and hence the values of $\wh{K}_{_I}$) are not the same in the two terms, since the poles appear in different objects.

\paragraph{Pole in a gluon propagator} Let us start with the first term, that involves the more familiar pole in a gluon propagator,
\begin{align}
{\cal A}_{4a}(0^+1^*2^*3^*)
=
{\cal A}_3(0^+-\wh{I}^-\wh{3}^*)
\frac{i}{K_{_I}^2}
{\cal A}_3(\wh{I}^+1^*\wh{2}^*)
\end{align}
The momenta $p_0$ and $p_3$ are represented by the following $2\times 2$ matrices
\begin{align}
\overline{P}_0
=
\big|0\big>\big[0\big|,
\quad
\overline{P}_3
=
\frac{q\cdot p_3}{q\cdot v_2}
\big|\v_2\big>\big[\v_2\big|
-\kappa_3\frac{\big|\v_2\big>\big[\q\big|}{\sbk{\v_2\q}}
+
\overline{\kappa}_3
\frac{\big|\q\big>\big[\v_2\big|}{\abk{\q\v_2}},
\end{align}
where $q^\mu$ is an on-shell auxiliary vector that we can choose freely.  The value of $z$ at which $(p_0+p_3-zv_2)^2=0$ is 
\begin{align}
z_* = 
\frac{p_3^2+2p_0\cdot p_3}{2v_2\cdot p_0}.
\end{align}
A convenient choice of $q^\mu$ is $q=p_0$. With this choice, we have at the pole
\begin{align}
\wh{\overline{K}}_{_I}
&=
\big|0\big>\big[0\big|
-\kappa_3\frac{\big|\v_2\big>\big[0\big|}{\sbk{\v_2 0}}
+
\overline{\kappa}_3
\frac{\big|0\big>\big[\v_2\big|}{\abk{0\v_2}}
+\frac{\kappa_3\overline{\kappa}_3}{\abk{0\v_2}\sbk{0\v_2}}
\big|\v_2\big>\big[\v_2\big|
\nonumber\\
&=
\Big(
\underbrace{
\big|0\big>+\frac{\kappa_3}{\sbk{0\v_2}}\big|\v_2\big>
}_{\big|\wh{I}\big>}
\Big)
\Big(
\underbrace{
\big[0\big|+\frac{\overline{\kappa}_3}{\abk{0\v_2}}\big[\v_2\big|
}_{\big[\wh{I}\big|}
\Big).
\end{align} 
This implies 
\begin{align}
\abk{\wh{I}\v_2}=\abk{0\v_2},\quad
\abk{0\wh{I}}=\kappa_3\frac{\abk{0\v_2}}{\sbk{0\v_2}},
\quad
\abk{\wh{I}\v_1}
=
\abk{0\v_1}
+\kappa_3\frac{\abk{\v_2\v_1}}{\sbk{0\v_2}},
\end{align} 
and the factors in the first term read
\begin{align}
{\cal A}_3(0^+\!\!-\!\wh{I}^-\wh{3}^*)
=
\frac{ig}{\big|\kappa_3\big|^2}\abk{0\v_2}\sbk{0\v_2},\quad
{\cal A}_3(\wh{I}^+1^*\wh{2}^*)
=
-\frac{g}{\sqrt{2}\oka_1\oka_2}\frac{\abk{\v_1\v_2}^3\sbk{0\v_2}}
{\abk{\v_20}(\abk{0\v_1}\sbk{0\v_2}+\kappa_3\abk{\v_2\v_1})}.
\end{align}
Therefore, the first term in ${\cal A}_{4}(0^+1^*2^*3^*)$ is
\begin{align}
{\cal A}_{4a}(0^+1^*2^*3^*)
=
\frac{g^2}{\sqrt{2}\oka_1\oka_2\big|\kappa_3\big|^2}
\frac{1}{(p_0+p_3)^2}
\frac{\abk{\v_1\v_2}^3\sbk{0\v_2}^2}
{\abk{\v_10}\sbk{0\v_2}+\kappa_3\abk{\v_1\v_2}}.
\label{eq:A4a}
\end{align}

\paragraph{Pole in an eikonal propagator} Let us consider now the second term, coming from a pole in an eikonal propagator,
\begin{align}
{\cal A}_{4b}(0^+1^*2^*3^*)
=
{\cal A}_ 3(0^+-\wh{I}^*_{_E}\wh{3}^*)
\frac{1}{K_{_I}\cdot v_1}
{\cal A}_3(\wh{I}^*_{_E}1^*\wh{2}^*).
\end{align} 
(The subscript $E$ on the line $I^*$ is a reminder that this an eikonal line.) 
At the pole, we have by definition $\wh{K}_{_I}\cdot v_1=0$, and the momentum $\wh{K}_{_I}$ is therefore a valid argument for a source $\rho_1^{(0)}$. The value of $z$ at which this pole occurs is
\begin{align}
z_*=(p_0+p_3)\cdot v_1,
\end{align}
and at this $z_*$ we have
\begin{align}
\wh{\overline{K}}_{_I}
=
\big|0\big>\big[0\big|
+
\Big(\frac{p_0\cdot p_3}{p_0\cdot v_2}
-(p_0+p_3)\cdot v_1\Big)\big|\v_2\big>\big[\v_2\big|
-\ka_3\frac{\big|\v_2\big>\big[0\big|}{\sbk{\v_20}}
+\oka_3\frac{\big|0\big>\big[\v_2\big|}{\abk{0\v_2}}.
\end{align}
By using the Schouten identities,
\begin{align}
\big|0\big>
=
\frac{\abk{0\v_2}}{\abk{\v_1\v_2}}\big|\v_1\big>
+
\frac{\abk{\v_10}}{\abk{\v_1\v_2}}\big|\v_2\big>,\quad
\big|0\big]
=
\frac{\sbk{0\v_2}}{\sbk{\v_1\v_2}}\big|\v_1\big]
+
\frac{\sbk{\v_10}}{\sbk{\v_1\v_2}}\big|\v_2\big],
\end{align}
this can be rewritten as
\begin{align}
\wh{\overline{K}}_{_I}
=
\underbrace{\tfrac{1}{2}\abk{0\v_2}\sbk{0\v_2}}_{=\;p_0^+}
\big|\v_1\big>\big[\v_1\big|
&+
\Big(
\underbrace{
\frac{\oka_3}{\abk{\v_1\v_2}}+\frac{\abk{0\v_2}\sbk{\v_10}}{2}
}_{-\wh{\kappa}_{_I}/\sbk{\v_1\v_2}}\Big)\big|\v_1\big>\big[\v_2\big|
\nonumber\\
&\quad
+
\Big(
\underbrace{
\frac{\ka_3}{\sbk{\v_1\v_2}}+\frac{\abk{\v_10}\sbk{0\v_2}}{2}
}_{\wh{\oka}_{_I}/\abk{\v_2\v_1}}\Big)
\big|\v_2\big>\big[\v_1\big|.
\end{align}
The fact that this matrix has no term in $\big|\v_2\big>\big[\v_2\big|$ is just the statement that $\wh{K}_{_I}\cdot v_1=0$ at the pole. Note that, since $\wh{K}_{_I}^\mu$ is the momentum coming from a source $\rho_1^{(0)}$, we must parameterize its transverse components consistently with the other momenta of longitudinal component along $v_1^\mu$ (because we are going to reuse results obtained with this convention). From this formula, we read the transverse components of the momentum $\wh{K}_{_I}^\mu$,
\begin{align}
\wh{\kappa}_{_I}
=
-\sbk{\v_1\v_2}\Big(
\frac{\oka_3}{\abk{\v_1\v_2}}+\frac{\abk{0\v_2}\sbk{\v_10}}{2}\Big),
\quad
\wh{\overline{\kappa}}_{_I}
=-\abk{\v_1\v_2}\Big(
\frac{\ka_3}{\sbk{\v_1\v_2}}+\frac{\sbk{0\v_2}\abk{\v_10}}{2}\Big).
\end{align}

Since $\wh{K}_{_I}\cdot v_1=0$, the left factor in ${\cal A}_{4b}(0^+1^*2^*3^*)$ is a standard 3-point amplitude with one on-shell and two off-shell gluons,
\begin{align}
{\cal A}_{3}(0^+-\wh{I}^*_{_E}\wh{3}^*)
=
\frac{g}{\sqrt{2}\,\wh{\overline{\kappa}}_{_I}{\overline{\kappa}}_3}
\frac{\abk{\v_1\v_2}^3}{\abk{\v_20}\abk{0\v_1}}.
\end{align}
(Note that the transverse components of $-\wh{K}_{_I}^\mu$ are $-\wh{\kappa}_{_I}$ and $-\wh{\overline{\kappa}}_{_I}$, and that $\wh{\overline{\kappa}}_3=\oka_3$.) The right factor in ${\cal A}_{4b}(0^+1^*2^*3^*)$ is the amplitude for a single eikonal scattering of the source $\rho_1^{(0)}$ off a gluon produced by a source $\rho_2^{(0)}$. It comes from a single graph, and reads
\setbox1\hbox to 10mm{\includegraphics[width=10mm]{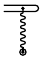}}
\begin{equation}
{\cal A}_3(\wh{I}^*_{_E}1^*\wh{2}^*)
=\raise -7mm\box1
=
\frac{g}{\wh{p}_ 2^2}
=
-\frac{g}{\big|\kappa_2\big|^2}.
\end{equation}	
Therefore, ${\cal A}_{4b}(0^+1^*2^*3^*)$ reads
\begin{align}
{\cal A}_{4b}(0^+1^*2^*3^*)
=
\frac{2g^2}{\sqrt{2}\,\big|\kappa_2\big|^2\overline{\kappa}_3}
\frac{1}{p_0^-+p_3^-}
\frac{\abk{\v_1\v_2}^2}{\abk{\v_20}\abk{0\v_1}\big(\abk{\v_10}\sbk{0\v_2}+{\ka_3}{\abk{\v_1\v_2}}\big)}.
\label{eq:A4b}
\end{align}

\paragraph{Full amplitude}
We can now add eqs.~(\ref{eq:A4a}) and (\ref{eq:A4b}) in order to obtain the full amplitude. By factoring out the common factors, we first obtain
\begin{align}
{\cal A}_4(0^+1^*2^*3^*)
&=
\frac{g^2\abk{\v_1\v_2}^3}{\sqrt{2}\,\big|\ka_1\big|^2\big|\ka_2\big|^2\big|\ka_3\big|^2 \abk{\v_20}\abk{0\v_1}}
\frac{\ka_1\abk{0\v_2}\sbk{0\v_2}}{\abk{\v_10}\sbk{0\v_2}+\kappa_3\abk{\v_1\v_2}}
\nonumber\\
&\qquad\qquad\times
\Bigg\{
\frac{\ka_2\abk{\v_10}\sbk{0\v_2}}{(p_0+p_3)^2}
+
\frac{\oka_1\ka_3\sbk{\v_1\v_2}}{2p_0^+(p_0^-+p_3^-)}
\Bigg\}.
\label{eq:A4-1}
\end{align}
Using also 
\begin{align}
(p_0+p_3)^3=
2p_0^+(p_0^-+p_3^-)-(\p_{0\perp}+\p_{3\perp})^2,
\end{align}
the curly bracket in eq.~(\ref{eq:A4-1}) can be rewritten as
\begin{align}
\Big\{\cdots\Big\}
=
\frac{\ka_2\abk{\v_10}\sbk{0\v_2}+\oka_1\ka_3\sbk{\v_1\v_2}}{(p_0+p_3)^2}
-
\frac{(\p_{0\perp}+\p_{3\perp})^2)\oka_1\ka_3\sbk{\v_1\v_2}}{2p_0^+(p_0^-+p_3^-)(p_0+p_3)^2}.
\end{align}
The second term in this expression has two simple poles in the variable $p_3^-$, with opposite residues. Recall that the sources $\rho_2^{(0)}$ do not depend on the minus component of the momentum, so this is the only dependence on $p_3^-$ (after we have used the factor $\delta(p_0^-+p_2^-+p_3^-)$ from momentum conservation to perform the integration over $p_2^-$), and therefore the integral over $p_3^-$ of this term is zero. Thus, we drop this term from now on\footnote{This simplification was implicit when solving directly the classical Yang-Mills equations. Indeed, the solutions derived in \cite{Dumitru:2001ux,Blaizot:2004wu,Gelis:2005pt} all used the ''shockwave approximation'', i.e., the assumption that $\rho_2(x)$ is proportional to $\delta(x^+)$ for a high energy projectile (this is equivalent to $\rho_2(p)$ being independent of $p^-$ in momentum space).}.

From momentum conservation, we have
\begin{align}
\abk{\v_10}\sbk{0\v_2}
=
\oka_1\sbk{\v_1\v_2}-(\ka_2+\ka_3)\abk{\v_1\v_2},
\end{align}
and the curly bracket (amputated of the term that vanishes when integrated over $p_3^-$, which we denote by using the symbol $\doteq$) becomes
\begin{align}
\Big\{\cdots\Big\}
&\doteq
\frac{\ka_2\abk{\v_10}\sbk{0\v_2}+\ka_3\big(\abk{\v_10}\sbk{0\v_2}+(\ka_2+\ka_3)\abk{\v_1\v_2}\big)}{(p_0+p_3)^2}
\nonumber\\
&=
\frac{\ka_2+\ka_3}{(p_0+p_3)^2}\left(\abk{\v_10}\sbk{0\v_2}+\ka_3\abk{\v_1\v_2}\right)
.
\end{align}
The factor $\abk{\v_10}\sbk{0\v_2}+\ka_3\abk{\v_1\v_2}$ cancels against a similar factor in the denominator of the amplitude, and we therefore obtain the following compact expression
\begin{align}
{\cal A}_4(0^+1^*2^*3^*)
&=
-\frac{g^2}{\big|\ka_1\big|^2\big|\ka_2\big|^2\big|\ka_3\big|^2 }
\frac{\abk{0\v_2}\sbk{0\v_2}}
{(p_0+p_3)^2}
\underbrace{
\frac{\ka_1(\ka_2+\ka_3)\abk{\v_1\v_2}^3}{\sqrt{2}\abk{0\v_1}\abk{0\v_2}}
}_{L\cdot\epsilon_+(\p_0,\q)}
.
\label{eq:A4-final}
\end{align}
Note that the last factor is the contraction of the Lipatov vertex (for one  $\rho_1$ and two $\rho_2$'s, as can be seen from the factor $\ka_1(\ka_2+\ka_3)$ in the numerator) contracted with the polarization vector of the produced gluon. The factor in the middle of the expression comes from the momentum dependence of a gluon vertex and the denominator of the additional gluon propagator. Finally, the first factor comprises the coupling constants, and the denominators of the external off-shell gluon propagators. 

The final expression of ${\cal A}_4$ is a rather natural extension of the expression (\ref{eq:A3-1}): besides an extra off-shell denominator $\big|\ka_3\big|^2$, and the propagator of the additional intermediate gluon, the only change is $\ka_2\to \ka_2+\ka_3$ in the Lipatov vertex, to account for the total transverse momentum  coming from sources $\rho_2$.  Interestingly, this lean structure of ${\cal A}_4$ is manifest only after we have combined the contributions of the gluon and eikonal poles, ${\cal A}_{4a}$ and ${\cal A}_{4b}$.

\begin{figure}[htbp]
\begin{center}
\includegraphics[width=0.8\textwidth]{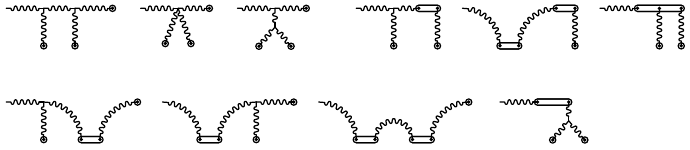}
\end{center}
\caption{\label{fig:A4-PT}Feynman graphs contributing to the production of a gluon from one source $\rho_1$ and two sources $\rho_2$. The graphs containing eikonal propagators come from the evolution of the sources due to current conservation (in certain gauges, some of these graphs vanish).}
\end{figure}

From the point of view of the perturbative expansion, the simplicity of the final expression in eq.~(\ref{eq:A4-final}) is quite remarkable. Indeed, in this approach, it would be obtained by summing the contributions of the Feynman graphs listed in Figure \ref{fig:A4-PT}, and combining all the pieces to reach such a compact expression would be rather challenging. Solving the Yang-Mills equations instead of calculating Feynman graphs would take care automatically of the bookkeeping, but one would still face the difficulties associated with the fact that the field $A^\mu(x)$ is gauge dependent. In contrast, in the BCFW approach, we arrive directly at a gauge invariant amplitude.

\subsection{Order $\rho_1\rho_2^n$}
Our next goal is to study gluon production at arbitrary order in the sources $\rho_2$. The gluon labeled $0$ is the gluon produced on-shell, the gluon labeled $1$ comes from a source of direction $v_1^\mu$, and the gluons labeled $2$ to $n+1$ come from sources of direction $v_2^\mu$.

From the already studied cases of one and two gluons produced by sources of direction $v_2^\mu$, we can guess the following ansatz for the general case where there are $n$ sources $\rho_2$:
\begin{align}
{\cal A}_{n+2}(0^+1^*2^*\cdots (n+1){}^*)
&=
-\frac{g^{n}\ka_1(\ka_2+\cdots+\ka_{n+1})}{\big|\ka_1\big|^2\big|\ka_2\big|^2\cdots\big|\ka_{n+1}\big|^2 }
\frac{\abk{\v_1\v_2}^3}{\sqrt{2}\abk{\v_20}\abk{0\v_1}}
\nonumber\\
&\qquad\qquad\times
\frac{\abk{\v_20}\sbk{0\v_2}}
{(p_0+p_{n+1})^2}\cdots \frac{\abk{\v_20}\sbk{0\v_2}}{(p_0+p_{n+1}+...+p_3)^2}.
\label{eq:An-ansatz}
\end{align}
Our goal in this subsection is to prove the correctness of this ansatz, possibly up to a term whose integral over the minus components $p_3^-,p_4^-,\cdots$ is zero.

\paragraph{BCFW recursion} To prove the validity of the ansatz in eq.~(\ref{eq:An-ansatz}), we use BCFW to obtain a relationship between ${\cal A}_{n+2}$ and ${\cal A}_{n+1}$. We  apply a shift $e^\mu=v_2^\mu$ on the lines $n$ and $n+1$. Like in the calculation of ${\cal A}_4(0^+1^*2^*3^*)$, this shift produces two poles: one in a gluon propagator and one in an eikonal propagator:
\begin{align}
{\cal A}_{n+2}(0^+1^*2^*...(n+1){}^*)
&=
\underbrace{
{\cal A}_{3}(0^+ -\wh{I}^- \wh{n+1}{}^*)
\frac{i}{K_{_I}^2}
{\cal A}_{n+1}(\wh{I}^+1^*...\wh{n}{}^*)
}_{{\cal A}_{n+2,a}(0^+1^*2^*...(n+1){}^*)}
\nonumber\\
&+
\underbrace{
{\cal A}_{3}(0^+ -\wh{I}^*\wh{n+1}{}^*)
\frac{1}{K_{_I}\cdot v_1}
{\cal A}_{n+1}(\wh{I}^*1^*...\wh{n}{}^*)
}_{{\cal A}_{n+2,b}(0^+1^*2^*...(n+1){}^*)}
,
\end{align}
with $K_{_I}= p_0+p_{n+1}$.

\paragraph{Pole in a gluon propagator} This pole is realized when $(p_0+p_{n+1}-zv_2)^2=0$, i.e., at
\begin{align}
z_*=\frac{p_{n+1}^2+2p_0\cdot p_{n+1}}{2p_0\cdot v_2}.
\end{align}
At this value $\wh{\overline{K}}_{_I}$ factorizes as
\begin{align}
\wh{\overline{K}}_{_I}
=
\big|\wh{I}\big>\big[\wh{I}\big|,\quad
\big|\wh{I}\big>=\big|0\big>+\frac{\ka_{n+1}}{\sbk{0\v_2}}\big|\v_2\big>,\quad
\big[\wh{I}\big|
=
\big[0\big|+\frac{\oka_{n+1}}{\abk{0\v_2}}\big[\v_2\big|.
\end{align}
This pole produces the following contribution,
\begin{align}
{\cal A}_{n+2,a}(0^+1^*2^*...(n+1){}^*)
&=
\frac{g}{\big|\ka_{n+1}\big|^2}\frac{\abk{\v_20}\sbk{0\v_2}}{(p_0+p_{n+1})^2}\,
{\cal A}_{n+1}(\wh{I}{}^*1^*...\wh{n}{}^*)
\nonumber\\
&=
-\frac{g^n \ka_1(\ka_2+...+\ka_n)}{\sqrt{2}\big|\ka_1\big|^2...\big|\ka_{n+1}\big|^2}
\frac{\abk{\v_1\v_2}^3\sbk{0\v_2}}{\abk{0\v_2}\big(\abk{\v_10}\sbk{0\v_2}+\ka_{n+1}\abk{\v_1\v_2}\big)}\nonumber\\
&\qquad\times
\frac{\abk{\v_20}\sbk{0\v_2}}{(p_0+p_{n+1})^2}\cdots
\frac{\abk{\v_20}\sbk{0\v_2}}{(p_0+p_{n+1}+...+p_3)^2}.
\end{align}
The last equality assumes that our ansatz is true for ${\cal A}_{n+1}$.

\paragraph{Pole in an eikonal propagator} The pole in the eikonal propagator arises when $\wh{K}_{_I}\cdot v_1=0$. This happens at
\begin{align}
z_*=(p_0+p_{n+1})\cdot v_1.
\end{align}
By explicitly calculating $\wh{\overline{K}}_{_I}$ at this $z_*$, we obtain its transverse components
\begin{align}
\wh{\ka}_{_I}=-\sbk{\v_1\v_2}
\left(\frac{\oka_{n+1}}{\abk{\v_1\v_2}}+\frac{\sbk{\v_10}\abk{0\v_2}}{2}\right),\quad
\wh{\oka}_{_I}=-\abk{\v_1\v_2}
\left(\frac{\ka_{n+1}}{\sbk{\v_1\v_2}}+\frac{\abk{\v_10}\sbk{0\v_2}}{2}\right).
\end{align}
Using also eq.~(\ref{eq:eik-n-scatt}) in Appendix \ref{sec:A-eik}, the contribution of this pole is
\begin{align}
{\cal A}_{n+2,b}(0^+1^*2^*...(n+1){}^*)
&=
-\frac{g^n \ka_1\oka_1\ka_{n+1}}{\sqrt{2}\big|\ka_1\big|^2...\big|\ka_{n+1}\big|^2}
\frac{\abk{\v_1\v_2}^3\sbk{\v_1\v_2}}{\abk{\v_20}\abk{0\v_1}
\big(\abk{\v_10}\sbk{0\v_2}+\ka_{n+1}\abk{\v_1\v_2}\big)}
\nonumber\\
&\qquad\times
\underbrace{
\frac{-1}{p_0^-+p_{n+1}^-}\cdots \frac{-1}{p_0^-+p_{n+1}^-+...+p_3^-}
}_{\doteq\frac{\abk{\v_20}\sbk{0\v_2}}{(p_0+p_n)^2}\cdots\frac{\abk{\v_20}\sbk{0\v_2}}{(p_0+p_{n+1}+...+p_3)^2}}.
\end{align}
Under the braces in the second line, we have indicated an alternate equivalent expression, that gives the same result after the integrations over $p_3^-, p_4^-,\cdots, p_{n+1}^-$. When we combine ${\cal A}_{n+2,b}$ with the contribution of the gluon pole, we use this alternate expression.

\paragraph{Total color ordered amplitude}
By adding ${\cal A}_{n+2,a}$ and ${\cal A}_{n+2,b}$, we obtain the full amplitude
\begin{align}
{\cal A}_{n+2}(0^+1^*2^*...(n+1){}^*)
&=
-\frac{g^n\,\ka_1}{\big|\ka_1\big|^2...\big|\ka_{n+1}\big|^2}
\frac{\abk{\v_1\v_2}^3}{\sqrt{2}\abk{\v_20}\abk{0\v_1}}
\nonumber\\
&\qquad\times
\frac{\abk{\v_20}\sbk{0\v_2}}{(p_0+p_{n+1})^2}\cdots
\frac{\abk{\v_20}\sbk{0\v_2}}{(p_0+p_{n+1}+...+p_3)^2}
\nonumber\\
&\qquad\times\frac{\abk{\v_10}\sbk{0\v_2}(\ka_2+...+\ka_n)+\sbk{\v_1\v_2}\oka_1\ka_{n+1})}{\abk{\v_10}\sbk{0\v_2}+\ka_{n+1}\abk{\v_1\v_2}}.
\end{align}
Using momentum conservation,
\begin{align}
\abk{\v_10}\sbk{0\v_2}=\oka_1\sbk{\v_1\v_2}-(\ka_2+...+\ka_{n+1})\abk{\v_1\v_2},
\end{align}
we see that the factor on the third line is equal to $\ka_2+...+\ka_{n+1}$, which proves the validity of our ansatz for ${\cal A}_{n+2}$.

\paragraph{Full-color amplitude} We can take for granted the Kleiss-Kuijf relations, since they just come from a change of color basis (from the basis made of traces of products of generators, to the basis made of products of structure constants). Therefore, the color decomposition in terms of structure constants can still be written as
\begin{align}
&{\cal M}_{n+2}(0^+_{a_0}1^*_{a_1}2^*_{a_2}...(n+1)^*_{a_{n+1}})
\nonumber\\
&\qquad
=
-(-i)^{n} \!\!\!\!\!\!\!
\sum_{\sigma\in{\mathfrak S}(2...n+1)}\!\!\!\!\!
f^{a_0 a_{\sigma_2}d_1}
f^{d_1 a_{\sigma_3}d_2}
\cdots
f^{d_{n-1} a_{\sigma_{n+1}}a_1}
{\cal A}_{n+2}(0^+\sigma_2^*...\sigma_{n+1}^*1^*),
\end{align}
despite the fact that the gluon $1$ is now off-shell and comes from a source that has a different direction from the sources of the other off-shell gluons. The Kleiss-Kuijf relations also imply
\begin{align}
{\cal A}_{n+2}(0^+\{\sigma^*\}1^*)
=
(-1)^{n}
{\cal A}_{n+2}(0^+1^*\{\sigma^*\}^t).
\end{align}
Therefore, the partial amplitudes of the form ${\cal A}_{n+2}(0^+1^*2^*...(n+1){}^*)$  that we have calculated above are sufficient to construct the full color amplitude (we only need to permute the off-shell lines $2,\cdots,n+1$).  In appendix \ref{sec:long}, we perform the integration over the longitudinal momenta carried by the color sources in order to arrive at a more familiar form for a Wilson line.

\section{Summary}
In this paper, we have shown how on-shell methods -- in particular the BCFW recursion -- can be extended in order to deal with the off-shell external gluons encountered in Color Glass Condensate calculations at tree level. Indeed, in the CGC, all the gluons connected to color sources are space-like (their propagators are not amputated, and these external legs are contracted with the direction $v^\mu$ along which the source is moving). An additional complication in the CGC is that the source density $\rho(x)$ must satisfy the covariant conservation equation  $[v\cdot D, \rho]=0$, which implies that $\rho$ cannot be given once for all in all spacetime. Instead, $\rho$ is specified on some intial surface, and then evolves dynamically along with the gauge field $A^\mu$. Enforcing this conservation law is in fact crucial for the final results to be gauge invariant.

In the CGC at tree level, the traditional approach is to first solve the classical Yang-Mills equations to obtain the gauge potential $A^\mu$, and then to feed $A^\mu$ into LSZ reduction formulas to obtain observables. In this approach, one must choose a gauge to make $A^\mu$ unique, and the solution of Yang-Mills equations differ considerably upon the gauge choice. In particular, the solution contain terms that are pure gauge artifacts, that drop out completely when used to calculate observable quantities.

In this paper, we use the fact that contributions to the classical $A^\mu$ can be represented as tree Feynman diagrams. Upon application of the LSZ reduction, these sum of diagrams can be viewed as amplitudes with the gluons attached to color sources off-shell. These amplitudes can be decomposed on a basis of color structures in the same way as fully on-shell amplitudes, with gauge invariant partial (color-ordered) amplitudes. 

In this paper, we used and slightly extended the methods proposed in \cite{vanHameren:2014iua,vanHameren:2012if,Bury:2017jxo,Bury:2015dla,vanHameren:2015bba,Kutak:2016goj}
in order to calculate these amplitude recursively by using BCFW. We restricted ourselves to the case where the color sources move at the speed of light (hence their direction $v^ \mu$ is light-like and their support is squeezed to an infinitely thin shockwave by Lorentz contraction).  We first recalled how to obtain the 3-point amplitudes with one or two off-shell gluons, from the fully on-shell 3-point amplitude by using BCFW. This is already considerably simpler than solving the Yang-Mills equations at lowest order in the color sources, and it offers the advantage of providing directly a gauge invariant result. Notably, even though the perturbative calculation of these  amplitudes require special graphs to encode the evolution of the sources due to current conservation, these graphs are not explicitly needed in the BCFW derivation of the 3-gluon off-shell amplitudes. (In perturbation theory, these graphs provide a crucial contribution to restore gauge invariance. In contrast, the BCFW approach works in terms of already gauge invariant objects.). 

Then, we used the BCFW recursion in order to increase the number of scatterings off the second nucleus. We first calculated the scattering amplitude of an on-shell gluon off a nucleus, at all orders in the color density of this nucleus. And then, we calculated the amplitude for the production of an on-shell gluon, in the collision between two colored projectiles, at lowest order in the color density of the first projectile and to all orders in the color density of the second projectile. At all steps of these calculations, all the objects we manipulate are gauge invariant amplitudes, and we never encounter the type of gauge dependent terms one would have when solving the Yang-Mills equations. Furthermore, we do not need to fix the gauge, at the exception of the gauge at the initial time (that we choose so that the initial gauge field is zero, rather than a non-zero pure gauge). Our results obtained without gauge fixing in the bulk are equivalent to earlier results obtained with the standard approach (gauge fixing, gauge dependent solution of Yang-Mills equations, LSZ), but are obtained at a fraction of the calculational cost.  As another illustration of this approach, in appendix \ref{sec:primakoff}, we use BCFW to obtain a simple expression of the amplitude of gluon splitting in an external color field.

Let us mention one downside of this approach: the Wilson line that appears in  the final all-orders result is obtained in expanded form, order-by-order in the source of the dense projectile, and not yet integrated over the momenta carried  by the color sources. It is only after the integrals over the longitudinal momenta carried by the color sources have been performed that one recognizes the standard form of a Wilson line.

Although the final result for gluon production was already known, our novel derivation illustrates the power of on-shell methods, mainly in eliminating all the gauge dependent intermediate steps that one would encounter when solving the Yang-Mills equations. Thanks to these simplifications, it may be possible to go beyond these results by calculating the next order in the color density of the dilute projectile. Another direction for future exploration could be to depart from the assumption that the sources move at the speed of light, in order to incorporate non eikonal corrections in the result.

\section*{Acknowledgements}
I would like to thank I. Castelli for many discussions during the infancy of this project, as well as A. van Hameren and K. Kutak.

\appendix

\section{Generic off-shell momenta}
\label{sec:generic}
Unlike the momenta produced by a single source ($\rho_1^{(0)}(p)$ or $\rho_2^{(0)}(p)$), a generic off-shell momentum $p^\mu$ has both $p\cdot v_1\not=0$ and $p\cdot v_2\not=0$. For this reason, its parameterization must be more general than the one discussed in Section \ref{sec:offshell}, to account for the fact that the two longitudinal components are non-zero. We now parameterize $p^\mu$ as
\begin{align}
p^\mu = p^+ v_1^\mu + p^- v_2^\mu + p_\perp^\mu. 
\label{eq:gen1}
\end{align}
In this decomposition, the transverse component $p_\perp^\mu$ should be orthogonal to both $v_1^\mu$ and $v_2^\mu$. This can be accomplished by writing it as a linear combination of the polarization vectors $\epsilon_\pm^\mu({\bs v}_1,{\bs v}_2)$,
\begin{align}
p_\perp^\mu \equiv
-\frac{\kappa}{\sqrt{2}}\epsilon_-^\mu({\bs v}_1,{\bs v}_2)
+
\frac{\overline{\kappa}}{\sqrt{2}} \epsilon_+^\mu({\bs v}_1,{\bs v}_2).
\label{eq:gen2}
\end{align}
For a momentum $p^\mu$ with real valued components, we have $\overline{\kappa}=\kappa^*$. Moreover the squared norm of the momentum reads
\begin{align}
p_\mu p^\mu = 2p^+p^--\kappa \overline{\kappa}.
\end{align}
The $2\times 2$ complex matrix $\overline{P}\equiv -p_\mu \overline{\sigma}^\mu$ is given by
\begin{align}
\overline{P}
=
p^+ \big|\v_1\big>\big[\v_1\big| 
+
p^-\big|\v_2\big>\big[\v_2\big|
-\kappa \frac{\big|\v_1\big>\big[\v_2\big|}{\sbk{\v_1\v_2}}
+
\overline{\kappa}\frac{\big|\v_2\big>\big[\v_1\big|}{\abk{\v_2\v_1}}.
\label{eq:gen3}
\end{align}	

Eqs.~(\ref{eq:gen1}) and (\ref{eq:gen2}) are a parameterization of a completely arbitrary momentum. It should also be valid when the momentum is on-shell. When this is the case, there is a way to connect eq.~(\ref{eq:gen3}) with the factorized form $\overline{P}=\big|\p\big>\big[\p\big|$ that holds when $p^\mu$ is on-shell. Since the space of 2-component spinors is 2-dimensional, we can use $\big|\v_1\big>,\big|\v_2\big>$ as a basis for expressing $\big|\p\big>$. We thus look for linear combinations of the form
\begin{align}
\big|\p\big>=\alpha\big|\v_1\big>+\beta\big|\v_2\big>,\quad
\big|\p\big]=\overline{\alpha}\big|\v_1\big]+\overline{\beta}\big|\v_2\big].
\end{align}
(Here, we are not assuming that $p_\mu$ has real valued components, and the coefficients $\overline{\alpha},\overline{\beta}$ in $\big|\p\big]$ are not necessarily the complex conjugates of the $\alpha,\beta$ in $\big|\p\big>$.) By equating $\overline{P}=\big|\p\big>\big[\p\big|$ and the expression obtained by multiplying the above spinors, we obtain
\begin{align}
\alpha\overline{\alpha}=p^+,\quad
\beta\overline{\beta}=p^-,\quad
\alpha\overline{\beta}=-\frac{\kappa}{\sbk{\v_1\v_2}},\quad
\overline{\alpha}\beta=-\frac{\overline{\kappa}}{\abk{\v_1\v_2}}.
\end{align}
Firstly, note that these equations are consistent only if $2p^+p^-=\kappa\overline{\kappa}$, which is nothing but the on-shellness condition for $p^\mu$. Note also that these equations constrain $\alpha$ and $\beta$ only up to an overall irrelevant phase. We can choose this overall phase such that
\begin{align}
\alpha=\overline{\alpha}=\sqrt{p^+},
\end{align}
(note that $\overline{\alpha}\not=\alpha^*$ if $p^+$ is complex valued)
which then implies
\begin{align}
\beta=-\frac{\overline{\kappa}}{\sqrt{p^+}\abk{\v_1\v_2}},
\quad
\overline{\beta}=-\frac{\kappa}{\sqrt{p^+}\sbk{\v_1\v_2}}.
\end{align}
The spinors $|\p\big>,\big|\p\big]$, expressed in the $\v_1,\v_2$ basis, therefore read
\begin{align}
\big|\p\big>
=
\sqrt{p^+}\big|\v_1\big>
-\frac{\overline{\kappa}}{\sqrt{p^+}\abk{\v_1\v_2}}\big|\v_2\big>,\quad
\big|\p\big]
=
\sqrt{p^+}\big|\v_1\big]
-\frac{{\kappa}}{\sqrt{p^+}\sbk{\v_1\v_2}}\big|\v_2\big].
\end{align}

\section{Equivalence of eqs.~(\ref{eq:3p-a}) and (\ref{eq:3p-b})}
\label{sec:app-3p-ab}
        In order to see that the expressions (\ref{eq:3p-a}) and (\ref{eq:3p-b}) are equal, start
        from the explicit formulas for the transverse coordinates:
          \begin{align}
            \kappa_2 = \frac{\big<\q\big|\overline{P}_2\big|\v_2\big]}{\abk{\q\v_2}},\quad
      \overline\kappa_2 = -\frac{\big<\v_2\big|\overline{P}_2\big|\q\big]}{\sbk{\v_2\q}}.
          \end{align}
          (Recall that these expressions do not depend on $\q$, that
          we may therefore choose to our convenience.) $\kappa_2$ and
          $\overline{\kappa}_2$ can be expressed in terms of brackets
          by starting from momentum conservation,
          $p_0^\mu+p_1^\mu+p_2^\mu=0$, which is equivalent to
          \begin{align}
            \big|0\big>\big[0\big|
              +
            \big|1\big>\big[1\big|
              +
              p_2^-\big|\v_2\big>\big[\v_2\big|
              -\kappa_2\frac{\big|\v_2\big>\big[\q\big|}{\sbk{\v_2\q}}
              +\overline{\kappa}_2\frac{\big|\q\big>\big[\v_2\big|}{\abk{\q \v_2}} =0.
          \end{align}
          Inserting this between $\big<\q|\cdots\big|\v_2\big]$, we get
\begin{align}
\kappa_2=-\frac{\abk{\q 0}}{\abk{\q\v_2}}\sbk{0\v_2}-\frac{\abk{\q1}}{\abk{\q\v_2}}\sbk{1\v_2}.
\end{align}
By contracting the Schouten identity,
\begin{align}
\big|0\big>\abk{1\v_2}+\big|1\big>\abk{\v_20}+\big|\v_2\big>\abk{01}=0,
\end{align} 
with the spinor $\big<\q\big|$, we obtain
\begin{align}
\frac{\abk{\q0}}{\abk{\q\v_2}}
=
-
\frac{\abk{\q1}}{\abk{\q\v_2}}
\frac{\abk{\v_20}}{\abk{1\v_2}}
-\frac{\abk{01}}{\abk{1\v_2}}.
\end{align}
Inserting this identity in the previous expression of $\kappa_2$, it becomes
\begin{align}
	\kappa_2 =
	\frac{\abk{01}\sbk{0\v_2}}{\abk{1\v_2}}
	-
	\frac{\abk{\q1}}{\abk{\q\v_2}\abk{1\v_2}}
	\big(\underbrace{\abk{0\v_2}\sbk{0\v_2}+\abk{1\v_2}\sbk{1\v_2}}_{2(p_0^++p_1^+)=0}\big)
	=
	\frac{\abk{01}\sbk{0\v_2}}{\abk{1\v_2}}.
\end{align}
(We see explicitly that the transverse component $\kappa_2$ does not depend on the choice of auxiliary vector $\q$. Note that we could have made the calculation simpler by anticipating this independence with respect to $\q$ in order to set $\q=\p_1$.) By using $\kappa_2\overline{\kappa}_2=-p_2^2=-(p_0+p_1)^2=-2p_0\cdot p_1 =-\abk{01}\sbk{01}$, we also obtain
            \begin{align}
              \overline{\kappa}_2 = \frac{\abk{1\v_2}\sbk{01}}{\sbk{\v_20}}.
              \label{eq:kappabar2}
            \end{align}
            This leads to the following expression for the amplitude, 
            \begin{align}
              {\cal A}_3(0^+1^-2^*)
              =
              ig
              \frac{\abk{1\v_2}\sbk{\v_20}}{\abk{01}\sbk{01}}
              \frac{\abk{1\v_2}}{\abk{\v_20}}.
            \end{align}
To prove that this is equal to eq.~(\ref{eq:3p-b}), we use
the fact that $0=v_2\cdot p_2=-v_2\cdot(p_0+p_1)$, which is equivalent to
\begin{align}
  \frac{\abk{1\v_2}}{\abk{\v_20}}+\frac{\sbk{\v_20}}{\sbk{1\v_2}}=0.
\end{align}
Thus, the amplitude is also given by
\begin{align}
{\cal A}_3(0^+1^-2^*)
=
              -ig\frac{\abk{1\v_2}\sbk{\v_20}}{\abk{01}\sbk{01}}
              \frac{\sbk{\v_20}}{\sbk{1\v_2}}
              =
              \frac{ig}{\kappa_2}
              \frac{\sbk{\v_20}^3}{\sbk{01}\sbk{1\v_2}}.
              \label{eq:A3-alt1}
\end{align}

\section{$|z|\to \infty$ for shifts of two off-shell gluons}
\label{sec:high-z}
Let us consider here a shift applied to two off-shell gluons produced
by sources sharing a common light-like direction $v^\mu$. In this
case, the shift vector $e^\mu=\tfrac{1}{2}\big<\v\big|\overline{\sigma}^\mu\big|\v\big]$ equals the direction $v^\mu$, and the
off-shell propagators directly attached to the sources are
unchanged. Indeed, if $p^\mu$ is the momentum carried by one of these
gluons, we have $p\cdot v =0$, which implies
\begin{align}
\wh{p}^2=(p+zv)^2=p^2+2z \underbrace{p\cdot v}_{=\;0} + z^2 \underbrace{v^2}_{=\;0} = p^2.
\end{align}
Thus, these off-shell propagators are independent of $z$. The $z$
dependence of the shifted amplitude comes from 3-gluon vertices along
the line of shifted momenta, that are generically linear in $z$. It
also comes from the internal shifted propagators, that behave like
$1/z$. However, in the worst case, there is one more 3-gluon vertex
than internal propagators, and this power counting suggests that the shifted
amplitude could grow as $z$. Note that each insertion of a 4-gluon
vertex reduces the degree in $z$ by one unit.

In this appendix, we show that this graph-by-graph power counting is
too pessimistic, and that the actual large-$z$ behavior of the shifted
amplitude is $1/z$.  Let us split the gluon field $A^\mu$ into a hard
part $a^\mu$, that carries the large momentum proportional to $z$ due
to the shift, and a soft part ${\cal A}^\mu$ carrying a momentum whose
magnitude is set by the typical momenta of the amplitude under
consideration,
\begin{align}
  A^\mu\equiv {\cal A}^\mu+ a^\mu.
\end{align}
This decomposition of the color field leads to the following
expression for the field strength,
\begin{align}
  F^{\mu\nu}=
  {\cal F}^{\mu\nu}
  +{\cal D}^\mu a^\nu-{\cal D}^\nu a^\mu -ig[a^\mu,a^\nu],
\end{align}
where ${\cal D}^\mu\equiv \partial^\mu-ig {\cal A}^\mu$ and ${\cal
  F}^{\mu\nu}\equiv \partial^\mu{\cal A}^\nu-\partial^\nu{\cal
  A}^\mu-ig[{\cal A}^\mu,{\cal A}^\nu]$ are the covariant derivative
and field strength associated to the soft field ${\cal A}^\mu$.  Since
the shift we perform affects only two external lines, we need only the
terms of the Yang-Mills action that are quadratic in the hard field
$a^\mu$. These terms can be written as
\begin{align}
  {\cal L}_{_{YM}}
  &=-\tfrac{1}{2}{\rm tr}\,\big(({\cal D}_\mu a_\nu-{\cal D}_\nu a_\mu)({\cal D}^\mu a^\nu-{\cal D}^\nu a^\mu)\big)
  + ig\,{\rm tr}\,\big({\cal F}_{\mu\nu}[a^\mu,a^\nu]\big)+\cdots\nonumber\\
  &=
  -{\rm tr}\,\big(({\cal D}^\mu a^\nu)({\cal D}_\mu a_\nu)\big)
  +{\rm tr}\,\big(({\cal D}_\mu a^\mu)({\cal D}_\nu a^\nu)\big)
  + ig\,{\rm tr}\,\big({\cal F}_{\mu\nu}[a^\mu,a^\nu]\big)+\cdots
\end{align}
In the second line, we have anticipated an integration by parts in the
integral over space-time that gives the action, in order to reorganize
some of the terms. For the subsequent discussion, it is convenient to
adopt the background field gauge condition, ${\cal D}_\mu
a^\mu=0$. With this particular gauge condition, the gauge fixing term
added to the Lagrangian simply cancels the second term in the above
expression, so that the gauge fixed Lagrangian reads
\begin{align}
  {\cal L}_{_{YM}, \rm{gauge\ fixed}}
  &=
  -{\rm tr}\,\big(({\cal D}^\mu a^\nu)({\cal D}_\mu a_\nu)\big)
  + ig\,{\rm tr}\,\big({\cal F}_{\mu\nu}[a^\mu,a^\nu]\big)+\cdots
\end{align}
The first term in this Lagrangian exhibits an extended Lorentz
symmetry. Indeed, it is invariant under separate Lorentz transformations of the
background field covariant derivatives, and of the hard field $a^\nu$.
The second term is antisymmetric in the two hard field, because of the
commutator $[a^\mu,a^\nu]$. However, this term is a local interaction
between two $a$'s and two ${\cal A}$'s, and it therefore comes from the
4-gluon vertex, which is momentum independent and therefore does not
bring a power of $z$. 

Consider now an amputated Feynman graph with two hard fields
(corresponding to the two shifted external gluons) and any number of
soft ones, ${\cal M}^{\mu\nu}$ (the Lorentz indices $\mu,\nu$ are those carried by the hard fields, and we do not
write explicitly the Lorentz indices carried by the soft fields). The above
considerations lead to the following structure
\begin{align}
  {\cal M}^{\mu\nu}
  =
  (c_1 z+ c_0 + c_{-1}z^{-1}+\cdots)\, g^{\mu\nu}
  +
  A^{\mu\nu}
  +
  R^{\mu\nu}.
\end{align}
The first term is the result one would obtain by keeping only the first term
of the Lagrangian. It is proportional to $g^{\mu\nu}$ because of the
extended Lorentz symmetry mentioned above, and it grows like $z$
according to the naive power counting (this term does not involve any
4-gluon vertex).  The second term, $A^{\mu\nu}$, comes from a single
insertion of the term in $[a^\mu,a^\nu]{\cal
  F}_{\mu\nu}$. Consequently, $A^{\mu\nu}$ is antisymmetric, and its
leading behavior is $z^0$ because it involves a 4-gluon vertex.  The
remainder, $R^{\mu\nu}$, contains at least two 4-gluon vertices. This
term has not specific symmetry, but is guaranteed to be at most of
order $z^{-1}$.

A bona fide scattering amplitude is obtained by contracting the
amputated graph ${\cal M}^{\mu\nu}$ with the propagators of two
off-shell gluons and with the directions of the sources, as in
\begin{align}
  (v^\alpha G_{\alpha\mu}){\cal M}^{\mu\nu} (G_{\nu\beta}v^\beta)
  &\propto
  v_\mu {\cal M}^{\mu\nu} v_\nu\nonumber\\
  &
  =
  (c_1 z+ c_0 + c_{-1}z^{-1}+\cdots) \underbrace{v^2}_{=\;0}
  + \underbrace{A^{\mu\nu}v_\mu v_\nu}_{=\;0}
  +\underbrace{R^{\mu\nu}v_\mu v_\nu}_{{\cal O}(z^{-1})}.
\end{align}
The first term is zero because $v^\mu$ is light-like, and the second
one vanishes because $A^{\mu\nu}$ is antisymmetric. The first terms
that do not cancel are of order $z^{-1}$, and therefore fulfill the
criterion of applicability of this particular BCFW shift.

\section{Derivation of ${\cal A}_3(0^+1^*2^*)$ from perturbation theory}
\label{sec:A3-eik-der}
In this appendix, we derive from perturbation theory the 3-point amplitude with an on-shell gluon of positive helicity, and two off-shell gluons, produced by sources of directions $v_1^\mu$ and $v_2^\mu$, respectively. Diagrammatically, this object receives the following contributions
\setbox1\hbox to 90mm{\includegraphics[width=90mm]{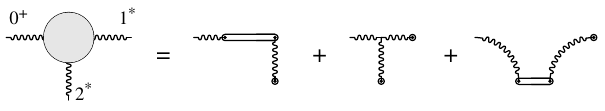}}
\begin{equation}
{\cal A}_3(0^+1^*2^*)
\equiv
\raise -7mm\box1.
\end{equation}
Using the Feynman rules, the expression for this amplitude takes the form
$
{\cal A}_3(0^+1^*2^*)
=
g\,
L_\mu\epsilon^\mu_+(\p_0,\q)
$,
where the
 vector $L_\mu$ is given by
\begin{align}
L^\mu\equiv
\frac{v_1^\mu}{\big|\kappa_2\big|^2}
\Big(\frac{1}{p_0^-}-\frac{2p_0^+}{\big|\kappa_1\big|^2}\Big)
-
\frac{v_2^\mu}{\big|\kappa_1\big|^2}
\Big(\frac{1}{p_0^+}-\frac{2p_0^-}{\big|\kappa_2\big|^2}
\Big)
+
\frac{p_2^\mu-p_ 1^\mu}{\big|\kappa_1\big|^2\big|\kappa_2\big|^2}.
\label{eq:Lmu}
\end{align}
Note that this vector satisfies $p_0\cdot L=0$ (recall that $p_0+p_1+p_2=0$), which ensures that the amplitude is independent of the choice of the auxiliary vector $q^\mu$ in the polarization vector. 

We sketch now the main steps in the derivation of the amplitude, starting from the expression (\ref{eq:Lmu}). Firstly, the contractions of the polarization vector with the light-cone directions $v_{1,2}^\mu$ are given by
\begin{align}
v_1\cdot \epsilon_+(\p_0,\q)
=
-\frac{\abk{\v_1\q}\sbk{\v_10}}{\sqrt{2}\,\abk{\q0}},
\quad
v_2\cdot \epsilon_+(\p_0,\q)
=
-\frac{\abk{\q\v_2}\sbk{\v_20}}{\sqrt{2}\,\abk{\q0}}.
\end{align}
In this calculation, we keep the auxiliary vector $\q$ unspecified. This will provide a consistency check at the end; indeed, the final result should be independent of $\q$. The off-shell momenta $p_1^\mu$ and $p_2^\mu$ read
\begin{align}
p_1^\mu
&=
p_1^+v_1^\mu 
-
\kappa_1
\frac{\big<\v_11\big|\overline{\sigma}^\mu\big|\v_2\big]}{2\sbk{\v_1\v_2}}
+
\overline{\kappa}_1
\frac{\big<\v_2\big|\overline{\sigma}^\mu\big|\v_1\big]}{2\abk{\v_2\v_1}}
,\nonumber\\
p_2^\mu
&=
p_2^-v_2^\mu 
-
\kappa_2
\frac{\big<\v_2\big|\overline{\sigma}^\mu\big|\v_1\big]}{2\sbk{\v_2\v_1}}
+
\overline{\kappa}_2
\frac{\big<\v_1\big|\overline{\sigma}^\mu\big|\v_2\big]}{2\abk{\v_1\v_2}}
.
\end{align}
Here, we have made explicit choices for the auxiliary vectors used in the polarization vectors that serve as a basis for the transverse momentum plane.  Given these expressions for the momenta, we have
\begin{align}
(p_1-p_2)\cdot\epsilon_+(\p_0,\q)
&=p_2^-
\frac{\abk{\q\v_2}\sbk{\v_20}}{\sqrt{2}\,\abk{\q0}}
-
p_1^+
\frac{\abk{\q\v_1}\sbk{\v_10}}{\sqrt{2}\,\abk{\q0}}
\nonumber\\
&
+
\Big(\frac{\kappa_1}{\sbk{\v_1\v_2}}
+
\frac{\overline{\kappa}_2}{\abk{\v_1\v_2}}\Big)
\frac{\abk{\q\v_1}\sbk{\v_20}}{\sqrt{2}\,\abk{\q0}}
+
\Big(
\frac{\overline{\kappa}_1}{\abk{\v_1\v_2}}
+
\frac{\kappa_2}{\sbk{\v_1\v_2}}
\Big)
\frac{\abk{\q\v_2}\sbk{\v_10}}{\sqrt{2}\,\abk{\q0}}.
\end{align}
From momentum conservation, $p_0+p_1+p_2=0$, we get
\begin{align}
\frac{1}{2}\abk{\v_10}\sbk{0\v_2}
=
\frac{\overline{\kappa}_1}{\abk{\v_1\v_2}}
-
\frac{\kappa_2}{\sbk{\v_1\v_2}},\quad
\frac{1}{2}\sbk{\v_10}\abk{0\v_2}
=
\frac{{\kappa}_1}{\sbk{\v_1\v_2}}
-
\frac{\overline{\kappa}_2}{\abk{\v_1\v_2}}.
\end{align}
This leads to the following expression for the amplitude,
\begin{align}
&
{\cal A}_3(0^+1^*2^*)
=
-\frac{g}{\sqrt{2}\big|\kappa_1\big|^2\big|\kappa_2\big|^2}
\nonumber\\
&\quad\times
\Bigg\{
\frac{\abk{\q\v_1}}{\abk{\q0}\abk{0\v_1}}
\Big[2p_0^+p_0^--2\big|\kappa_1\big|^2
+2\Big(
\frac{\oka_1}{\abk{\v_1\v_2}}-\frac{\ka_2}{\sbk{\v_1\v_2}}\Big)
\Big(
\frac{\ka_1}{\sbk{\v_1\v_2}}
+
\frac{\oka_2}{\abk{\v_1\v_2}}
\Big)
\Big]
\nonumber\\
&\quad\quad+
\frac{\abk{\q\v_2}}{\abk{\q0}\abk{\v_20}}
\Big[2p_0^+p_0^--2\big|\kappa_2\big|^2
+2\Big(
\frac{\oka_2}{\abk{\v_1\v_2}}-\frac{\ka_1}{\sbk{\v_1\v_2}}\Big)
\Big(
\frac{\ka_2}{\sbk{\v_1\v_2}}
+
\frac{\oka_1}{\abk{\v_1\v_2}}
\Big)
\Big]
\Bigg\}.
\end{align}
It turns out that the expressions inside the square brackets multiplying $\abk{\q\v_1}$ and $\abk{\q\v_2}$ are both equal to
\begin{align}
\Big[2p_0^+p_0^--\cdots\Big]
&=
2p_0^+p_0^-
-\big|\ka_1\big|^2
-\big|\ka_2\big|^2
+2\Big(
\frac{\oka_1\oka_2}{\abk{\v_1\v_2}^2}
-
\frac{\ka_1\ka_2}{\sbk{\v_1\v_2}^2}
\Big)
=
-
\ka_1\ka_2\abk{\v_1\v_2}^2
.
\end{align}
From the Schouten identity,
\begin{align}
\big|0\big>\abk{\v_1\v_2}
+
\big|\v_1\big>\abk{\v_20}
+
\big|\v_2\big>\abk{0\v_1}
=0
\end{align}	
(the space of 2-component spinors is 2-dimensional, and this identity encodes the linear relationship  that exits within any set of three such spinors), we also obtain
\begin{align}
\frac{\abk{\q\v_1}}{\abk{\q0}\abk{0\v_1}}
+
\frac{\abk{\q\v_2}}{\abk{\q0}\abk{\v_20}}
=
-\frac{\abk{\v_1\v_2}}{\abk{\v_20}\abk{0\v_1}}.
\end{align}
(Incidentally, this identity implies the cancellation of the  dependence on the auxiliary vector $q^\mu$ in the final result.)
This leads immediately to 
\begin{align}
{\cal A}_3(0^+1^*2^*)=
-\frac{g}{\sqrt{2}\,\oka_1\oka_2}
\frac{\abk{\v_1\v_2}^3}{\abk{\v_20}\abk{0\v_1}}
.
\end{align}
This derivation using perturbation theory is arguably more involved than the one using BCFW recursion, especially if the starting point of the recursion (the amplitude with three on-shell gluons) is obtained from little group scaling. In particular, the requirement to include graphs with internal eikonal propagators in order to account for the covariant conservation of the color current never appears explicitly in the BCFW calculation of this amplitude.

\section{Kleiss-Kuijf relations}
\label{sec:KK}
In this appendix, we justify the Kleiss-Kuijf relations on a few examples. Firstly, note that the kinematical factors ${\cal K}$ that appear in the right hand side of the recursion (\ref{eq:K1}) have some of their momenta shifted. Consider for instance the factor ${\cal K}(p_0,\{\alpha_1... \alpha_k\},\{\wh{\beta}_q\})$ that appears in eq.~(\ref{eq:K1}). The on-shell condition for the intermediate momentum reads
            \begin{align}
            &(p_0+p_{\beta_q}-zv_2+p_{\alpha_1}+\cdots +p_{\alpha_k})^2=0,
            \end{align}
            that gives the following value of $z$ at the pole,
            \begin{align}
            2z_* v_2\cdot p_0 = (p_0+p_{\beta_q}+p_{\alpha_1}+\cdots +p_{\alpha_k})^2.
            \end{align}
            From the above examples, we expect that ${\cal K}$ contains denominators of the form
            \begin{align}
            	(p_0+\wh{p}_{\beta_q}+\sum_l p_{\alpha_l})^2,
            \end{align}	
            that involve the shifted momentum $\wh{p}_{\beta_q}$ evaluated at $z_*$. The index ${l}$ in the sum belongs to a strict subset of $[1,k]$.  At $z_*$, this quantity can be written as 
            \begin{align}
            	(p_0+{p}_{\beta_q}-z_*v_2+\sum_l p_{\alpha_l})^2
            	&=
            	(p_0+{p}_{\beta_q}+\sum_l p_{\alpha_l})^2-2z_*v_2\cdot p_0
            	\nonumber\\
 &           	=
            	(p_0+{p}_{\beta_q}+\sum_l p_{\alpha_l})^2
            	-
            	(p_0+p_{\beta_q}+p_{\alpha_1}+\cdots +p_{\alpha_k})^2,
            \end{align}
            i.e., a difference between the unshifted denominator and the unshifted denominator of the propagator joigning the left and right factors.
            
Let us now consider a few examples of orderings where the on-shell gluons $0$ and $1$ are not adjacent. Firstly, let us calculate ${\cal K}(p_0,\{2\},\{34\})$, that enters in the amplitude ${\cal A}_5(0^+2^* 1^- 3^* 4^*)$. The recursion formula  (\ref{eq:K1}), combined with the results for lower-point amplitudes, gives
            \begin{align}
            {\cal K}(p_0,\{2\},\{34\})
            =&
            -\Big[\frac{1}{(p_0+\wh{p}_4)^2}+\frac{1}{(p_0+p_2)^2}\Big] \frac{1}{(p_0+p_2+p_4)^2}
            \nonumber\\
            &- \frac{1}{(p_0+p_4)^2}\Big[\frac{1}{(p_0+\wh{p}_4+\wh{p}_3)^2}+\frac{1}{(p_0+\wh{p}_4+p_2)^2}\Big].
            \end{align}
            Then, by the method explained above to calculate the denominators containing shifted momenta, some simple algebra leads to
            \begin{align}
            {\cal K}(p_0,\{2\},\{34\})
            =
            -&\Big[
			\frac{1}{(p_0+p_2)^2}\frac{1}{(p_0+p_2+p_4)^2}
			+
			\frac{1}{(p_0+p_4)^2}\frac{1}{(p_0+p_4+p_2)^2}
			\nonumber\\&+
			\frac{1}{(p_0+p_4)^2}\frac{1}{(p_0+p_4+p_3)^2}    
            \Big],
			\end{align}
which is consistent with eq.~(\ref{eq:kin}).

			The 5-point amplitude ${\cal A}_5(0^+2^* 3^* 1^- 4^*)$ that differs by the ordering of the off-shell gluons relative to the on-shell ones involves the following kinematical factor,
			\begin{align}
			{\cal K}(p_0,\{23\},\{4\})
			&=\frac{1}{(p_0+p_4)^2}\frac{1}{(p_0+p_4+p_2)^2}
			+
			\frac{1}{(p_0+p_2)^2}\frac{1}{(p_0+p_2+p_3)^2}
			\nonumber\\&+
			\frac{1}{(p_0+p_2)^2}\frac{1}{(p_0+p_2+p_4)^2},   
			\end{align}
also a special case of eq.~(\ref{eq:kin}).

			Consider as a last example the 6-point amplitude ${\cal A}_6(0^+2^* 1^- 3^* 4^* 5^*)$. By the same method, we obtain the following kinematical factor,
			\begin{align}
				{\cal K}(p_0,\{2\},\{345\})
				=
				-&\Big[
				\frac{1}{(p_0+p_2)^2}\frac{1}{(p_0+p_2+p_5)^2}\frac{1}{(p_0+p_2+p_5+p_4)^2}
				\nonumber\\
				&+
				\frac{1}{(p_0+p_5)^2}\frac{1}{(p_0+p_5+p_2)^2}\frac{1}{(p_0+p_5+p_2+p_4)^2}
				\nonumber\\
				&+
				\frac{1}{(p_0+p_5)^2}\frac{1}{(p_0+p_5+p_4)^2}\frac{1}{(p_0+p_5+p_4+p_2)^2}
				\nonumber\\
				&+
				\frac{1}{(p_0+p_5)^2}\frac{1}{(p_0+p_5+p_4)^2}\frac{1}{(p_0+p_5+p_4+p_3)^2}
				\Big].
\end{align}

\section{Solution of Yang-Mills equations in Landau gauge}
\label{sec:YM}
In this appendix, we summarize the result of  \cite{Blaizot:2004wu} regarding the solution of Yang-Mills equations in Landau gauge, $\partial_\mu A^\mu=0$.
This solution reads 
\begin{align}
  A_a^\mu(p)=A_{1a}^\mu(p) + \frac{ig}{k^2}\int \frac{d^2\k_{2\perp}}{(2\pi)^2}
  \Big\{
  &L^\mu\big[U(\k_{2\perp})-(2\pi)^2\delta(\k_{2\perp})\big]_{ab}
  \nonumber\\
  +&
  D^\mu\big[V(\k_{2\perp})-(2\pi)^2\delta(\k_{2\perp})\big]_{ab}
  \Big\}
  \frac{\rho_{1b}^{(0)}(\k_{1\perp})}{\k_{1\perp}^2}.
  \label{eq:YM_Lorenz}
\end{align}
In this formula, $\k_{1\perp}\equiv \p_\perp-\k_{2\perp}$ (note that in this expression, the momentum $p^\mu$ is defined to be outgoing), and {the momentum $p^\mu$ is outgoing}.  The field $A_1^\mu$ is the field produced  by the source $\rho_1^{(0)}$,
\begin{align}
A_{1a}^\mu(k)=2\pi\delta^{\mu+}\delta(k^-)\frac{\rho_{1a}^{(0)}(\k_\perp)}{\k_\perp^2}.
\end{align}
Likewise, from the field $A_2^\mu$ produced  by the source $\rho_2^{(0)}$,
\begin{align}
A_{2a}^\mu(k)=2\pi\delta^{\mu-}\delta(k^+)\frac{\rho_{2a}^{(0)}(\k_\perp)}{\k_\perp^2},
\end{align}
we define two Wilson lines $U$ and $V$,
\begin{align}
  &U(\x_\perp)\equiv {\rm T}\exp\Big\{ig\int dx^+ A_2^-(x^+,\x_\perp)\Big\},\nonumber\\
  &V(\x_\perp)\equiv {\rm T}\exp\Big\{\frac{ig}{2}\int dx^+ A_2^-(x^+,\x_\perp)\Big\}.
\end{align}
(Note the factor $\tfrac{1}{2}$ in the exponential in $V$. This unusual feature is a peculiarity of the solution in this gauge.)

The momentum dependent vectors $L^\mu$ and $D^\mu$ read,
\begin{align}
&
L^+=-\frac{\k_{1\perp}^2}{p^-},\quad 
L^-=\frac{\k_{2\perp}^2-\p_\perp^2}{p^+},\quad
L^i= -2\k_{1\perp}^i,
\nonumber\\
&
D^+=2p^+,\quad
D^-=2\frac{\p_\perp^2}{p^+}-2p^-,\quad
D^i=2p^i.
\end{align}
They 
have the following
important properties,
\begin{align}
  &p\cdot L=p\cdot D=0 \qquad\mbox{(for any $p^\mu$)},\nonumber\\
  &D^\mu=2p^\mu,\quad L^2=-4\frac{\k_{1\perp}^2\k_{2\perp}^2}{\p_\perp^2}\quad \mbox{(for $p^2=0$)}.
\end{align}
$L^\mu$ is known as the Lipatov vertex. The fact that $D^\mu$ is proportional to $p^\mu$ when $p^\mu$ is on-shell ensures that the gauge specific Wilson line $V$ does not contribute when we contract $A^\mu$ with physical polarization vectors.

\section{Lipatov vertex in the spinor-helicity language}
\label{sec:lipatov}
In order to facilitate the comparison between the result obtained from the solution of Yang-Mills equations and the result obtained  via BCFW recursion, it is useful to express the Lipatov vertex $L^\mu$ in the spinor-helicity language. Firstly, note that $L^\mu$ can be written as
\begin{align}
L^\mu
=
-2k_1^\mu 
+\Big(2p^+-\frac{\k_{1\perp}^2}{p^-}\Big)v_1^\mu 
+\Big(\frac{\k_{2\perp}^2-\p_\perp^2}{p^+}\Big)v_2^\mu,
\end{align}
where $v_1^\mu\equiv \delta^{\mu+}$ and $v_2^\mu\equiv \delta^{\mu-}$. Next, we calculate the $2\times 2$ complex matrix $\overline{L}\equiv -L_\mu \overline{\sigma}^\mu$,
\begin{align}
\overline{L}
=
-2 \overline{K}_1
+
\Big(2p^+-\frac{\k_{1\perp}^2}{p^-}\Big)\big|\v_1\big>\big[\v_1\big|
+
\Big(\frac{\k_{2\perp}^2-\p_\perp^2}{p^+}\Big)\big|\v_2\big>\big[\v_2\big|.
\end{align}
Since $k_1^\mu$ is off-shell, we use eq.~(\ref{eq:off-shell-k}) with the auxiliary vector $q^\mu=v_2^\mu$ in order to obtain
\begin{align}
\overline{K}_1
=
p^+ \big|\v_1\big>\big[\v_1\big|
-
\kappa_1 \frac{\big|\v_1\big>\big[\v_2\big|}{\sbk{\v_1\v_2}}
+
\overline{\kappa}_1\frac{\big|\v_2\big>\big[\v_1\big|}{\abk{\v_2\v_1}}.
\end{align}

In order to contract $L^\mu$ with the polarization vectors, we use
\begin{align}
L\cdot \epsilon_+(\p_0,\q) = -\frac{\big<\q\big|\overline{L}\big|0\big]}{\sqrt{2}\abk{\q0}},\quad
L\cdot\epsilon_-(\p_0,\q) = - \frac{\big<0\big|\overline{L}\big|\q\big]}{\sqrt{2}\sbk{0\q}}.
\end{align}
Here, we also choose $q^\mu=v_2^\mu$. This leads to\footnote{Note that the outgoing momentum $p^\mu$ is the opposite of the incoming $p_0^\mu$ used elsewhere in this paper. Therefore $2p^-=-2p_0^- = -2p_0\cdot v_1 = -\abk{0\v_1}\sbk{0\v_1}$. }
\begin{align}
L\cdot\epsilon_+(\p_0,\q)
=\sqrt{2}\,\kappa_1\frac{\abk{\v_1\v_2}\,\big(\sbk{0\v_2}\abk{0\v_1}+\overline{\kappa}_1\sbk{\v_1\v_2}\big)}{\abk{0\v_2}\abk{0\v_1}\sbk{\v_1\v_2}}.
\end{align}
The numerator in this expression can be written in a simpler way as follows. Start from momentum conservation,
\begin{align}
p_0+k_1+k_2=0,
\end{align}
(note that here, $k_2$ is the total momentum coming from the sources $v_2^\mu$ if there are multiple scatterings)
which is equivalent to the following equation in the spinor-helicity language,
\begin{align}
\big|0\big>\big[0\big|
&+
\frac{q_1\cdot k_1}{q_1\cdot v_1}\big|\v_1\big>\big[\v_1\big|
-
\kappa_1\frac{\big|\v_1\big>\big[\q_1\big|}{\sbk{\v_1\q_1}}
+
\overline{\kappa}_1\frac{\big|\q_1\big>\big[\v_1\big|}{\abk{\q_1\v_1}}
\nonumber\\
&\quad
+
\frac{q_2\cdot k_2}{q_2\cdot v_2}\big|\v_2\big>\big[\v_2\big|
-
\kappa_2\frac{\big|\v_2\big>\big[\q_2\big|}{\sbk{\v_2\q_2}}
+
\overline{\kappa}_2\frac{\big|\q_2\big>\big[\v_2\big|}{\abk{\q_2\v_2}}
=
0.
\end{align}
Then, insert this identity inside $\big<\v_1\big|\cdots\big|\v_2\big]$. This gives
\begin{align}
\kappa_2\abk{\v_1\v_2} = \sbk{0\v_2}\abk{0\v_1}+\overline{\kappa}_1\sbk{\v_1\v_2},
\end{align}
and therefore
\begin{align}
L\cdot\epsilon_+(\p_0,\q)
=
\sqrt{2}\,\kappa_1\kappa_2\frac{\abk{\v_1\v_2}^2}{\abk{0\v_1}\abk{0\v_2}\sbk{\v_1\v_2}}
=
\frac{\kappa_1\kappa_2}{\sqrt{2}}\frac{\abk{\v_1\v_2}^3}{\abk{0\v_1}\abk{0\v_2}}
.
\label{eq:lipatov1}
\end{align}
(The last equality is obtained using $\abk{\v_1\v_2}\sbk{\v_1\v_2}=2v_1\cdot v_2=2$.)
Note that in this equation, $\kappa_2$ is the transverse component of the {\sl total} momentum transferred from sources $v_2^\mu$. Likewise, we have
\begin{align}
L\cdot\epsilon_-(\p_0,\q)
=
-\sqrt{2}\,\overline{\kappa}_1\overline{\kappa}_2
\frac{\sbk{\v_1\v_2}^2}{\sbk{0\v_1}\sbk{0\v_2}\abk{\v_1\v_2}}
=
-\frac{\overline{\kappa}_1\overline{\kappa}_2}{\sqrt{2}}
\frac{\sbk{\v_1\v_2}^3}{\sbk{0\v_1}\sbk{0\v_2}}
,
\label{eq:lipatov2}
\end{align}
which is consistent with $\epsilon_-^\mu(\p_0,\q)=-\epsilon_+^{\mu*}(\p_0,\q)$ and $\abk{\a\b}=\sbk{\a\b}^*$. Remarkably, the rather unilluminating components of the Lipatov vertex $L^\mu$ lead to very compact and symmetric formulas when contracted with polarization vectors and expressed in the spinor-helicity formalism.

\section{Calculation of the amplitudes ${\cal A}_{n+1}(0^*_{_E}1^*2^*...n^*)$}
\label{sec:A-eik} 
In the calculation of gluon production, we used a shift $e^\mu=v_2^\mu$, which can lead to a pole in an eikonal propagator $1/(p\cdot v_1)$. The residue of this pole factorizes into two smaller amplitudes that have an eikonal line as external leg. For instance, in the calculation of ${\cal A}_{n+2}(0^+1^*...(n+1)^*)$, with a shift $v_2^\mu$ on the legs $n$ and $n+1$, the pole coming from an eikonal propagator gives the term
\begin{align}
{\cal A}_3(0^+-\wh{I}^*_{_E}\wh{n+1}{}^*)\frac{1}{K_{_I}\cdot v_1}
{\cal A}_{n+1}(\wh{I}^*_{_E}1^*...\wh{n}{}^*),
\end{align}
where the subscript $E$ denotes an eikonal line. At the pole, the shifted momentum $\wh{K}_{_I}$ satisfies $\wh{K}_{_I}\cdot v_1=0$. This momentum is off-shell but has a null minus component, and it is therefore a valid argument for a source $\rho_1^{(0)}$. Thanks to this property, the amplitude ${\cal A}_3(0^+-\wh{I}^*_{_E}\wh{n+1}{}^*)$ on the left is in fact an ordinary 3-gluon amplitude
\begin{align}
{\cal A}_3(0^+-\wh{I}^*_{_E}\wh{n+1}{}^*)
=
{\cal A}_3(0^+-\wh{I}^*\wh{n+1}{}^*)
=
\frac{g}{\sqrt{2}\,\wh{\oka}_{_I}\wh{\oka}_{n+1}}
\frac{\abk{\v_1\v_2}^3}{\abk{\v_20}\abk{0\v_1}}
.
\end{align}
With a  shift $v_2^\mu$, the transverse component $\wh{\oka}_{n+1}$ is equal to $\oka_{n+1}$, and $\wh{\oka}_{_I}$ is determined by an explicit calculation of $\wh{K}_{_I}^\mu$ at the pole.

\begin{figure}[htbp]
\begin{center}
\includegraphics[width=25mm]{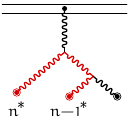}
\end{center}
\caption{\label{fig:bcfw-tree}Configuration where the shifted momenta (the propagators carrying a shifted momentum are shown in red) do not reach the eikonal line.}
\end{figure}

In order to calculate the factor ${\cal A}_{n+1}(\wh{I}^*_{_E}1^*...\wh{n}{}^*)$ on the right, let us start from the case $n=2$. In this case, a source $\rho_1^{(0)}$ undergoes a single scattering off a gluon produced by a  $\rho_2^{(0)}$. This amplitude is trivially calculated from the Feynman rules:
\setbox1\hbox to 10mm{\includegraphics[width=10mm]{A3-I12}}
\begin{equation}
{\cal A}_3(\wh{I}^*_{_E}1^*\wh{2}^*)
=\raise -7mm\box1
=
\frac{g}{\wh{p}_ 2^2}
=
-\frac{g}{\big|\kappa_2\big|^2}.
\label{eq:eik-simple}
\end{equation}	
Note that the results depends only on the unshifted momenta. This property is in fact general.
For higher values of $n$, we use BCFW recursion with a shift $v_2^\mu$ applied on the legs $n-1$ and $n$.
With a shift $v_2^\mu$ applied to two adjacent gluons with momenta that obey $p\cdot v_2=0$, gluon propagators cannot have a pole. In order to see this, consider as an example the situation illustrated in Figure \ref{fig:bcfw-tree}, in which the shift traverses only gluon propagators. All the  gluon momenta affected by the shift obey $p\cdot v_2=0$, which implies $\wh{p}^2=p^2-2zp\cdot v_2=p^2$. They are thus independent of $z$ and cannot have a pole.
The only non-zero contribution is obtained when the shift traverses the eikonal line, since the shifted eikonal propagator depends on $z$. This gives the following contribution,
\begin{align}
{\cal A}_{n+1}(\wh{I}{}^*_{_E}1^*...\wh{n}^*)
=
{\cal A}_n({I}{}^*_{_E}1^*...{n}^*)
=
\underbrace{
{\cal A}_3(I^*_{_E}-\wh{J}^*_{_E}\wh{n}{}^*)
}_{-g/\big|{\ka}_{n}\big|^2}
\frac{1}{K_{_J}\cdot v_1}
{\cal A}_{n}(\wh{J}{}^*_{_E}1^*...\wh{n-1}{}^*),
\end{align}
with $K_{_J}=K_{_I}+p_{n}$.
With a single source of type $\rho_1$, the external eikonal line $I$ must in fact start at the external leg $1$, where the source $\rho_1^{(0)}$ is attached.

The factor ${\cal A}_3(0^*_{_E}-\wh{J}^*_{_E}\wh{n-1}{}^*)$ on the left is given by a unique Feynman graph, like in eq.~(\ref{eq:eik-simple}). The factor on the right, ${\cal A}_{n}(\wh{J}{}^*_{_E}1^*...\wh{n-1}{}^*)$, depends on the momenta $\wh{K}_{_J}$ and $\wh{p}_{n-2}$ only through their sum $\wh{K}_{_J}+\wh{p}_{n-1}=K_{_J}+p_{n-1}$, and we have in fact
\begin{align}
{\cal A}_{n}(\wh{J}{}^*_{_E}1^*...\wh{n-1}{}^*)
=
{\cal A}_{n}({J}{}^*_{_E}1^*...(n-1)^*).
\end{align}
The recursion therefore reads
\begin{align}
{\cal A}_{n+1}(I^*_{_E}1^*...n^*)
=
-\frac{g}{\big|{\ka}_{n}\big|^2\,((K_{_I}+p_{n})\cdot v_1)}
{\cal A}_{n}((K_{_I}+p_{n})^*_{_E}1^*...(n-1)^*).
\end{align}
In this form, we see that each extra scattering off sources $\rho_2$ brings a simple factor (that includes a coupling constant, an eikonal propagator and the propagator of an off-shell gluon). This recursion can be solved easily, leading to
\begin{align}
{\cal A}_{n+1}(I^*_{_E}1^*...n^*)
=
\frac{(-g)^{n-1}}{\big|\ka_{n}\big|^2...\big|\ka_2\big|^2}
\frac{1}{(K_{_I}+p_{n})\cdot v_1}\cdots\frac{1}{(K_{_I}+p_{n}+...+p_3)\cdot v_1}.
\label{eq:eik-n-scatt}
\end{align}

\section{Integration over the longitudinal momenta}
\label{sec:long}
In this appendix, we perform the integrations over the longitudinal components of the source momenta, starting from eq.~(\ref{eq:scatt-final}), in order to bring this expression to the more familiar form for eikonal scattering. Firstly, recall that
\begin{align}
\rho_{2a}^{(0)}(p)=2\pi\delta(p^+)\rho_{2a}(\p_\perp).
\end{align}
(In order to keep the notation simple, we use the same symbols $\rho_{2}$ to denote the function that carries the transverse momentum dependence of the sources.)  $\rho_{2a}^{(0)}(p)$) is independent of $p^-$ because we assumed that the sources move at the speed of light and  are localized at $z=-t$ (a consequence of Lorentz contraction), and it is proportional to $\delta(p^+)$ because the sources have no $x^-$ dependence.  

The amplitude in eq.~(\ref{eq:scatt-final}) should be multiplied by
a factor 
\begin{align}
(2\pi)^4\delta(p_0+p_1+\sum_{i\ge 2}p_i)
=
(2\pi)^4\delta(p_0^++p_1^+)
\delta(p_0^-+p_1^-+\sum_{i\ge2}p_i^-)
\delta(\p_{0\perp}+\p_{1\perp}+\sum_{i\ge 2}\p_{i\perp})
\end{align}
in order to enforce momentum conservation. In eq.~(\ref{eq:scatt-final}), let us consider the term where $\sigma$ is the identity  (the integrations for the other $\sigma$'s are done in the same way). The only factors that depend on the longitudinal momenta of the sources are
\begin{align}
&2\pi\delta\big(p_0^-+p_1^-+\sum_{i= 2}^{n+1}p_i^-\big)\,
\frac{\big(\sbk{1\v_2}\abk{1\v_2}\big)^n}{(p_0+p_{2})^2\cdots(p_0+p_{2}+...+p_{n})^2}
\nonumber\\
&\qquad=
2\pi\delta\big(p_0^-+p_1^-+\sum_{i= 2}^{n+1}p_i^-\big)
\frac{(-1)^n 2p_0^+}{(p_{2}^-+\Omega_2)\cdots(p_{2}^-+...+p_{n}^-+\Omega_n)},
\end{align}
where we denote
\begin{align}
\Omega_2\equiv p_0^--\frac{(\p_{0\perp}+\p_{2\perp})^2}{2p_0^+},
\cdots,
\Omega_n\equiv p_0^--\frac{(\p_{0\perp}+\p_{2\perp}+...+\p_{n\perp})^2}{2p_0^+}.
\end{align}
Since we are solving the Yang-Mills equations with a retarded boundary condition, the internal gluon propagators should be retarded. This means that one should view the denominators in the previous equation as
\begin{align}
p_{2}^-+\Omega_2\to p_{2}^-+\Omega_2+i0^+,\cdots,
p_{2}^-+...+p_{n}^-+\Omega_n \to p_{2}^-+...+p_{n}^-+\Omega_n+i0^+.
\end{align}
Note also that the $\Omega_i$'s are independent of the longitudinal momenta of the sources, and that $p_{{n+1}}^-$ appears only in the delta function.  When we integrate over the longitudinal momenta of the sources, the $p_i^+$ integrals are trivial thanks to the fact that the sources are proportional to $2\pi\delta(p_i^+)$.

Thanks to the nested way in which the minus components appear in the integrand, one may think of integrating first over $p_{n+1}^-$ (using the delta function), then over $p_n^-$, followed by $p_{n-1}^-$,... However, if we proceed like this, we get seemingly divergent integrals. The integrals are in fact convergent, but some reorganization of the integrand is necessary to make this manifest. For this, we write
\begin{align}
\frac{2\pi\delta\big(p_0^-+p_1^-+\sum_{i= 2}^{n+1}p_i^-\big)}{(p_{2}^-+\Omega_2)\cdots(p_{2}^-+...+p_{n}^-+\Omega_n)}
&=-
\frac{(p_0^-\!+\!p_1^-)\,2\pi\delta\big(p_0^-+p_1^-+\sum_{i= 2}^{n+1}p_i^-\big)}{(p_{2}^-+\Omega_2)\cdots(p_{2}^-+...+\Omega_n)(p_2^-+...+p_n^-+p_{n+1}^-)}\nonumber\\
&=
-
\frac{(p_0^-\!+\!p_1^-)\,2\pi\delta\big(p_0^-+p_1^-+\sum_{i= 2}^{n+1}p_i^-\big)}{k_2^-(k_2^-+k_3^-)\cdots(k_2^-+...+k_n^-)},
\end{align}
where we have defined $k_i^-\equiv p_i^-+\omega_i$, $\omega_i\equiv \Omega_i-\Omega_{i-1}$, with $\Omega_1=\Omega_{n+1}=0$. The numerator is invariant under permutations of the indices $[2,n+1]$. Since the $k_i^-$ are dummy integration variables, we can symmetrize the expression by the following substitution,
\begin{align}
&\frac{1}{k_2^-(k_2^-+k_3^-)\cdots(k_2^-+...+k_n^-)}
\nonumber\\
&\qquad\qquad
\to
\frac{1}{n!}
\sum_{\sigma\in{\mathfrak{S}}(2...n+1)}
\frac{1}{k_{\sigma_2}^-(k_{\sigma_2}^-+k_{\sigma_3}^-)\cdots(k_{\sigma_2}^-+...+k_{\sigma_n}^-)}
=
\frac{1}{n!}
\frac{1}{k_2^-k_3^-\cdots k_{n+1}^-}.
\end{align}
Once the integrand is rewritten in this form, we use the delta function to integrate over $k_{n+1}^-$. This gives
\begin{align}
\frac{(-1)^{n+1}2p_0^+(p_0^-+p_1^-)}{n!}
\frac{1}{k_2^-+i0^+}\cdots \frac{1}{k_n^-+i0^+}
\frac{1}{-(p_0^-+p_1^-+k_2^-+...+k_n^-)+i0^+}.
\end{align}
Now, the denominator is quadratic in every variable $k_i^-$, and the integrals converge. Moreover, the two poles in $k_n^-$ are on opposite sides of the real axis. Using the theorem of residues,
the integral over $k_n^-$ of this expression gives
\begin{align}
\frac{i(-1)^{n+1}2p_0^+(p_0^-+p_1^-)}{n!}\frac{1}{k_2^-+i0^+}\cdots \frac{1}{k_{n-1}^-+i0^+}
\frac{1}{-(p_0^-+p_1^-+k_2^-+...+k_{n-1}^-)+i0^+}.
\end{align}
This pattern repeats with all the subsequent variables, and after having integrated over all the $k_i^-$, we eventually obtain
\begin{align}
\frac{i^{n-1}(-1)^{n}2p_0^+}{n!}
.
\end{align}
Note that this results does not depend on the ordering of the off-shell gluons $2$ to $n+1$ in the amplitude. The summation over the $n!$ permutations in eq.~(\ref{eq:scatt-final}) simply removes the $1/n!$ prefactor in the previous equation. Discarding an overall constant phase, the scattering amplitude integrated over the momenta of the sources reads
\begin{align}
&\int \prod_{i=2}^{n+1}\frac{d^4p_i}{(2\pi)^4}\rho_{2a_i}(p_i)
\,
(2\pi)^4\delta\big(p_0+p_1+\sum_{i\ge2}p_i\big)
\,
{\cal M}_{n+2}(0_{a_0}^+1_{a_1}^-2^*_{a_2}...(n+1)^*_{a_{n+1}})
\nonumber\\
&\qquad=
-2\pi\delta(p_0^++p_1^+)
2p_0^+ \,f^{a_0a_2d_1}f^{d_1a_3d_2}...f^{d_{n-1}a_{n+1}a_1}
\nonumber\\
&\qquad\quad\times g^n
\int \prod_{i=2}^{n+1}\frac{d^2\p_{i\perp}}{(2\pi)^3}\frac{\rho_{2a_i}^{(0)}(\p_{i\perp})}{\big|\ka_i\big|^2}
\,
(2\pi)^2\delta\big(\p_{0\perp}+\p_{1\perp}+\sum_{i\ge2}\p_{i\perp}\big).
\end{align}
There is no factor $1/n!$ in this final expression, but the external fields are ordered by the fact that their color indices $a_2,a_3,...$ appear in a definite order inside the product of structure constants. Up to the prefactor $-2\pi\delta(p_0^++p_1^+)
2p_0^+$ (common to all values of $n$), we have the $n$th term of a Wilson line in the external field produced by the sources $\rho_2$.

\section{Primakoff effect}
\label{sec:primakoff}
In this appendix, we study the QCD analogue of the Primakoff effect, namely the conversion of a gluon into two gluons via the interaction with an external field.

Let us consider the amplitude ${\cal A}_4(0^+1^+2^-3^*)$, where the off-shell gluon $3$ comes from a source of direction $v^\mu$. In order to calculate this amplitude, the shifts $e^\mu=\tfrac{1}{2}\big<2\big|\overline{\sigma}^\mu\big|\v\big]$ (on the lines $2$ and $3$) and $e^\mu=\tfrac{1}{2}\big<\v\big|\overline{\sigma}^\mu\big|0\big]$ (on the lines $0$ and $3$) are both allowed because they produce shifted polarization vectors that behave as $z^{-1}$ at large $|z|$. However, the shift $e^\mu=\tfrac{1}{2}\big<\v\big|\overline{\sigma}^\mu\big|0\big]$ produces a term where the off-shell gluon $3^*$ becomes an on-shell gluon of positive helicity; this term is zero because it is a $+++-$ amplitude. If we choose this shift, the amplitude  ${\cal A}_4(0^+1^+2^-3^*)$ has only one term
\begin{align}
 {\cal A}_4(0^+1^+2^-3^*)
 =
 {\cal A}_3(\wh{0}^+1^+-\wh{I}^-)
 \frac{i}{k_{_I}^2}
 {\cal A}_3(\wh{I}^+2^-\wh{3}^*),
\end{align}
where $K_{_I}\equiv p_0+p_1$. The factor on the left is a 3-point MHV amplitude given by
\begin{align}
{\cal A}_3(\wh{0}^+1^+-\wh{I}^-)
=
-g\sqrt{2}\frac{\sbk{\wh{0}1}^3}{\sbk{1-\wh{I}}\sbk{-\wh{I}\wh{0}}},
\end{align}
and the factor on the right is a scattering amplitude we have already encountered
\begin{align}
{\cal A}_3(\wh{I}^+2^-\wh{3}^*)
=
\frac{ig}{\wh{\overline{\kappa}}_3}
\frac{\abk{2\wh{\v}}^3}{\abk{\wh{\v}\wh{I}}\abk{\wh{I}2}}.
\end{align}
The spinor $\big|\wh{\v}\big>$ is equal to $\big|\v\big>$ since the direction of the source does not change in the shift.
The denominator of the intermediate gluon propagator is $K_{_I}^2=(p_0+p_1)^2=\abk{01}\sbk{01}$, and we have $\wh{\overline{\kappa}}_3=\overline{\kappa}_3$ for this shift. The value of $z$ at the pole 
is given by
\begin{align}
z_*=\frac{(p_0+p_1)^2}{2e\cdot(p_0+p_1)}
=\frac{\abk{01}}{\abk{\v1}}.
\end{align}
At this $z_*$, we have
\begin{align}
\big|\wh{0}\big>\big[\wh{0}\big|
=
\big|{0}\big>\big[{0}\big|-z_*\big|\v\big>\big[0\big|=\frac{\abk{\v0}}{\abk{\v1}}\,\big|1\big>\big[0\big|,
\label{eq:s4}
\end{align}
from which we read $\big|\wh{0}\big]=\big|0\big]$, $\big|\wh{0}\big>=\abk{\v0}\big|1\big>/\abk{\v1}$ (The Schouten identity is used to obtain the last equality in eq.~(\ref{eq:s4})). Likewise, we have
\begin{align}
\big|\wh{I}\big>\big[\wh{I}\big|
=
\big|\wh{0}\big>\big[\wh{0}\big|
+
\big|{1}\big>\big[{1}\big|
=
\big|1\big>\Big(\big[1\big|+\frac{\abk{\v0}}{\abk{\v1}}\big[0\big|\Big),
\end{align}
implying
\begin{align}
\big|\wh{I}\big>=\big|1\big>,\quad
\big|\wh{I}\big]
=
\big|1\big]+\frac{\abk{\v0}}{\abk{\v1}}\big|0\big].
\end{align}
By combining these building blocks, we obtain the following compact expression,
\begin{align}
{\cal A}_4(0^+1^+2^-3^*)
=
-\frac{g^2\sqrt{2}}{\overline{\kappa}_3}
\frac{\abk{2\v}^3}{\abk{\v0}\abk{01}\abk{12}}.
\label{eq:primakoff1}
\end{align}


\begin{thebibliography}{10}

\bibitem{Collins:1989gx}
John~C. Collins, Davison~E. Soper, and George~F. Sterman.
\newblock {Factorization of Hard Processes in QCD}.
\newblock {\em Adv. Ser. Direct. High Energy Phys.}, 5:1--91, 1989.

\bibitem{Catani:1990eg}
S.~Catani, M.~Ciafaloni, and F.~Hautmann.
\newblock {High-energy factorization and small x heavy flavor production}.
\newblock {\em Nucl. Phys. B}, 366:135--188, 1991.

\bibitem{Collins:1985ue}
John~C. Collins, Davison~E. Soper, and George~F. Sterman.
\newblock {Factorization for Short Distance Hadron - Hadron Scattering}.
\newblock {\em Nucl. Phys. B}, 261:104--142, 1985.

\bibitem{Bauer:2001yt}
Christian~W. Bauer, Dan Pirjol, and Iain~W. Stewart.
\newblock {Soft collinear factorization in effective field theory}.
\newblock {\em Phys. Rev. D}, 65:054022, 2002.

\bibitem{Gribov:1983ivg}
L.~V. Gribov, E.~M. Levin, and M.~G. Ryskin.
\newblock {Semihard Processes in QCD}.
\newblock {\em Phys. Rept.}, 100:1--150, 1983.

\bibitem{Mueller:1985wy}
Alfred~H. Mueller and Jian-wei Qiu.
\newblock {Gluon Recombination and Shadowing at Small Values of x}.
\newblock {\em Nucl. Phys. B}, 268:427--452, 1986.

\bibitem{Iancu:2000hn}
Edmond Iancu, Andrei Leonidov, and Larry~D. McLerran.
\newblock {Nonlinear gluon evolution in the color glass condensate. 1.}
\newblock {\em Nucl. Phys. A}, 692:583--645, 2001.

\bibitem{Ferreiro:2001qy}
Elena Ferreiro, Edmond Iancu, Andrei Leonidov, and Larry McLerran.
\newblock {Nonlinear gluon evolution in the color glass condensate. 2.}
\newblock {\em Nucl. Phys. A}, 703:489--538, 2002.

\bibitem{Iancu:2003xm}
Edmond Iancu and Raju Venugopalan.
\newblock {\em {The Color glass condensate and high-energy scattering in QCD}},
  pages 249--3363.
\newblock 3 2003.

\bibitem{Weigert:2005us}
Heribert Weigert.
\newblock {Evolution at small x(bj): The Color glass condensate}.
\newblock {\em Prog. Part. Nucl. Phys.}, 55:461--565, 2005.

\bibitem{Gelis:2010nm}
Francois Gelis, Edmond Iancu, Jamal Jalilian-Marian, and Raju Venugopalan.
\newblock {The Color Glass Condensate}.
\newblock {\em Ann. Rev. Nucl. Part. Sci.}, 60:463--489, 2010.

\bibitem{McLerran:1993ni}
Larry~D. McLerran and Raju Venugopalan.
\newblock {Computing quark and gluon distribution functions for very large
  nuclei}.
\newblock {\em Phys. Rev. D}, 49:2233--2241, 1994.

\bibitem{McLerran:1993ka}
Larry~D. McLerran and Raju Venugopalan.
\newblock {Gluon distribution functions for very large nuclei at small
  transverse momentum}.
\newblock {\em Phys. Rev. D}, 49:3352--3355, 1994.

\bibitem{McLerran:1994vd}
Larry~D. McLerran and Raju Venugopalan.
\newblock {Green's functions in the color field of a large nucleus}.
\newblock {\em Phys. Rev. D}, 50:2225--2233, 1994.

\bibitem{Gelis:2008rw}
Francois Gelis, Tuomas Lappi, and Raju Venugopalan.
\newblock {High energy factorization in nucleus-nucleus collisions}.
\newblock {\em Phys. Rev. D}, 78:054019, 2008.

\bibitem{Gelis:2008ad}
Francois Gelis, Tuomas Lappi, and Raju Venugopalan.
\newblock {High energy factorization in nucleus-nucleus collisions. II.
  Multigluon correlations}.
\newblock {\em Phys. Rev. D}, 78:054020, 2008.

\bibitem{Gelis:2008sz}
Francois Gelis, Tuomas Lappi, and Raju Venugopalan.
\newblock {High energy factorization in nucleus-nucleus collisions. 3. Long
  range rapidity correlations}.
\newblock {\em Phys. Rev. D}, 79:094017, 2009.

\bibitem{Gelis:2006yv}
Francois Gelis and Raju Venugopalan.
\newblock {Particle production in field theories coupled to strong external
  sources}.
\newblock {\em Nucl. Phys. A}, 776:135--171, 2006.

\bibitem{Gelis:2006cr}
Francois Gelis and Raju Venugopalan.
\newblock {Particle production in field theories coupled to strong external
  sources. II. Generating functions}.
\newblock {\em Nucl. Phys. A}, 779:177--196, 2006.

\bibitem{Krasnitz:1999wc}
Alex Krasnitz and Raju Venugopalan.
\newblock {The Initial energy density of gluons produced in very high-energy
  nuclear collisions}.
\newblock {\em Phys. Rev. Lett.}, 84:4309--4312, 2000.

\bibitem{Krasnitz:1998ns}
Alex Krasnitz and Raju Venugopalan.
\newblock {Nonperturbative computation of gluon minijet production in nuclear
  collisions at very high-energies}.
\newblock {\em Nucl. Phys. B}, 557:237, 1999.

\bibitem{Krasnitz:2000gz}
Alex Krasnitz and Raju Venugopalan.
\newblock {The Initial gluon multiplicity in heavy ion collisions}.
\newblock {\em Phys. Rev. Lett.}, 86:1717--1720, 2001.

\bibitem{Krasnitz:2001qu}
Alex Krasnitz, Yasushi Nara, and Raju Venugopalan.
\newblock {Coherent gluon production in very high-energy heavy ion collisions}.
\newblock {\em Phys. Rev. Lett.}, 87:192302, 2001.

\bibitem{Krasnitz:2002mn}
Alex Krasnitz, Yasushi Nara, and Raju Venugopalan.
\newblock {Gluon production in the color glass condensate model of collisions
  of ultrarelativistic finite nuclei}.
\newblock {\em Nucl. Phys. A}, 717:268--290, 2003.

\bibitem{Lappi:2003bi}
T.~Lappi.
\newblock {Production of gluons in the classical field model for heavy ion
  collisions}.
\newblock {\em Phys. Rev. C}, 67:054903, 2003.

\bibitem{Dumitru:2001ux}
Adrian Dumitru and Larry~D. McLerran.
\newblock {How protons shatter colored glass}.
\newblock {\em Nucl. Phys. A}, 700:492--508, 2002.

\bibitem{Blaizot:2004wu}
Jean~Paul Blaizot, Francois Gelis, and Raju Venugopalan.
\newblock {High-energy pA collisions in the color glass condensate approach. 1.
  Gluon production and the Cronin effect}.
\newblock {\em Nucl. Phys. A}, 743:13--56, 2004.

\bibitem{Gelis:2005pt}
Francois Gelis and Yacine Mehtar-Tani.
\newblock {Gluon propagation inside a high-energy nucleus}.
\newblock {\em Phys. Rev. D}, 73:034019, 2006.

\bibitem{Dixon:1996wi}
Lance~J. Dixon.
\newblock {Calculating scattering amplitudes efficiently}.
\newblock In {\em {Theoretical Advanced Study Institute in Elementary Particle
  Physics (TASI 95): QCD and Beyond}}, pages 539--584, 1 1996.

\bibitem{Elvang:2013cua}
Henriette Elvang and Yu-tin Huang.
\newblock {Scattering Amplitudes}.
\newblock 8 2013.

\bibitem{Parke:1986gb}
Stephen~J. Parke and T.~R. Taylor.
\newblock {An Amplitude for $n$ Gluon Scattering}.
\newblock {\em Phys. Rev. Lett.}, 56:2459, 1986.

\bibitem{Berends:1987me}
Frits~A. Berends and W.~T. Giele.
\newblock {Recursive Calculations for Processes with n Gluons}.
\newblock {\em Nucl. Phys. B}, 306:759--808, 1988.

\bibitem{Britto:2005fq}
Ruth Britto, Freddy Cachazo, Bo~Feng, and Edward Witten.
\newblock {Direct proof of tree-level recursion relation in Yang-Mills theory}.
\newblock {\em Phys. Rev. Lett.}, 94:181602, 2005.

\bibitem{vanHameren:2014iua}
A.~van Hameren.
\newblock {BCFW recursion for off-shell gluons}.
\newblock {\em JHEP}, 07:138, 2014.

\bibitem{vanHameren:2012if}
A.~van Hameren, P.~Kotko, and K.~Kutak.
\newblock {Helicity amplitudes for high-energy scattering}.
\newblock {\em JHEP}, 01:078, 2013.

\bibitem{Bury:2017jxo}
Marcin Bury, Andreas van Hameren, Hannes Jung, Krzysztof Kutak, Sebastian
  Sapeta, and Mirko Serino.
\newblock {Calculations with off-shell matrix elements, TMD parton densities
  and TMD parton showers}.
\newblock {\em Eur. Phys. J. C}, 78(2):137, 2018.

\bibitem{Bury:2015dla}
M.~Bury and A.~van Hameren.
\newblock {Numerical evaluation of multi-gluon amplitudes for High Energy
  Factorization}.
\newblock {\em Comput. Phys. Commun.}, 196:592--598, 2015.

\bibitem{vanHameren:2015bba}
Andreas van Hameren and Mirko Serino.
\newblock {BCFW recursion for TMD parton scattering}.
\newblock {\em JHEP}, 07:010, 2015.

\bibitem{Kutak:2016goj}
Krzysztof Kutak, Andreas Hameren, and Mirko Serino.
\newblock {QCD amplitudes with 2 initial spacelike legs via generalised BCFW
  recursion}.
\newblock {\em JHEP}, 02:009, 2017.

\bibitem{DelDuca:1999rs}
Vittorio Del~Duca, Lance~J. Dixon, and Fabio Maltoni.
\newblock {New color decompositions for gauge amplitudes at tree and loop
  level}.
\newblock {\em Nucl. Phys. B}, 571:51--70, 2000.

\end{thebibliography}
\end{document}